\documentclass{article}
\usepackage{amstex}
\usepackage{mapleenv}

\begin{document}
\begin{sloppypar}
\title{Algorithms for computer algebra calculations in spacetime\\
	I. The calculation of curvature}
\renewcommand{\thefootnote}{\fnsymbol{footnote}}
\author{Denis Pollney, Peter Musgrave, Kevin Santosuosso and 
	Kayll Lake\footnotemark\\
	{\small\em Deptartment of Physics, Queen's University, Kingston, Ontario}}
\date{22 January 1996}
\maketitle
\addtocounter{footnote}{1}
\makeatletter
\footnotetext{e-mail: {\tt pollney@astro.queensu.ca},
	{\tt musgrave@astro.queensu.ca},
	{\tt kevin@astro.queensu.ca}, and
	{\tt lake@astro.queensu.ca}}
\makeatother
\renewcommand{\thefootnote}{\arabic{footnote}}
\setcounter{footnote}{0}
%-----------------------------------------------------------------------------
\begin{abstract}
We examine the relative performance of algorithms for the calculation of 
curvature in spacetime. The classical coordinate component method is compared 
to two distinct versions of the Newman-Penrose tetrad approach for a variety 
of spacetimes, and distinct coordinates and tetrads for a given spacetime. 
Within the system GRTensorII, we find that there is no single preferred 
approach on the basis of speed. Rather, we find that the fastest algorithm is 
the one that minimizes the amount of time spent on simplification. This means 
that arguments concerning the theoretical superiority of an algorithm need 
not translate into superior performance when applied to a specific spacetime 
calculation. In all cases it is the global simplification strategy which is of
paramount importance. An appropriate simplification strategy can change an 
untractable problem into one which can be solved essentially instantaneously.
\end{abstract}
%-----------------------------------------------------------------------------
\section{Introduction}
 It is fair to say that tetrad calculations are generally considered superior 
to classical coordinate methods for the calculation of curvature in spacetime. 
Experiments by Campbell and Wainwright \cite{campbell/wainwright:1977}
now dating back many years showed that tetrad methods are faster than 
coordinate methods by factors of $2 \sim 4$. 
Even larger factors have been obtained by MacCallum \cite{maccallum:1989}
in a Euclidean context. 
Within the well known system SHEEP, for example, the advice to the 
beginner is to always use frame versions of the metric (e.g. MacCallum
and Skea \cite{maccallum/skea:1994} p. 23). On the commercial side, within the
system MACSYMA2 \cite{macsyma} the demonstration 
CTENSOR4 begins with an explanation that ``frame fields'' (orthonormal bases) 
allow the computations to run much more quickly. The demonstration calculates 
the bases components of the Ricci tensor for the Kerr-Newman spacetime in 
Boyer-Lindquist coordinates, and is a good place to begin our discussion.\\ 

In Table 1 we have reproduced this demonstration within the system 
GRTensorII \cite{musgrave/pollney/lake:1994}\cite{grhome} running under
MapleV Release 3 \cite{maple}, and have included the 
calculation of the Weyl tensor.
The Table demonstrates some interesting 
properties. The theoretical advantage of the frame approach is clearly 
demonstrated in the Boyer-Lindquist coordinates (Column BKN). However, 
under the elementary coordinate transformation $u=a \cos\theta$ this advantage 
fails to deliver superior performance (Column BKNU). The importance of
strategic application of simplification at intermediate steps is
illustrated in Column BKNS. For this test simplification of components
has been carried out only after the components of the Ricci and Weyl tensors
are calculated. It is worth noting that without some optimization in the 
simplification strategy (e.g. post-calculation simplification only) this 
calculation cannot be executed in MapleV on a 32 bit machine.

%------------------------------------------------------------------------------
\begin{table}
\centering
\begin{tabular}{lrrr}
\hline\hline
Calculation             & BKN        & BKNU  & BKNS \\ \hline
$R_{(a)(b)}$            & 6.7              & 5.6         & 30.0 \\
$C_{(a)(b)(c)(d)}$      & 1.4              & 0.8         &  9.5 \\ \hline
Total                   & 8.1              & 6.4         & 39.5 \\ \hline
                        &                  &             &      \\
$R_{ab}$                & 8.4              & 2.1         & 18.9 \\
$C_{abcd}$              & 21.1             & 4.3         & 67.5 \\ \hline
Total                   & 29.5             & 6.4         & 86.4 \\ 
\hline\hline
\end{tabular}\\
\vspace{\baselineskip}
\begin{minipage}{.9\linewidth}
Table~1: Average CPU time\footnotemark~in seconds for the calculation and 
simplification of the bases components of the Ricci and Weyl tensors
($R_{(a)(b)}$ and $C_{(a)(b)(c)(d)}$) compared to the same for the coordinate 
components ($R_{ab}$ and  $C_{abcd}$). BKN refers to the Kerr-Newman
spacetime in Boyer-Lindquist coordinates, and BKNU the same but with the 
transformation $u=a\cos\theta$. For both BKN and BKNU the 
simplification procedures have been structured for optimum 
performance\footnotemark $^,$\footnotemark. For the Column BKNS the tetrad
components of BKN are used, but simplification procedures are not applied
to intermediate calculation steps, only to the final results.
\addtocounter{footnote}{-2}
\end{minipage}\\
\end{table}
%------------------------------------------------------------------------------
Clearly one could dismiss the findings in Table 1 if the implementation 
of the bases algorithms in GRTensorII were particularly inefficient compared 
to coordinate methods. We do not believe this to be the case
\cite{lake/musgrave/pollney:1995b}. Rather, as we attempt to show in what
follows, we believe that Table 1 reflects the 
fact that bases methods are not fundamentally superior to classical 
coordinate methods. We find that the most important criterion for speed is 
the minimization of the time spent on simplification. This underlines the 
importance of the choice of coordinates or tetrad in a computer algebra 
calculation and, more importantly,  points out the fact that the user must be 
able to select the style of simplification which is most appropriate for the 
particular problem.
\footnotetext{All times are in seconds as
returned by the MapleV {\tt status} function and are
the average of four runs on a Sun Sparc 5 (see Section \ref{sec:2.4}). The
maximum deviation from the average is less than 5\% for times exceeding 
2 seconds and about 10\% for shorter times.}
\addtocounter{footnote}{1}
\footnotetext{We consider a worksheet (a sequence of calculation 
and simplification procedures) to be optimized when the execution time
has reached a minimum.}
\addtocounter{footnote}{1}
\footnotetext{Due to their length, the complete text
of worksheets used for these tests have not been included in this
report (except for an example in Appendix \ref{app:C}), however they
have been made publicly available \cite{worksheets}.}

%------------------------------------------------------------------------------
\section{Protocol for comparisons}
%------------------------------------------------------------------------------
\subsection{Choice of algorithms}\label{sec:2.1}
 Within the framework of tetrad methods, the formalism of Newman and Penrose 
\cite{newman/penrose:1962} has proven most useful for calculations in spacetime
(see e.g. \cite{campbell/wainwright:1977}). Some 
of the earliest applications of computer algebra to relativity stressed the 
efficiency of this formalism (e.g. \cite{campbell/wainwright:1977}). 
McLenaghan \cite{mclenaghan:1994} (see also Allen et al. \cite{allenetal:1994})
has emphasized two distinct approaches within this formalism.
These are distinguished as the methods of {\em Cartan} and {\em Debever} in 
\cite{mclenaghan:1994}, and as Methods A and B in \cite{allenetal:1994}, a 
notation which we adopt here. The methods are outlined in
\cite{mclenaghan:1994} and \cite{allenetal:1994} with references and we do not
repeat this material here\footnote{
The curvature component $\Phi_{12}$ is consistently incorrect in
\cite{mclenaghan:1994} and \cite{allenetal:1994}. In particular, the
coefficients of $\mu\tau$ and $\nu\sigma$ are -1 and 1 respectively,
{\em not} -2 and +2. This error is also present in the {\tt debever} package
in MapleV Releases 2 and 3.}.
We simply note that Method A uses the definitions of Newman and 
Penrose explicitly, while Method B essentially uses definitions constructed 
so as to avoid inversion of coordinate indicies. 

In this paper we compare these two approaches to classical coordinate methods
(suitably optimized).

\subsection{Basis for comparison}
The null tetrad formalism is sufficiently distinct from the classical 
coordinate approach that a basis for the comparison of the two methods is 
not clearly defined. In the classical approach the ``curvature'' of a spacetime
is usually considered evaluated when the coordinate components of the Ricci 
and Weyl tensors have been evaluated. In the Newman-Penrose (NP)
formalism it is the tetrad
components of these tensors (the $\Phi$s and $\Psi$s) that display the 
``curvature''. The complication that arises in a comparison of such different
methods is the fact that the natural output of each method is distinct. Now
given the coordinate components, and the null tetrad, the tetrad components 
follow in the usual way \cite{newman/penrose:1962}. One could then form a
basis for comparison 
by defining the ``curvature'' as the tetrad components of the Ricci and Weyl 
tensors . This is the comparison used in \cite{campbell/wainwright:1977}.
Naturally, this puts the 
classical component method at a disadvantage since the extra sums involved 
are not a natural part of the method. In this paper we have tried to cover 
all possibilities by having both NP approaches output the tetrad components 
of the Ricci and Weyl tensors, and the coordinate approach output both the 
coordinate and tetrad components of these tensors.

\subsection{Choice of spacetimes}
 To compare the performance of algorithms for spacetime calculations it is 
essential that a variety of spacetimes be considered, and that within a given
spacetime different tetrads (coordinates) be examined. Campbell and 
Wainwright \cite{campbell/wainwright:1977} chose to examine the spacetimes of 
Griffiths \cite{griffiths:1975}, Lewis-Papapetrou \cite{ernst:1968},
Bondi \cite{bondi/vanderburg/metzner:1962}, and Debever 
\cite{debever/mclenaghan/tariq:1979}.
More recently, the examination by Allen et al. \cite{allenetal:1994} 
(which is a comparison of the NP approaches) included these spacetimes 
(with a more general form of the Debever metric, the 
Debever-McLenaghan-Tariq metric \cite{debever/mclenaghan/tariq:1979}) and
also the plane wave, 2x2 decomposable, and static spherically symmetric
spacetimes \cite{krameretal:1980}. We have found these last 
three spacetimes to be too simple  since the associated calculation times
are too short to form a reliable basis for comparison. They also include 
one form of the Kerr-Newman metric\footnote{
Campbell and Wainwright \cite{campbell/wainwright:1977} also examined the
Kerr-Newman metric. However, their description involves intermediate hand
simplification in this case and does not include comparison to the component
method. In this Section we take the position that a fair comparison of computer
algorithm efficiency can only be carried out if no hand simplification of
tensor components is permitted.} and a
general tetrad (their Case 9). Here we examine the Kerr-Newman
metric in a variety of forms. We do not include the tetrad 9 given in 
\cite{allenetal:1994} since it does not conform to the requirements of a null
tetrad in the Newman-Penrose formalism \cite{krameretal:1980}.

\subsection{Method of comparison}\label{sec:2.4}
 The comparisons were made by way of 26 re-executable Maple worksheets, and 
the calculations were performed with GRTensorII
\cite{musgrave/pollney/lake:1994} under MapleV 
Release 3 \cite{maple} (in the X-Windows interface) with patchlevel 3 on a 
Sun SPARC5 running SunOS 4.1.4
and equipped with a 75MHz CPU and 64Mb of RAM\footnote{This configuration was
chosen for its reliability and reproducibility of CPU times, not its speed. 
By way of a comparison, a Pentium 133 is about 1.5 times faster, however the 
DOS/Windows implementation of MapleV Release 3  reports only integer CPU
times.}. 

The Maple worksheets and associated input files used in these tests are
summarized 
in Appendix \ref{app:A}. The associated tetrads, and line elements are shown
in Appendix \ref{app:B}. The reproduction of tetrads is prone to errors, and
references \cite{campbell/wainwright:1977}, \cite{mclenaghan:1994} and 
\cite{allenetal:1994} all contain misprints in the tetrad components. Appendix 
\ref{app:B} has been produced directly from the input files and so an error
constitutes not a misprint, but an actual error in the input which would 
invalidate our conclusions in the case concerned. The worksheets are 
available for detailed examination and execution \cite{worksheets}.

The worksheets are constructed in the following way. Either the contravariant
or covariant tetrad is loaded into GRTensorII from an input file. The choice 
determines NP algorithm to be used for the tetrad part of the calculation. 
The tetrad components of the Ricci and Weyl tensors are then evaluated, 
usually in simplified form\footnote{A tensor component is considered to
be fully simplified when its size, as measured in Maple `words', reaches
a minimum.} (see Section \ref{sec:3}). The metric is then 
generated from the tetrad, and the covariant coordinate components of the
Ricci and Weyl tensors are evaluated and simplified when necessary (again
see Section \ref{sec:3}). We have been careful to 
ensure that the metric, though generated from the tetrad, is presented in 
optimal form. Assuming that the tetrads are given, the worksheets then 
evaluate the tetrad components from the covariant coordinate 
components. These are then simplified to the exact form of the tetrad 
calculation.

We believe that each worksheet has been fully optimized\footnotemark[2].
That is, the procedures (e.g. precalculation of the spin coefficients) and
simplification procedures (type and order of simplification) at each step 
of the calculation have been constructed so as to present each approach in 
its best performance mode. It should be pointed out that partial optimization 
of a worksheet is straightforward (see Appendix \ref{app:D}), but the full 
optimization of a worksheet is a somewhat involved task. Full optimization is
necessary if a real comparison of approaches is to be given.

A sample worksheet is given in Appendix \ref{app:C} along with the
associated input file. In 
GRTensorII the input files contain no information beyond the components of the
basis or metric. In particular, no simplification information is contained in
the input file. The simplification strategy used is read from the worksheet.

%------------------------------------------------------------------------------
\section{Comparisons}\label{sec:3}

Summarized in Table 2 are the results of our tests\footnotemark[1]
and the results
available from the previous tests in \cite{campbell/wainwright:1977} and 
\cite{allenetal:1994}. Before we discuss the 
comparisons it is appropriate to emphasize the importance of simplification 
procedures. The fastest calculation in the Kerr-Newman metric is the 
covariant NP tetrad calculation (Method B) in Row 9. If even only part of the 
simplification strategy is altered (e.g. the background simplification
procedure used before the components are more fully simplified) the 
execution time can increase by a factor well over two orders of magnitude. 
The global simplification strategy is of paramount importance.

\begin{table}
\vspace*{-2\baselineskip}
\centering
\begin{tabular}{rlrrrrrrrrrrrrr}\hline\hline
1. & Coord's & $A$    & $B$   & $C$     & $D$    & $E$     & $F$     & $A/D$
   & \cite{allenetal:1994}  & $A/B$   & $A/C$   & 
\cite{campbell/wainwright:1977}   & $D/E$   & $D/F$ \\ \hline
 2. & {\bf Grif}     & 2.0  & 1.9 &  2.9  & 1.6 & 2.3  &  3.5  & 1.25  & 1.65
  & 1.05  & 0.69  & 0.39  & 0.70  & 0.46 \\
 3. & {\bf L-P}     & 1.9  & 2.8 &  4.9  & 2.6 & 2.9  &  6.1  & 0.73  & 1.44
  & 0.68  & 0.39  & 0.40  & 0.90  & 0.43 \\
 4. & {\bf Bondi1}   & 3.5  & 6.2 & 10.5  & 4.4 & 6.2  & 10.4  & 0.80  & 0.83
  & 0.56  & 0.33  & 0.27  & 0.71  & 0.42 \\
 5. & {\bf Bondi2}   & 1.5  & 1.8 &  4.7  & 2.6 & 1.8  &  4.6  & 0.58  &  ?  
  & 0.83  & 0.32  &  ?    & 1.44  & 0.57 \\
 6. & {\bf Deb}      &11.1  &13.3 & 39.8  &24.0 &13.6  & 39.7  & 0.46  &      
 & 0.84  & 0.28  &       & 1.76  & 0.60 \\
 7. & {\bf DMT1}     &10.9  &13.5 & 40.6  &22.3 &13.5  & 40.7  & 0.49  &      
 & 0.81  & 0.27  &       & 1.65  & 0.55 \\
 8. & {\bf DMT2}     &35.3  &32.1 &109.0  &55.0 &32.6  &105.0  & 0.64  &0.93  
 & 1.10  & 0.32  &       & 1.69  & 0.52 \\
 9. & {\bf KN-Euc1}  &22.7  &13.1 & 19.9  & 6.3 &13.7  & 20.0  & 3.60  &5.80 
  & 1.68  & 1.14  &       & 0.46  & 0.32 \\
10. & {\bf KN-Euc2}  &54.4  &13.2 & 17.2  &24.4 &13.1  & 16.2  & 2.23  &      
 & 4.12  & 3.16  &       & 1.86  & 1.51 \\
11. & {\bf KN-BL1}   &26.6  &27.5 & 38.4  &30.0 &30.0  & 34.6  & 0.89  &      
 & 0.97  & 0.69  &       & 1.00  & 0.87 \\
12. & {\bf KN-EF1}   &32.5  &22.7 & 30.2  & 7.8 &22.7  & 25.4  & 4.17  &      
 & 1.43  & 1.08  &       & 0.34  & 0.31 \\
13. & {\bf KN-BL2}   &12.6  & 6.7 & 14.9  &18.2 & 9.1  & 15.8  & 0.69  &      
 & 1.88  & 0.84  &       & 2.00  & 1.15 \\
14. & {\bf KN-EF2}   &41.0  &22.9 & 38.9  &20.8 &22.8  & 26.0  & 1.97  &      
 & 1.79  & 1.05  &       & 0.91  & 0.80 \\
\hline\hline
\end{tabular}\\
\vspace{\baselineskip}
\begin{minipage}{.9\linewidth}
Table 2: CPU times and comparisons for optimized calculations. Columns $A$
through $F$ give CPU times in seconds\protect\footnotemark[1].\\
\vspace*{-2\baselineskip}
\begin{center}
\begin{tabular}{cl}
Column & \\
$A$ & -- total CPU time to generate and simplify the curvature 
components ($\Phi$s and $\Psi$s)\\
&  \hspace*{5mm}from a contravariant tetrad using the standard NP approach 
(`Method A').\\
$B$ & -- time for the simplified (covariant) coordinate components of
the Ricci and Weyl\\
&  \hspace*{5mm}tensors for the metric generated from the tetrad $A$.\\
$C$ & -- time (including $B$) to generate and simplify the tetrad
curvature components from\\
&  \hspace*{5mm}the coordinate components calculated in $B$, given the tetrad 
$A$.\\
$D$ & -- the time to generate and simplify the curvature components from
a covariant\\
&  \hspace*{5mm}tetrad using the modified NP approach of \cite{allenetal:1994} 
(`Method B'). \\
$E$ & -- time for the simplified (covariant) coordinate components of the
Ricci and Weyl\\
&  \hspace*{5mm}tensors for the metric generated from the covariant tetrad 
$D$.\\
$F$ & -- time (including $E$) to generate and simplify the tetrad curvature
components from\\
&  \hspace*{5mm}the coordinate components calculated in $E$ given the tetrad 
$D$.\\
\end{tabular}
\end{center}
The differences in Columns $B$ and $E$ are due to differences in the exact
form of the metric generated from the tetrad. The spacetimes and form of the
output are distinguished by the following abbreviations:
\begin{center}
\begin{tabular}{ll}
{\bf Grif} & Griffiths metric \cite{griffiths:1975}. Output in factored form.\\
{\bf L-P} & Lewis-Papapetrou metric \cite{ernst:1968}. Output in factored 
	form.\\
{\bf Bondi1} & Bondi metric \cite{bondi/vanderburg/metzner:1962}. Output in 
	factored form.\\
{\bf Bondi2} & Bondi metric. Output in expanded form.\\
{\bf Deb} & Debever metric \cite{debever/mclenaghan/tariq:1979}. Output in
	normal form (does not simplify further).\\
{\bf DMT1} & Debever-McLenaghan-Tariq metric 
	\cite{debever/mclenaghan/tariq:1979}. Output in normal form (does not\\
	& simplify further).\\
{\bf DMT2} & Debever-McLenaghan-Tariq metric in general form
	\cite{debever/mclenaghan/tariq:1979}. Output in normal\\
	& form (does not simplify further).\\
{\bf KN-Euc1} & Kerr-Newman metric \cite{allenetal:1994}. Output in factored
	form.\\
{\bf KN-Euc2} & Modified form of 9.  Output in factored form.\\
{\bf KN-BL1} & Kerr-Newman metric in Boyer-Lindquist coordinates
	\cite{boyer/lindquist:1967}. Output in\\
	& factored form.\\
{\bf KN-EF1} & Kerr-Newman metric in advanced Eddington-Finkelstein coordinates
	\cite{mtw}.\\
	& Output in factored form.\\
{\bf KN-BL2} & Kerr-Newman metric in modified Boyer-Lindquist coordinates
	using\\
	& $u=a\cos\theta$. Output in factored form.\\
{\bf KN-EF2} & Modified form of 12. Output in factored form.\\
\end{tabular}
\end{center}
\end{minipage}\\
\vspace{\baselineskip}
\end{table}

A number of interesting points emerge from Table 2.

{\bf i)} It is appropriate to begin by comparing our results with previously 
published tests.  Starting with the work of Campbell and Wainwright 
\cite{campbell/wainwright:1977}, although the exact
form of their output is unavailable, we observe a notable agreement for
the ratio $A/C$ and their results for the Lewis-Papapetrou and Bondi metrics
as shown in Table 2. For the Griffiths metric our component times 
are somewhat faster\footnote{The fourth component of $m^a$ for this metric
given in \cite{campbell/wainwright:1977} is wrong. We 
believe this to be a misprint which would not alter the time reported.}.
Quite naturally, a concern at the time was the storage requirements for the  
calculations. They report storage requirements for the component method a 
factor $2 \sim 5$ times that of the tetrad approach. Whereas storage is no 
longer the concern that it once was, we note that we have observed storage 
requirements for the component calculations only about $1.1 \sim 1.5$ times
that of the tetrad method.

The paper by Allen et al. \cite{allenetal:1994} is concerned with the ratio
$A/D$, 
that is, the relative performance of the two null tetrad methods. The exact 
form of their output is unavailable. In general we find that the performance
of the {\em Debever} approach (Method B) is overestimated in 
\cite{allenetal:1994}\footnote{The errors in $\Phi_{12}$ within the MapleV
{\tt debever} package affects only their result for the 
Debever-McLenaghan-Tariq metric (See footnote 4).}.
Although the central thesis of \cite{allenetal:1994} is the superiority of the 
{\em Debever} approach, this rests principally on their analysis of the 
Kerr-Newman metric. It is clear from Table 2 that this superior
performance is tetrad (coordinate) dependent.

{\bf ii)} For metrics of a general type (Rows 3 through 8 in Table
2) we find that the standard NP approach is faster than the
alternative proposed in \cite{mclenaghan:1994} and \cite{allenetal:1994}.
We find that this superior performance is not uniformly maintained 
if the general functions are replaced by specific ones. We have used the 
Kerr-Newman metric as an example, and as can be seen from Rows 9 through 14 
of Table 2, the relative performance of the two tetrad approaches
is highly dependent on the tetrads (coordinates). However, whereas the
{\em Debever} approach can significantly outpace the NP approach (Rows 9 
and 12), it is never far behind.

{\bf iii)} For metrics of a general type (Rows 3 through 8 in Table
2) both tetrad approaches are superior to a calculation of the
tetrad components from the coordinate components. This is exactly as one
would expect. The extra sums involved slow the coordinate approach down 
(compare Columns $B$ and $C$ as well as $E$ and $F$ ). Again, however, this 
superior performance is not 
uniformly maintained if the general functions are replaced by specific ones. 
Columns $A/C$ and $D/F$ for Rows 9 through 14 show that the classical approach 
can rival the tetrad approaches even for a calculation of the tetrad 
components.

{\bf iv)} It could be argued that the natural output of the coordinate
approach, as 
regards the calculation of curvature,  is simply the coordinate components of 
the Ricci and Weyl tensors. For metrics of a general type the standard NP 
approach retains its superiority over the coordinate calculation (Column 
$A/B$). This is not true for the modified tetrad approach (Column $D/E$). 
Columns $A/B$ and $D/E$ indicate that at least in the Kerr-Newman metric the 
calculation of the coordinate components of the Ricci and Weyl tensors is 
usually faster than the calculation of the tetrad components.

%------------------------------------------------------------------------------
\section{Discussion}
Our central conclusion is that the best algorithm, as regards speed, for the
computer algebra calculation of curvature in spacetime is the one that 
minimizes the amount of time spent on simplification. This underlines the 
importance of the careful choice of coordinates or tetrad in a computer 
calculation and, more importantly, demonstrates that the user must be able to
style the global simplification strategy in a manner most appropriate for the 
particular problem being studied. An appropriate simplification strategy 
can change an untractable problem into one which can be solved essentially 
instantaneously\footnote{Some simplification strategies appropriate to 
GRTensorII in MapleV are given in Appendix \ref{app:D}.}. Our comparisons 
(Table 2) indicate that there is no
uniformly superior algorithm. In the development of these comparisons we
observed that the differences between procedures optimized with respect to the
global simplification strategy for the procedure were less than the variations
within a given procedure for different simplification strategies.

%------------------------------------------------------------------------------
\begin{table}
\centering
\begin{tabular}{lrrrr}\hline\hline
Calculation        & {\bf Mix} & {\bf Mix1} & {\bf Mix2} & {\bf Mix3}\\ \hline
$R_{(a)(b)}$       & 8.7       & 7.5        & 4.2        & 2.5 \\
$C_{(a)(b)(c)(d)}$ & 8.5       & 1.1        & 7.4        & 1.1 \\ \hline
Total              &17.2       & 8.6        &11.6        & 3.6 \\ \hline
                   &           &            &            &     \\
$R_{ab}$           & 8.5       & 8.5        & 8.5        & 8.5 \\
$C_{abcd}$         &52.5       &13.7        &52.5        &13.7 \\ \hline
Total              &61.0       &22.2        &61.0        &22.2 \\
\hline\hline
\end{tabular}\\
\vspace{\baselineskip}
\begin{minipage}{.9\linewidth}
Table 3: Average CPU time in seconds for the calculation and 
simplification of the bases components of the Ricci and Weyl tensors
($R_{(a)(b)}$ and $C_{(a)(b)(c)(d)}$) compared to the same for the coordinate
components ($R_{ab}$ and $C_{abcd}$). {\bf Mix} refers to the standard
1-forms with trigonometric functions \cite{mtw2} and a time dependent basis
inner product. The same inner product is used for {\bf Mix1} but the 
trigonometric functions have been transformed away. For {\bf Mix2} a
constant basis inner product has been used,
and in {\bf Mix3} the trigonometric functions have been transformed away. 
In all cases the simplification procedures have been structured for optimum 
performance.
\end{minipage}
\end{table}
%------------------------------------------------------------------------------
Although we have not found any algorithm to be uniformly superior, there
certainly are cases where the appropriate algorithm stands out. A case in
point is the mixmaster spacetime \cite{mtw}\footnote{This
example was suggested to us by Prof. C. W. Misner.}.
Here an appropriate choice of basis
removes any angular dependence in the bases components of both the Ricci and
Weyl tensors. The tetrad approach (not a null tetrad in this case) would
certainly be expected to outperform a coordinate calculation in this case.
This is confirmed in Table 3. We have considered both a constant
basis inner product (for {\bf Mix1} and {\bf Mix3}) and a time dependent one
(for {\bf Mix} and {\bf Mix2}). The coordinate transformations are simply 
$\Theta=\cos\theta$ and 
$\Psi=\sin\psi$. It is clear that the classical coordinate calculation is no 
match for the basis approach exactly as one would guess. Interestingly, it is 
the Weyl tensor calculation that improves under coordinate transformation, 
and the Ricci tensor under change in the basis inner product.

\subsection*{Acknowledgements}
It is a pleasure to thank Malcolm MacCallum for a number of useful
discussions and suggestions. This work was supported in part by grants
(to KL) from the Natural Sciences and Engineering Research Council of Canada
and the Advisory Research Committee of Queen's University. PM and DP would
like to thank the Ministry of Colleges and Universities of Ontario for
financial support through Ontario Graduate Scholarships.
%------------------------------------------------------------------------------

%------------------------------------------------------------------------------
% Appendices
%------------------------------------------------------------------------------
\appendix
%------------------------------------------------------------------------------
\pagebreak
\section{Simplification strategies for GRTensorII in MapleV}\label{app:D}
%------------------------------------------------------------------------------

We have found that the global simplification strategy is of central 
importance for the computer algebra calculation of curvature in spacetime. 
Summarized here are some general rules of simplification appropriate to 
GRTensorII running under MapleV. Although the general philosophy described
here is more widely applicable, our comments are specific to the system used
and so we have relegated this information to an Appendix.

\begin{enumerate}
\item There should be some default simplification procedure applied to every 
step of a calculation. Failure to do this can make a simple calculation 
intractable. The Maple routine normal is a good starting point for the 
default. If the calculation involves exponentials (e.g. the Bondi metric) 
the routine expand may be more appropriate. Only if very general functions 
are involved is it appropriate to consider no default simplification. In
general cases this may be the optimal choice.

\item For both coordinate and tetrad calculations the removal of trigonometric
and like functions via elementary coordinate transformations will often improve
performance.

\item Simplification of the metric tensor or tetrad components before further
calculation will improve performance. 

\item Precalculation and further simplification of the spin coefficients (and 
their complex conjugates) will improve performance only in more complicated
cases. The same holds for the Christoffel symbols in the coordinate approach.

\item For further simplification after an object has been calculated, the Maple
routine simplify is seldom a good first choice. The routine expand followed 
by factor is often more appropriate. If the situation is sufficiently general
(e.g. the Debever-McLenaghan-Tariq metric) there will be no further 
simplification if normal has been used as default.

\item If complicated functions are involved, it can be advantageous to 
substitute the explicit forms of the functions after a more general 
calculation is completed.

\item When a calculation is proceeding slowly, it should be halted, the 
simplification strategy altered, and the worksheet reexecuted.
\end{enumerate}

For most situations these general rules will give adequate performance, and
reduce the calculation of curvature for even complex spacetimes to an 
essentially trivial exercise. Usually, it is the answer that is of interest  
and not the fact that the simplification strategy is optimal. When optimal
simplification strategies are the prime concern ( as in this paper ) the
problem is more involved because of the large number of simplification
procedures available\footnote{GRTensorII provides a menu of 12 distinct 
commonly used predefined 
simplification routines with the ability to introduce customized constraints 
and simplification routines. Any single parameter routine can be applied 
with {\tt gralter}, and any multiple parameter routine can be applied with the
command {\tt grmap}.} and the size of the resultant parameter space to be
explored.

%------------------------------------------------------------------------------
\pagebreak
\section{Worksheets and input files}\label{app:A}
%------------------------------------------------------------------------------
This Appendix is simply a bookkeeping 
operation which relates the input files necessary to run the worksheets. The 
first three worksheets are associated with Table 1, the last four with Table 3,
and all the rest with Table 2. The worksheet {\tt appc.ms} generates Appendix
\ref{app:B} from the input files. Abbreviations which appear twice in the 
last column of this Table refer to contravariant and covariant tetrad
components, respectively.\\

\noindent Worksheets used to generate Table 1: \\
\begin{center}
\begin{tabular}{|l|l|l|}\hline\hline
Test		& Worksheet	& Input file \\ \hline
BKN 		& {\mbox{\tt bkn1.ms}} & {\mbox{\tt bkn.mpl}} \\
BKNU		& {\mbox{\tt bkn2.ms}} & {\mbox{\tt bknu.mpl}} \\
BKNS		& {\mbox{\tt bkn3.ms}} & {\mbox{\tt bkn.mpl}} \\
\hline\hline
\end{tabular}
\end{center} \vspace{\baselineskip}

\noindent Worksheets used to generate Table 2: \\
\begin{center}
\begin{tabular}{|l|l|l|l|l|}\hline\hline
 & \multicolumn{2}{c|}{Columns $A$--$C$}  	
 & \multicolumn{2}{c|}{Columns $D$--$F$} \\ \cline{2-5}
Test		& Worksheet		& Input file
		& Worksheet		& Input file \\ \hline
{\bf Grif}	& {\mbox{\tt griff1.ms}}          & {\mbox{\tt npupgrif.mpl}}
		& {\mbox{\tt griff2.ms}}          & {\mbox{\tt npdngrif.mpl}}\\
{\bf L-P} 	& {\mbox{\tt lewis1.ms}}          & {\mbox{\tt npuplew.mpl}}
		& {\mbox{\tt lewis2.ms}}          & {\mbox{\tt npdnlew.mpl}}\\
{\bf Bondi1}	& {\mbox{\tt bondi1.ms}}          & {\mbox{\tt npupbon.mpl}}
		& {\mbox{\tt bondi2.ms}}          & {\mbox{\tt npdnbon.mpl}}\\
{\bf Bondi2}	& {\mbox{\tt bondi1.ms}}          & {\mbox{\tt npupbon.mpl}}
		& {\mbox{\tt bondi2.ms}}          & {\mbox{\tt npdnbon.mpl}}\\
{\bf Deb}	& {\mbox{\tt deb1.ms}}            & {\mbox{\tt npupdeb.mpl}}
		& {\mbox{\tt deb2.ms}}            & {\mbox{\tt npdndeb.mpl}}\\
{\bf DMT1}	& {\mbox{\tt dmts1.ms}}           & {\mbox{\tt npupsdmt.mpl}}
		& {\mbox{\tt dmts2.ms}}           & {\mbox{\tt npdnsdmt.mpl}}\\
{\bf DMT2}	& {\mbox{\tt dmt1.ms}}            & {\mbox{\tt npupdmt.mpl}}
		& {\mbox{\tt dmt2.ms}}            & {\mbox{\tt npdndmt.mpl}}\\
{\bf KN-Euc1}	& {\mbox{\tt kna1.ms}}            & {\mbox{\tt npupkn1.mpl}}
		& {\mbox{\tt kna2.ms}}            & {\mbox{\tt npdnkn1.mpl}}\\
{\bf KN-Euc2}	& {\mbox{\tt knb1.ms}}            & {\mbox{\tt npupkn2.mpl}}
		& {\mbox{\tt knb2.ms}}            & {\mbox{\tt npdnkn2.mpl}}\\
{\bf KN-BL1}	& {\mbox{\tt knc1.ms}}            & {\mbox{\tt npupkn3.mpl}}
		& {\mbox{\tt knc2.ms}}            & {\mbox{\tt npdnkn3.mpl}}\\
{\bf KN-EF1}	& {\mbox{\tt knd1.ms}}            & {\mbox{\tt npupkn4.mpl}}
		& {\mbox{\tt knd2.ms}}            & {\mbox{\tt npdnkn4.mpl}}\\
{\bf KN-BL2}	& {\mbox{\tt kne1.ms}}            & {\mbox{\tt npupkn5.mpl}}
		& {\mbox{\tt kne2.ms}}            & {\mbox{\tt npdnkn5.mpl}}\\
{\bf KN-EF2}	& {\mbox{\tt knf1.ms}}            & {\mbox{\tt npupkn6.mpl}}
		& {\mbox{\tt knf2.ms}}            & {\mbox{\tt npdnkn6.mpl}}\\
\hline\hline
\end{tabular}
\end{center} \vspace{\baselineskip}

\noindent Worksheets used to generate Table 3: \\
\begin{center}
\begin{tabular}{|l|l|l|}\hline\hline
Test		& Worksheet	& Input file \\ \hline
{\bf Mix}	& {\mbox{\tt mixmr.ms}}  & {\mbox{\tt mix.mpl}} \\
{\bf Mix1}	& {\mbox{\tt mixmr1.ms}} & {\mbox{\tt mix1.mpl}} \\
{\bf Mix2}	& {\mbox{\tt mixmr2.ms}} & {\mbox{\tt mix2.mpl}} \\
{\bf Mix3}	& {\mbox{\tt mixmr3.ms}} & {\mbox{\tt mix3.mpl}} \\
\hline\hline
\end{tabular}
\end{center}
%==============================================================================
% Appendix B: Tetrads and metrics produced from the input files
%==============================================================================
\pagebreak
%------------------------------------------------------------------------------
\section{Tetrads and metrics produced from the input files}\label{app:B}
%------------------------------------------------------------------------------
The following sections list tetrads and metrics used as inputs for the
tests listed in Tables 1--3. This list has been produced directly from
the input files which were used in the tests (listed in Appendix
\ref{app:A}) and converted to \LaTeX~using MapleV's \LaTeX~output
facility with only minor modifications to improve readability.

%==============================================================================
\subsection{Kerr-Newman ``frame field'' (Table 1)}
%==============================================================================
For this set of tests (whose output is given in Table 1), the Kerr-Newman
spacetime is described by a frame consisting of four independent covariant
vector fields whose inner product is the constant matrix
$\eta_{(a)(b)} = \text{diag}(-1,-1,-1,1)$. The basis vectors and
corresponding line element for each case are given below.\\

\begin{tabbing}
%== Kerr-Newman (Boyer-Lindquist coordinates) =================================
\rule{\linewidth}{.5mm}\\
{\large Spacetime: Kerr-Newman (Boyer-Lindquist coordinates) 
(BKN and BKNS) \cite{boyer/lindquist:1967} } \\
\rule{\linewidth}{.05mm}\\
Input file\hspace{10mm} \= Frame \\
\rule[.5\baselineskip]{\linewidth}{.05mm}\\
{\tt bkn.mpl}, \> ${ \omega_{{1a}}}  = $\=$ \left[ 
  \! \,{\displaystyle \frac {\sqrt {{r}
  ^{2} + {a}^{2}\,{\rm cos}(\,{ \theta}\,)^{2}}}{\sqrt {{r}^{2} - 2
  \,{m}\,{r} + {a}^{2} + {Q}^{2}}}}\, , 0\, , 0\, , 0\, \!  \right]$ \\
\> $ {{ \omega}_{{2a}}}  =$\>$  \left[ \! \,0\, ,
  \sqrt {{r}^{2} + {a}^{2}\, {\rm cos}(\,{ \theta}\,)^{2}}\, , 0\, , 0\, \! 
  \right] $ \\
\> $ {{ \omega}_{{3a}}}  =$\>$ \left[ \! \,0\, , 0\, , 
  {\displaystyle \frac {(\,{
  r}^{2} + {a}^{2}\,)\,{\rm sin}(\,{ \theta}\,)}{\sqrt {{r}^{2} + {
  a}^{2}\,{\rm cos}(\,{ \theta}\,)^{2}}}}\, , - \,{\displaystyle 
  \frac {{a}\,{\rm sin}(\,{ \theta}\,)}{\sqrt {{r}^{2} + {a}^{2}\,
  {\rm cos}(\,{ \theta}\,)^{2}}}}\, \!  \right] $ \\
\> ${{ \omega}_{{4a}}}  = $\>$ \left[ \! \,0 , \,0 , \, - \,{\displaystyle 
  \frac {{a}\,\sqrt {{r}^{2} - 2\,{m}\,{r} + {a}^{2} + {Q}^{2}}\,
  {\rm sin}(\,{ \theta}\,)^{2}}{\sqrt {{r}^{2} + {a}^{2}\,{\rm cos}
  (\,{ \theta}\,)^{2}}}}\, , {\displaystyle \frac {\sqrt {{r}^{2} - 2
  \,{m}\,{r} + {a}^{2} + {Q}^{2}}}{\sqrt {{r}^{2} + {a}^{2}\,{\rm 
  cos}(\,{ \theta}\,)^{2}}}}\, \!  \right] $ \\
\vspace{1mm} \> Corresponding line element: \\
\> ${\it \:ds}^{2}= - \,{\displaystyle \frac { \left( \! \,{
  r}^{2} + {a}^{2}\,{\rm cos}(\,{ \theta}\,)^{2}\, \!  \right) \,
  {\it \:d}\,{r}^{{\it 2\:}}}{{r}^{2} - 2\,{m}\,{r} + {a}^{2} + {Q}
  ^{2}}} +  \left( \! \, - {r}^{2} - {a}^{2}\,{\rm cos}(\,{ \theta}
  \,)^{2}\, \!  \right) \,{\it \:d}\,{ \theta}^{{\it 2\:}} + (\, - 
  1 + {\rm cos}(\,{ \theta}\,)\,)$ \\
\> $(\,{\rm cos}(\,{ \theta}\,) + 1\,) \left( {\vrule 
  height0.44em width0em depth0.44em} \right. \! \! {r}^{2}\,{a}^{2}
  \,{\rm cos}(\,{ \theta}\,)^{2} - 2\,{a}^{2}\,{m}\,{r}\,{\rm cos}(
  \,{ \theta}\,)^{2} + {a}^{4}\,{\rm cos}(\,{ \theta}\,)^{2} $\\
\> $ \mbox{} + {a}^{2}\,{Q}^{2}\,{\rm cos}(\,{ \theta}\,)^{2} + 2
  \,{a}^{2}\,{m}\,{r} + {r}^{4} - {a}^{2}\,{Q}^{2} + {r}^{2}\,{a}^{
  2} \! \! \left. {\vrule height0.44em width0em depth0.44em}
  \right) {\it \:d}\,{ \phi}^{{\it 2\:}} \left/ {\vrule 
  height0.44em width0em depth0.44em} \right. \! \!  \left( \! \,{r}
  ^{2} + {a}^{2}\,{\rm cos}(\,{ \theta}\,)^{2}\, \!  \right)$  \\
\> $\mbox{} + 2\,{\displaystyle \frac {{a}\,(\, - 1 + {\rm cos}(
  \,{ \theta}\,)\,)\,(\,{\rm cos}(\,{ \theta}\,) + 1\,)\,(\, - 2\,{
  m}\,{r} + {Q}^{2}\,)\,{\it \:d}\,{ \phi}^{{\:}}\,{\it d\:}\,{t}^{
  {\:}}}{{r}^{2} + {a}^{2}\,{\rm cos}(\,{ \theta}\,)^{2}}}$ \\
\> $\mbox{} + {\displaystyle \frac { \left( \! \,{a}^{2}\,{\rm 
  cos}(\,{ \theta}\,)^{2} + {r}^{2} - 2\,{m}\,{r} + {Q}^{2}\, \! 
  \right) \,{\it \:d}\,{t}^{{\it 2\:}}}{{r}^{2} + {a}^{2}\,{\rm 
  cos}(\,{ \theta}\,)^{2}}}$ \\
%
%== Kerr-Newman (Boyer-Lindquist coordinates) =================================
\rule{\linewidth}{.5mm}\\
{\large Spacetime: Kerr-Newman (Boyer-Lindquist coordinates, $u=a\cos\theta$) 
(BKNU) \cite{boyer/lindquist:1967} } \\
\rule{\linewidth}{.05mm}\\
Input file\hspace{10mm} \= Frame \\
\rule[.5\baselineskip]{\linewidth}{.05mm}\\
{\tt bknu.mpl} \> $ {{ \omega}_{{1a}}} = $ \> $ \left[ \! \,
  {\displaystyle \frac {\sqrt {{r}
  ^{2} + {u}^{2}}}{\sqrt {{r}^{2} - 2\,{m}\,{r} + {a}^{2} + {Q}^{2}
  }}}\, , 0\, , 0\, , 0\, \!  \right] $ \\
\> ${{ \omega}_{{2a}}} = $ \> $ \left[ \! \,0 , \, - \,{\displaystyle \frac {
  \sqrt {{r}^{2} + {u}^{2}}}{\sqrt {{a}^{2} - {u}^{2}}}}\, , 0\, , 0\,
  \!  \right] $ \\
\> $ {{ \omega}_{{3a}}} = $ \> $ \left[ \! \,0 , \,0 , \,
  {\displaystyle \frac {(\,{
  r}^{2} + {a}^{2}\,)\,\sqrt {{a}^{2} - {u}^{2}}}{\sqrt {{r}^{2} + 
  {u}^{2}}\,{a}}}\,  , - \,{\displaystyle \frac {\sqrt {{a}^{2} - {u}
  ^{2}}}{\sqrt {{r}^{2} + {u}^{2}}}}\, \!  \right] $ \\
\> ${{ \omega}_{{4a}}} = $ \> $ \left[ \! \,0\, , 0\, ,  - \,{\displaystyle 
  \frac {\sqrt {{r}^{2} - 2\,{m}\,{r} + {a}^{2} + {Q}^{2}}\,(\,{a}
  ^{2} - {u}^{2}\,)}{\sqrt {{r}^{2} + {u}^{2}}\,{a}}}\, , 
  {\displaystyle \frac {\sqrt {{r}^{2} - 2\,{m}\,{r} + {a}^{2} + {Q
  }^{2}}}{\sqrt {{r}^{2} + {u}^{2}}}}\, \!  \right] $ \\ 
\vspace{1mm} \> Corresponding line element: \\
\> ${\it \:ds}^{2}= - \,{\displaystyle \frac {(\,{r}^{2} + {
  u}^{2}\,)\,{\it \:d}\,{r}^{{\it 2\:}}}{{r}^{2} - 2\,{m}\,{r} + {a
  }^{2} + {Q}^{2}}} - {\displaystyle \frac {(\,{r}^{2} + {u}^{2}\,)
  \,{\it \:d}\,{u}^{{\it 2\:}}}{(\,{a} - {u}\,)\,(\,{a} + {u}\,)}}
  + (\,{a} - {u}\,)\,(\,{a} + {u}\,)$ \\
\>\> $(\, - {r}^{4} - {r}^{2}\,{a}^{2} - {r}^{2}\,{u}^{2} - 2\,{a}
  ^{2}\,{m}\,{r} + 2\,{m}\,{r}\,{u}^{2} - {a}^{2}\,{u}^{2} + {a}^{2
  }\,{Q}^{2} - {Q}^{2}\,{u}^{2}\,)\,{\it \:d}\,{ \phi}^{{\it 2\:}}
  \left/ {\vrule height0.37em width0em depth0.37em} \right. \! \! 
  ($ \\
\>\> $(\,{r}^{2} + {u}^{2}\,)\,{a}^{2}\,)\mbox{} - 2\,
  {\displaystyle \frac {(\,{a} - {u}\,)\,(\,{a} + {u}\,)\,(\, - 2\,
  {m}\,{r} + {Q}^{2}\,)\,{\it \:d}\,{ \phi}^{{\:}}\,{\it d\:}\,{t}
  ^{{\:}}}{(\,{r}^{2} + {u}^{2}\,)\,{a}}}
  + {\displaystyle \frac {(\,{u}^{2} + {r}^{2} - 2\,{m
  }\,{r} + {Q}^{2}\,)\,{\it \:d}\,{t}^{{\it 2\:}}}{{r}^{2} + {u}^{2
  }}}$\\ 
\rule{\linewidth}{.5mm}\\
\end{tabbing}
%==============================================================================
\subsection{Null tetrads (Table 2)}
%==============================================================================
This section lists the set of null tetrads for the test cases used
to generate Table 2. For each spacetime, four forms of input were
used. The times listed in Column $A$ of Table 2 are obtained using
a contravariant null tetrad as input. In Columns $B$ and $C$ the
metric is calculated from this tetrad and used for subsequent
calculation. In Column $D$ a covariant tetrad is loaded, and in 
$E$ and $F$, its corresponding metric is used. Though the metric of 
column $E$ is, of course, equivalent to that of $B$, there are often
differences in representation which can in principle alter calculation
times (though in practice we have found this effect to be minimal).
Thus the line elements used as inputs for each tests are listed along
with the tetrad used to calculate them.

\begin{tabbing}
%== Griffiths =================================================================
\rule{\linewidth}{.5mm}\\
{\large Spacetime: Griffiths ({\bf Grif}) \cite{griffiths:1975}}\\
\rule{\linewidth}{.05mm}\\
Input file\hspace{10mm} \= Contravariant tetrad \\
\rule[.5\baselineskip]{\linewidth}{.05mm}\\
{\tt npupgrif.mpl} \> ${l}^{{a}}=[\,0 , \,1 , \,0 , \,0\,]$ \\
\> ${n}^{{a}} =$ \= $[\,1 , \,0 , \,0 , \,0\,]$ \\
\> ${m}^{{a}} =$ \> $\left[ {\vrule 
  height0.79em width0em depth0.79em} \right. \! \! {\displaystyle 
  \frac {1}{3}}\,{\rm e}^{(\, - 2\,{I}\,{a}\,(\,{u} + {v}\,)\,)}\,
  \sqrt {3}\,(\, - {I} + 2\,{a}\,{v} + {a}\,{y}\,) , \,{\displaystyle 
  \frac {1}{3}}\,{\rm e}^{(\, - 2\,{I}\,{a}\,(\,{u} + {v}\,)\,)}\,
  \sqrt {3}\,(\, - {I} + 2\,{a}\,{u} + {a}\,{y}\,) , $ \\
\> \>${\displaystyle \frac {1}{3}}\,{\rm e}^{(\, - 2\,{I}\,{a}\,(
  \,{u} + {v}\,)\,)}\,\sqrt {3}\, , {\displaystyle \frac {1}{3}}\,
  {\rm e}^{(\, - 2\,{I}\,{a}\,(\,{u} + {v}\,)\,)}\,\sqrt {3}\,(\,{I
  } - 2\,{a}\,{u} - 2\,{a}\,{v}\,) \! \! \left. {\vrule 
  height0.79em width0em depth0.79em} \right] \mbox{\hspace{74pt}}$ \\
\> $\overline{m}^a =$ \> $ \left[ {\vrule 
  height0.79em width0em depth0.79em} \right. \! \! {\displaystyle 
  \frac {1}{3}}\,{\rm e}^{(\,2\,{I}\,{a}\,(\,{u} + {v}\,)\,)}\,
  \sqrt {3}\,(\,{I} + 2\,{a}\,{v} + {a}\,{y}\,) , \,{\displaystyle 
  \frac {1}{3}}\,{\rm e}^{(\,2\,{I}\,{a}\,(\,{u} + {v}\,)\,)}\,
  \sqrt {3}\,(\,{I} + 2\,{a}\,{u} + {a}\,{y}\,) ,  $ \\
\> \> ${\displaystyle \frac {1}{3}}\,{\rm e}^{(\,2\,{I}\,{a}\,(\,{u
  } + {v}\,)\,)}\,\sqrt {3}\, , {\displaystyle \frac {1}{3}}\,{\rm e}
  ^{(\,2\,{I}\,{a}\,(\,{u} + {v}\,)\,)}\,\sqrt {3}\,(\, - {I} - 2\,
  {a}\,{u} - 2\,{a}\,{v}\,) \! \! \left. {\vrule 
  height0.79em width0em depth0.79em} \right] \mbox{\hspace{79pt}}$ \\
\vspace{\baselineskip}
\vspace{1mm} \> Corresponding line element: \\
\> ${\it \:ds}^{2}=2\,{\it \:d}\,{u}^{{\:}}\,{\it d\:}\,{v}
  ^{{\:}} - 2\,{a}\,(\,{y} - 2\,{v}\,)\,{\it \:d}\,{u}^{{\:}}\,
  {\it d\:}\,{x}^{{\:}} + 2\,{\it \:d}\,{u}^{{\:}}\,{\it d\:}\,{y}
  ^{{\:}} + 2\,{a}\,(\,2\,{u} - {y}\,)\,{\it \:d}\,{v}^{{\:}}\,
  {\it d\:}\,{x}^{{\:}} $ \\
\> $\mbox{} + 2\,{\it \:d}\,{v}^{{\:}}\,{\it d\:}\,{y}^{{\:}} +  
  \left( \! \, - 4\,{a}^{2}\,{y}\,{u} + 2\,{a}^{2}
  \,{y}^{2} - 4\,{a}^{2}\,{v}\,{u} - 4\,{a}^{2}\,{v}\,{y} - 6\,{a}
  ^{2}\,{u}^{2} - {\displaystyle \frac {3}{2}} - 6\,{a}^{2}\,{v}^{2
  }\, \!  \right) \,{\it \:d}\,{x}^{{\it 2\:}}$ \\
\> $\mbox{} - 2\,{a}\,(\,{u} + 2\,{y} + {v}\,)\,{\it \:d}\,{x}^{
  {\:}}\,{\it d\:}\,{y}^{{\:}} + {\displaystyle \frac {1}{2}}\,
  {\it \:d}\,{y}^{{\it 2\:}}\mbox{\hspace{160pt}}$ \\
\rule{\linewidth}{.05mm}\\
%------------------------------------------------------------------------------
Input file\hspace{10mm} \> Covariant tetrad \\
\rule[.5\baselineskip]{\linewidth}{.05mm}\\
{\tt npdngrif.mpl} \> ${{l}_{{a}}} = $ \> $[\,1 , \,0 , \,{a}\,(\,2\,{u} - 
  {y}\,) , \,1\,]$ \\
\> ${{n}_{{a}}}= $ \> $[\,0 , \,1 , \,{a}\,(\,2\,{v} - {y}\,) , \,1\,]$ \\
\> ${{m}_{{a}}}= $ \> $\left[ \! \,0 , \,0 , \,-\,{\displaystyle \frac {1}{2}}
  \,\sqrt {3}\,{\rm e}^{(\, - 2\,{I}\,{a}\,(\,{u} + {v}\,)\,)}\,(\,
  1 + 2\,{I}\,{a}\,(\,{u} + {v}\,)\,)\,  , -\,{\displaystyle \frac {1
  }{2}}\,{I}\,\sqrt {3}\,{\rm e}^{(\, - 2\,{I}\,{a}\,(\,{u} + {v}\,
  )\,)}\, \!  \right] $ \\
\> $\overline{m}_a = $ \> $\left[ \! \,0 , \,0 , \,-\,{\displaystyle \frac {
  1}{2}}\,\sqrt {3}\,{\rm e}^{(\,2\,{I}\,{a}\,(\,{u} + {v}\,)\,)}\,
  (\,1 - 2\,{I}\,{a}\,(\,{u} + {v}\,)\,)\, , {\displaystyle \frac {1}{
  2}}\,{I}\,\sqrt {3}\,{\rm e}^{(\,2\,{I}\,{a}\,(\,{u} + {v}\,)\,)}
  \, \!  \right] $\\
\vspace{1mm} \> Corresponding line element: \\
\> $ \lefteqn{{\it \:ds}^{2}=2\,{\it \:d}\,{u}^{{\:}}\,{\it d\:}\,{v}
  ^{{\:}} - 2\,{a}\,(\,{y} - 2\,{v}\,)\,{\it \:d}\,{u}^{{\:}}\,
  {\it d\:}\,{x}^{{\:}} + 2\,{\it \:d}\,{u}^{{\:}}\,{\it d\:}\,{y}
  ^{{\:}} + 2\,{a}\,(\,2\,{u} - {y}\,)\,{\it \:d}\,{v}^{{\:}}\,
  {\it d\:}\,{x}^{{\:}}}$ \\
\> $ \mbox{} + 2\,{\it \:d}\,{v}^{{\:}}\,{\it d\:}\,{y}^{{\:}}
  +  \left( \! \, - 4\,{a}^{2}\,{y}\,{u} + 2\,{a}^{2}
  \,{y}^{2} - 4\,{a}^{2}\,{v}\,{u} - 4\,{a}^{2}\,{v}\,{y} - 6\,{a}
  ^{2}\,{u}^{2} - {\displaystyle \frac {3}{2}} - 6\,{a}^{2}\,{v}^{2
  }\, \!  \right) \,{\it \:d}\,{x}^{{\it 2\:}}$ \\
\> $ \mbox{} - 2\,{a}\,(\,{u} + 2\,{y} + {v}\,)\,{\it \:d}\,{x}^{
  {\:}}\,{\it d\:}\,{y}^{{\:}} + {\displaystyle \frac {1}{2}}\,
  {\it \:d}\,{y}^{{\it 2\:}}\mbox{\hspace{160pt}}$ \\
%
%== Lewis-Papapetrou ==========================================================
\rule{\linewidth}{.5mm}\\
{\large Spacetime: Lewis-Papapetrou ({\bf L-P}) \cite{ernst:1968}}\\
\rule{\linewidth}{.05mm}\\
Input file\hspace{10mm} \> Contravariant tetrad \\
\rule[.5\baselineskip]{\linewidth}{.05mm}\\
{\tt npuplew.mpl} \> ${l}^{{a}}=$ \> $ \left[ \! \, - \,{\displaystyle 
  \frac {1}{2}}\,\sqrt {
  2}\, \left( \! \, - {\rm e}^{(\, - {\rm s}(\,{x}, {y}\,)\,)} + 
  {\displaystyle \frac {{\rm w}(\,{x}, {y}\,)\,{\rm e}^{{\rm s}(\,{
  x}, {y}\,)}}{{\rm r}(\,{x}, {y}\,)}}\, \!  \right) \, , 0 , \,0 , \,
  {\displaystyle \frac {1}{2}}\,{\displaystyle \frac {\sqrt {2}}{
  {\rm r}(\,{x}, {y}\,)\,{\rm e}^{(\, - {\rm s}(\,{x}, {y}\,)\,)}}}
  \, \!  \right] $ \\
\> ${n}^{{a}}=$ \> $ \left[ \! \,{\displaystyle \frac {1}{2}}\,\sqrt {2}\,
  \left( \! \,{\rm e}^{(\, - {\rm s}(\,{x}, {y}\,)\,)} + 
  {\displaystyle \frac {{\rm w}(\,{x}, {y}\,)\,{\rm e}^{{\rm s}(\,{
  x}, {y}\,)}}{{\rm r}(\,{x}, {y}\,)}}\, \!  \right) ,  \,0 , \,0 , \, 
  -\,{\displaystyle \frac {1}{2}}\,{\displaystyle \frac {\sqrt {2}}{
  {\rm r}(\,{x}, {y}\,)\,{\rm e}^{(\, - {\rm s}(\,{x}, {y}\,)\,)}}}
  \, \!  \right] $ \\
\> ${m}^{{a}}=$ \> $ \left[ \! \,0 , \,{\displaystyle \frac {\sqrt {2}}{{\rm e
  }^{(\,{\rm k}(\,{x}, {y}\,) - {\rm s}(\,{x}, {y}\,)\,)}}} , \,0 , \,0\,
  \!  \right] $ \\
\> $\overline{m}^a = $ \> $ \left[ \! \,0 , \,0 , \,
  {\displaystyle \frac {\sqrt {
  2}}{{\rm e}^{(\,{\rm k}(\,{x}, {y}\,) - {\rm s}(\,{x}, {y}\,)\,)}
  }} , \,0\, \!  \right] $ \\
\vspace{1mm} \> Corresponding line element: \\
\> $\lefteqn{{\it \:ds}^{2}={\displaystyle \frac {{\it \:d}\,{t}^{
  {\it 2\:}}}{(\,{\rm e}^{(\, - {\rm s}(\,{x}, {y}\,)\,)}\,)^{2}}}
  + 2\,{\displaystyle \frac {{\rm w}(\,{x}, {y}\,)\,{\rm e}^{{\rm 
  s}(\,{x}, {y}\,)}\,{\it \:d}\,{t}^{{\:}}\,{\it d\:}\,{z}^{{\:}}}{
  {\rm e}^{(\, - {\rm s}(\,{x}, {y}\,)\,)}}} - (\,{\rm e}^{(\,{\rm 
  k}(\,{x}, {y}\,) - {\rm s}(\,{x}, {y}\,)\,)}\,)^{2}\,{\it \:d}\,{
  x}^{{\:}}\,{\it d\:}\,{y}^{{\:}}}$ \\
\> $\mbox{} - (\,{\rm e}^{(\, - {\rm s}(\,{x}, {y}\,)\,)}\,{\rm 
  r}(\,{x}, {y}\,) - {\rm w}(\,{x}, {y}\,)\,{\rm e}^{{\rm s}(\,{x}
  , {y}\,)}\,)\,(\,{\rm e}^{(\, - {\rm s}(\,{x}, {y}\,)\,)}\,{\rm r
  }(\,{x}, {y}\,) + {\rm w}(\,{x}, {y}\,)\,{\rm e}^{{\rm s}(\,{x}, 
  {y}\,)}\,)\,{\it \:d}\,{z}^{{\it 2\:}}$ \\
\rule{\linewidth}{.05mm}\\
%------------------------------------------------------------------------------
Input file\hspace{10mm} \> Covariant tetrad \\
\rule[.5\baselineskip]{\linewidth}{.05mm}\\
{\tt npdnlew.mpl} \> ${{l}_{{a}}} = $ \> $\left[ \! \,{\displaystyle 
  \frac {1}{2}}\,{\rm e}^{
  {\rm s}(\,{x}, {y}\,)}\,\sqrt {2}\, , 0\, , 0\, ,  - \,{\displaystyle 
  \frac {1}{2}}\,{\rm r}(\,{x}, {y}\,)\,{\rm e}^{(\, - {\rm s}(\,{x
  }, {y}\,)\,)}\,\sqrt {2} + {\displaystyle \frac {1}{2}}\,{\rm w}(
  \,{x}, {y}\,)\,{\rm e}^{{\rm s}(\,{x}, {y}\,)}\,\sqrt {2}\, \! \right] $ \\
\> ${{n}_{{a}}}=$\>$ \left[ \! \,{\displaystyle \frac {1}{2}}\,{\rm e}^{
  {\rm s}(\,{x}, {y}\,)}\,\sqrt {2}\, , 0\, , 0\, , {\displaystyle \frac {1
  }{2}}\,{\rm r}(\,{x}, {y}\,)\,{\rm e}^{(\, - {\rm s}(\,{x}, {y}\,
  )\,)}\,\sqrt {2} + {\displaystyle \frac {1}{2}}\,{\rm w}(\,{x}, {
  y}\,)\,{\rm e}^{{\rm s}(\,{x}, {y}\,)}\,\sqrt {2}\, \!  \right] $ \\
\> ${{m}_{{a}}}=$\>$ \left[ \! \,0 , \,0 , \, - \,{\displaystyle \frac {1}{2}}
  \,{\rm e}^{(\,{\rm k}(\,{x}, {y}\,) - {\rm s}(\,{x}, {y}\,)\,)}\,
  \sqrt {2}\, , 0\, \!  \right] $ \\
\> $\overline{m}_a = $ \> $\left[ \! \,0 , \, - \,{\displaystyle \frac {1}{
  2}}\,{\rm e}^{(\,{\rm k}(\,{x}, {y}\,) - {\rm s}(\,{x}, {y}\,)\,)
  }\,\sqrt {2}\, , 0\, , 0\, \!  \right] $ \\
\vspace{1mm} \> Corresponding line element: \\
\> $\lefteqn{{\it \:ds}^{2}={\rm e}^{(\,2\,{\rm s}(\,{x}, {y}\,)\,)}
  \,{\it \:d}\,{t}^{{\it 2\:}} + 2\,{\rm e}^{(\,2\,{\rm s}(\,{x}, {
  y}\,)\,)}\,{\rm w}(\,{x}, {y}\,)\,{\it \:d}\,{t}^{{\:}}\,{\it d\:
  }\,{z}^{{\:}} - {\rm e}^{(\,2\,{\rm k}(\,{x}, {y}\,) - 2\,{\rm s}
  (\,{x}, {y}\,)\,)}\,{\it \:d}\,{x}^{{\:}}\,{\it d\:}\,{y}^{{\:}}}$ \\
\> $\mbox{} - (\,{\rm e}^{(\, - {\rm s}(\,{x}, {y}\,)\,)}\,{\rm 
  r}(\,{x}, {y}\,) - {\rm w}(\,{x}, {y}\,)\,{\rm e}^{{\rm s}(\,{x}
  , {y}\,)}\,)\,(\,{\rm e}^{(\, - {\rm s}(\,{x}, {y}\,)\,)}\,{\rm r
  }(\,{x}, {y}\,) + {\rm w}(\,{x}, {y}\,)\,{\rm e}^{{\rm s}(\,{x}, 
  {y}\,)}\,)\,{\it \:d}\,{z}^{{\it 2\:}}$ \\
%
%== Bondi =====================================================================
\rule{\linewidth}{.5mm}\\
{\large Spacetime: Bondi ({\bf Bondi1}, {\bf Bondi2})
\cite{bondi/vanderburg/metzner:1962}}\\
\rule{\linewidth}{.05mm}\\
Input file\hspace{10mm} \> Contravariant tetrad \\
\rule[.5\baselineskip]{\linewidth}{.05mm}\\
{\tt npupbon.mpl} \> ${l}^{{a}}=$\>$[\,{\rm e}^{(\, - {\rm Q}(\,{r}, {u}, 
  { \theta}\,)\,)}\, , 0 , \,0 , \,0\,]$ \\
\> ${n}^{{a}}=$\>$ \left[ \! \, - \,{\displaystyle \frac {1}{2}}\,
  {\displaystyle \frac {{\rm e}^{(\, - {\rm Q}(\,{r}, {u}, { \theta
  }\,)\,)}\,{\rm V}(\,{r}, {u}, { \theta}\,)}{{r}}}\, , {\rm e}^{(\,
  -{\rm Q}(\,{r}, {u}, { \theta}\,)\,)}\, , {\rm e}^{(\, - {\rm Q}(
  \,{r}, {u}, { \theta}\,)\,)}\,{\rm U}(\,{r}, {u}, { \theta}\,)\, , 0
  \, \!  \right] $ \\
\> $ {m}^{{a}}= $\>$\left[ \! \,0 , \,0 , \, -\,{\displaystyle \frac {1}{2}}\,
  {\displaystyle \frac {\sqrt {2}\,{\rm e}^{(\, - { \gamma}(\,{r}, 
  {u}, { \theta}\,)\,)}}{{r}}}\, , -\,{\displaystyle \frac {1}{2}}\,
  {\displaystyle \frac {{I}\,{\rm e}^{{ \gamma}(\,{r}, {u}, { 
  \theta}\,)}\,\sqrt {2}}{{r}\,{\rm sin}(\,{ \theta}\,)}}\, \! 
  \right] $ \\
\> $\overline{m}^a =$\>$ \left[ \! \,0 , \,0 , \, -\,{\displaystyle \frac {1
  }{2}}\,{\displaystyle \frac {\sqrt {2}\,{\rm e}^{(\, - { \gamma}(
  \,{r}, {u}, { \theta}\,)\,)}}{{r}}}\, , {\displaystyle \frac {1}{2}}
  \,{\displaystyle \frac {{I}\,{\rm e}^{{ \gamma}(\,{r}, {u}, { 
  \theta}\,)}\,\sqrt {2}}{{r}\,{\rm sin}(\,{ \theta}\,)}}\, \! 
  \right] $ \\
\vspace{1mm} \> Corresponding line element: \\
\> $\lefteqn{{\it \:ds}^{2}=2\,(\,{\rm e}^{{\rm Q}(\,{r}, {u}, { 
  \theta}\,)}\,)^{2}\,{\it \:d}\,{r}^{{\:}}\,{\it d\:}\,{u}^{{\:}}
  +  \left( \! \,{\displaystyle \frac {(\,{\rm e}^{
  {\rm Q}(\,{r}, {u}, { \theta}\,)}\,)^{2}\,{\rm V}(\,{r}, {u}, { 
  \theta}\,)}{{r}}} - {r}^{2}\,(\,{\rm e}^{{ \gamma}(\,{r}, {u}, { 
  \theta}\,)}\,)^{2}\,{\rm U}(\,{r}, {u}, { \theta}\,)^{2}\, \! 
  \right) \,{\it \:d}\,{u}^{{\it 2\:}}}$ \\
\> $\mbox{} + 2\,{\rm U}(\,{r}, {u}, { \theta}\,)\,{r}^{2}\,(\,
  {\rm e}^{{ \gamma}(\,{r}, {u}, { \theta}\,)}\,)^{2}\,{\it \:d}\,{
  u}^{{\:}}\,{\it d\:}\,{ \theta}^{{\:}} - {r}^{2}\,(\,{\rm e}^{{ 
  \gamma}(\,{r}, {u}, { \theta}\,)}\,)^{2}\,{\it \:d}\,{ \theta}^{
  {\it 2\:}} - {\displaystyle \frac {{r}^{2}\,{\rm sin}(\,{ 
  \theta}\,)^{2}\,{\it \:d}\,{ \phi}^{{\it 2\:}}}{(\,{\rm e}^{{ 
  \gamma}(\,{r}, {u}, { \theta}\,)}\,)^{2}}}$ \\
\rule{\linewidth}{.05mm}\\
%------------------------------------------------------------------------------
Input file\hspace{10mm} \> Covariant tetrad \\
\rule[.5\baselineskip]{\linewidth}{.05mm}\\
{\tt npdnbon.mpl} \> ${{l}_{{a}}}=$\>$[\,0 , \,{\rm e}^{{\rm Q}(\,{r}, {u}, 
  { \theta}\,)} , \,0 , \,0\,]$ \\
\> ${{n}_{{a}}}= $\>$\left[ \,{\rm e}^{{\rm Q}(\,{r}, {u}, { \theta}\,
  )} , \,{\displaystyle \frac {{\rm e}^{{\rm Q}(\,{r}, {u}, { \theta}
  \,)}\,{\rm V}(\,{r}, {u}, { \theta}\,)}{{r}}} , \,0 , \,0\,  \right] $ \\
\> ${{m}_{{a}}}= $\>$ \left[ {\vrule 
  height0.79em width0em depth0.79em} \right. \,0 , \, - \,
  {\displaystyle \frac {1}{2}}\,{r}\,{\rm U}(\,{r}, {u}, { \theta}
  \,)\,{\rm e}^{{ \gamma}(\,{r}, {u}, { \theta}\,)}\,\sqrt {2} , \,
  {\displaystyle \frac {1}{2}}\,{r}\,{\rm e}^{{ \gamma}(\,{r}, {u}
  , { \theta}\,)}\,\sqrt {2} , \,{\displaystyle \frac {1}{2}}\,{I}\,{r
  }\,{\rm sin}(\,{ \theta}\,)\,{\rm e}^{(\, - { \gamma}(\,{r}, {u}
  , { \theta}\,)\,)}\,\sqrt {2}
  \left. {\vrule height0.79em width0em depth0.79em}
  \right] \mbox{\hspace{352pt}}$ \\
\> $ \overline{m}_a = $\>$ \left[ {\vrule 
  height0.79em width0em depth0.79em} \right. \! \! 0 , \, - \,
  {\displaystyle \frac {1}{2}}\,{r}\,{\rm U}(\,{r}, {u}, { \theta}
  \,)\,{\rm e}^{{ \gamma}(\,{r}, {u}, { \theta}\,)}\,\sqrt {2} , \,
  {\displaystyle \frac {1}{2}}\,{r}\,{\rm e}^{{ \gamma}(\,{r}, {u}
  , { \theta}\,)}\,\sqrt {2}  , 
  - \,{\displaystyle \frac {1}{2}}\,{I}\,{r}\,{\rm sin}(\,{ 
  \theta}\,)\,{\rm e}^{(\, - { \gamma}(\,{r}, {u}, { \theta}\,)\,)}
  \,\sqrt {2} \left. {\vrule 
  height0.79em width0em depth0.79em} \right] \mbox{\hspace{106pt}}$ \\
\vspace{1mm} \> Corresponding line element: \\
\> $\lefteqn{{\it \:ds}^{2}=2\,(\,{\rm e}^{{\rm Q}(\,{r}, {u}, { 
  \theta}\,)}\,)^{2}\,{\it \:d}\,{r}^{{\:}}\,{\it d\:}\,{u}^{{\:}}
  +  \left( \! \,2\,{\displaystyle \frac {(\,{\rm e}^{
  {\rm Q}(\,{r}, {u}, { \theta}\,)}\,)^{2}\,{\rm V}(\,{r}, {u}, { 
  \theta}\,)}{{r}}} - {r}^{2}\,(\,{\rm e}^{{ \gamma}(\,{r}, {u}, { 
  \theta}\,)}\,)^{2}\,{\rm U}(\,{r}, {u}, { \theta}\,)^{2}\, \! 
  \right) \,{\it \:d}\,{u}^{{\it 2\:}}} $\\
\> $ \mbox{} + 2\,{\rm U}(\,{r}, {u}, { \theta}\,)\,{r}^{2}\,(\,
  {\rm e}^{{ \gamma}(\,{r}, {u}, { \theta}\,)}\,)^{2}\,{\it \:d}\,{
  u}^{{\:}}\,{\it d\:}\,{ \theta}^{{\:}} - {r}^{2}\,(\,{\rm e}^{{ 
  \gamma}(\,{r}, {u}, { \theta}\,)}\,)^{2}\,{\it \:d}\,{ \theta}^{
  {\it 2\:}} - {\displaystyle \frac {{r}^{2}\,{\rm sin}(\,{ 
  \theta}\,)^{2}\,{\it \:d}\,{ \phi}^{{\it 2\:}}}{(\,{\rm e}^{{ 
  \gamma}(\,{r}, {u}, { \theta}\,)}\,)^{2}}}$ \\
%
%== Debever ===================================================================
\rule{\linewidth}{.5mm}\\
{\large Spacetime: Debever ({\bf Deb}) \cite{debever/mclenaghan/tariq:1979}}\\
\rule{\linewidth}{.05mm}\\
Input file\hspace{10mm} \> Contravariant tetrad \\
\rule[.5\baselineskip]{\linewidth}{.05mm}\\
{\tt npupdeb.mpl} \> ${{l}^{{a}}} =$ \> $\left[ {\vrule 
  height0.84em width0em depth0.84em} \right. \! \!  - \,
  {\displaystyle \frac {1}{2}}\,{\displaystyle \frac {{\rm P}(\,{x}
  , {y}\,)\,\sqrt {2}}{ - {\rm L}(\,{x}, {y}\,)\,{\rm P}(\,{x}, {y}
  \,) + {\rm M}(\,{x}, {y}\,)\,{\rm N}(\,{x}, {y}\,)}} ,  $ \\
\>\> $ {\displaystyle \frac {1}{2}}\,{\displaystyle \frac {{\rm N}(
  \,{x}, {y}\,)\,\sqrt {2}}{ - {\rm L}(\,{x}, {y}\,)\,{\rm P}(\,{x}
  , {y}\,) + {\rm M}(\,{x}, {y}\,)\,{\rm N}(\,{x}, {y}\,)}}\, ,  -\,
  {\displaystyle \frac {1}{2}}\,{\displaystyle \frac {{\rm Y}(\,{y}
  \,)\,\sqrt {2}}{{\rm S}(\,{x}, {y}\,)}}\, , 0 \! \! \left. {\vrule 
  height0.84em width0em depth0.84em} \right] $ \\
\>$ {{n}^{{a}}}=$\>$ \left[ {\vrule 
  height0.84em width0em depth0.84em} \right. \! \!  - \,
  {\displaystyle \frac {1}{2}}\,{\displaystyle \frac {{\rm P}(\,{x}
  , {y}\,)\,\sqrt {2}}{ - {\rm L}(\,{x}, {y}\,)\,{\rm P}(\,{x}, {y}
  \,) + {\rm M}(\,{x}, {y}\,)\,{\rm N}(\,{x}, {y}\,)}} ,  $ \\
\>\> $ {\displaystyle \frac {1}{2}}\,{\displaystyle \frac {{\rm N}(
  \,{x}, {y}\,)\,\sqrt {2}}{ - {\rm L}(\,{x}, {y}\,)\,{\rm P}(\,{x}
  , {y}\,) + {\rm M}(\,{x}, {y}\,)\,{\rm N}(\,{x}, {y}\,)}}\,
   , {\displaystyle \frac {1}{2}}\,{\displaystyle \frac {{\rm Y}(\,{y}
  \,)\,\sqrt {2}}{{\rm S}(\,{x}, {y}\,)}} , \,0 \! \! \left. {\vrule 
  height0.84em width0em depth0.84em} \right] $ \\
\> ${{m}^{{a}}}=$\>$ \left[ {\vrule 
  height0.84em width0em depth0.84em} \right. \! \! {\displaystyle 
 \frac {1}{2}}\,{\displaystyle \frac {{\rm M}(\,{x}, {y}\,)\,
  \sqrt {2}}{ - {\rm L}(\,{x}, {y}\,)\,{\rm P}(\,{x}, {y}\,) + 
  {\rm M}(\,{x}, {y}\,)\,{\rm N}(\,{x}, {y}\,)}} ,  $ \\
\>\> $  -\,{\displaystyle \frac {1}{2}}\,{\displaystyle \frac {
  {\rm L}(\,{x}, {y}\,)\,\sqrt {2}}{ - {\rm L}(\,{x}, {y}\,)\,{\rm 
  P}(\,{x}, {y}\,) + {\rm M}(\,{x}, {y}\,)\,{\rm N}(\,{x}, {y}\,)}}
   , \,0\,  , -\,{\displaystyle \frac {1}{2}}\,{\displaystyle \frac {{I}
  \,{\rm X}(\,{x}\,)\,\sqrt {2}}{{\rm S}(\,{x}, {y}\,)}} \! 
  \! \left. {\vrule height0.84em width0em depth0.84em} \right] $ \\
\> $\overline{m}^a = $\>$ \left[ {\vrule 
  height0.84em width0em depth0.84em} \right. \! \! {\displaystyle 
  \frac {1}{2}}\,{\displaystyle \frac {{\rm M}(\,{x}, {y}\,)\,
  \sqrt {2}}{ - {\rm L}(\,{x}, {y}\,)\,{\rm P}(\,{x}, {y}\,) + 
  {\rm M}(\,{x}, {y}\,)\,{\rm N}(\,{x}, {y}\,)}} ,  $ \\
\>\> $  -\,{\displaystyle \frac {1}{2}}\,{\displaystyle \frac {
  {\rm L}(\,{x}, {y}\,)\,\sqrt {2}}{ - {\rm L}(\,{x}, {y}\,)\,{\rm 
  P}(\,{x}, {y}\,) + {\rm M}(\,{x}, {y}\,)\,{\rm N}(\,{x}, {y}\,)}}
  \, , 0 , \,{\displaystyle \frac {1}{2}}\,{\displaystyle \frac {{I}\,
  {\rm X}(\,{x}\,)\,\sqrt {2}}{{\rm S}(\,{x}, {y}\,)}} \! 
  \! \left. {\vrule height0.84em width0em depth0.84em} \right] $ \\
\vspace{1mm} \> Corresponding line element: \\
\> $\lefteqn{{\it \:ds}^{2}=(\,{\rm L}(\,{x}, {y}\,) - {\rm N}(\,{x}
  , {y}\,)\,)\,(\,{\rm L}(\,{x}, {y}\,) + {\rm N}(\,{x}, {y}\,)\,)
  \,{\it \:d}\,{t}^{{\it 2\:}}} $ \\
\> $\mbox{} + 2\,(\, - {\rm P}(\,{x}, {y}\,)\,{\rm N}(\,{x}, {y}
  \,) + {\rm M}(\,{x}, {y}\,)\,{\rm L}(\,{x}, {y}\,)\,)\,{\it \:d}
  \,{t}^{{\:}}\,{\it d\:}\,{z}^{{\:}} $ \\
\> $\mbox{} - (\,{\rm P}(\,{x}, {y}\,) - {\rm M}(\,{x}, {y}\,)\,
  )\,(\,{\rm P}(\,{x}, {y}\,) + {\rm M}(\,{x}, {y}\,)\,)\,{\it \:d}
  \,{z}^{{\it 2\:}} - {\displaystyle \frac {{\rm S}(\,{x}, {y}\,)^{
  2}\,{\it \:d}\,{y}^{{\it 2\:}}}{{\rm Y}(\,{y}\,)^{2}}}
  - {\displaystyle \frac {{\rm S}(\,{x}, {y}\,)^{2}\,
  {\it \:d}\,{x}^{{\it 2\:}}}{{\rm X}(\,{x}\,)^{2}}} $ \\
\rule{\linewidth}{.05mm}\\
%------------------------------------------------------------------------------
Input file\hspace{10mm} \> Covariant tetrad \\
\rule[.5\baselineskip]{\linewidth}{.05mm}\\
{\tt npdndeb.mpl} \> ${{l}_{{a}}}=$\>$ \left[ \! \,{\displaystyle
  \frac {1}{2}}\,{\rm L}(\,
  {x}, {y}\,)\,\sqrt {2} , \,{\displaystyle \frac {1}{2}}\,{\rm M}(\,{
  x}, {y}\,)\,\sqrt {2}\ , ,{\displaystyle \frac {1}{2}}\,
  {\displaystyle \frac {{\rm S}(\,{x}, {y}\,)\,\sqrt {2}}{{\rm Y}(
  \,{y}\,)}}\, , 0\, \!  \right] $ \\
\> ${{n}_{{a}}}=$\>$ \left[ \! \,{\displaystyle \frac {1}{2}}\,{\rm L}(\,
  {x}, {y}\,)\,\sqrt {2} , \,{\displaystyle \frac {1}{2}}\,{\rm M}(\,{
  x}, {y}\,)\,\sqrt {2} , \, -\,{\displaystyle \frac {1}{2}}\,
  {\displaystyle \frac {{\rm S}(\,{x}, {y}\,)\,\sqrt {2}}{{\rm Y}(
  \,{y}\,)}}\, , 0\,  \right] $ \\
\> ${{m}_{{a}}}= $\>$ \left[ \! \, -\,{\displaystyle \frac {1}{2}}\,{\rm 
  N}(\,{x}, {y}\,)\,\sqrt {2} , \, -\,{\displaystyle \frac {1}{2}}\,
  {\rm P}(\,{x}, {y}\,)\,\sqrt {2} , \,0 , \,{\displaystyle \frac {1}{2}}
  \,{\displaystyle \frac {{I}\,{\rm S}(\,{x}, {y}\,)\,\sqrt {2}}{
  {\rm X}(\,{x}\,)}}\, \!  \right] $ \\
\> $\overline{m}_a =$\>$ \left[ \! \, -\,{\displaystyle \frac {1}{2}}
  \,{\rm N}(\,{x}, {y}\,)\,\sqrt {2} , \, -\,{\displaystyle \frac {1
  }{2}}\,{\rm P}(\,{x}, {y}\,)\,\sqrt {2} , \,0 , \, -\,{\displaystyle 
  \frac {1}{2}}\,{\displaystyle \frac {{I}\,{\rm S}(\,{x}, {y}\,)\,
  \sqrt {2}}{{\rm X}(\,{x}\,)}}\, \!  \right] $ \\
\vspace{1mm} \> Corresponding line element: \\
\> $\lefteqn{{\it \:ds}^{2}=(\,{\rm L}(\,{x}, {y}\,) - {\rm N}(\,{x}
  , {y}\,)\,)\,(\,{\rm L}(\,{x}, {y}\,) + {\rm N}(\,{x}, {y}\,)\,)
  \,{\it \:d}\,{t}^{{\it 2\:}}} $ \\
\> $ \mbox{} + 2\,(\, - {\rm P}(\,{x}, {y}\,)\,{\rm N}(\,{x}, {y}
  \,) + {\rm M}(\,{x}, {y}\,)\,{\rm L}(\,{x}, {y}\,)\,)\,{\it \:d}
  \,{t}^{{\:}}\,{\it d\:}\,{z}^{{\:}} $ \\
\> $ \mbox{} + (\,{\rm M}(\,{x}, {y}\,) - {\rm P}(\,{x}, {y}\,)\,
  )\,(\,{\rm P}(\,{x}, {y}\,) + {\rm M}(\,{x}, {y}\,)\,)\,{\it \:d}
  \,{z}^{{\it 2\:}} - {\displaystyle \frac {{\rm S}(\,{x}, {y}\,)^{
  2}\,{\it \:d}\,{y}^{{\it 2\:}}}{{\rm Y}(\,{y}\,)^{2}}}
  - {\displaystyle \frac {{\rm S}(\,{x}, {y}\,)^{2}\,
  {\it \:d}\,{x}^{{\it 2\:}}}{{\rm X}(\,{x}\,)^{2}}} $\\
%
%== Debever-McLenaghan-Tariq ==================================================
\rule{\linewidth}{.5mm}\\
{\large Spacetime: Debever-McLenaghan-Tariq ({\bf DMT1}) \cite{debever/mclenaghan/tariq:1979}}\\
\rule{\linewidth}{.05mm}\\
Input file\hspace{10mm} \> Contravariant tetrad \\
\rule[.5\baselineskip]{\linewidth}{.05mm}\\
{\tt npupsdmt.mpl} \> ${l}^{{a}}=$\>$ \left[ {\vrule 
  height0.84em width0em depth0.84em} \right.  {\displaystyle 
  \frac {1}{2}}\,{\displaystyle \frac {{\rm P}(\,{w}, {x}\,)\,
  \sqrt {2}}{{\rm L}(\,{w}, {x}\,)\,{\rm P}(\,{w}, {x}\,) - {\rm M}
  (\,{w}, {x}\,)\,{\rm N}(\,{w}, {x}\,)}} , 
  -\,{\displaystyle \frac {1}{2}}\,{\displaystyle \frac {
  {\rm N}(\,{w}, {x}\,)\,\sqrt {2}}{{\rm L}(\,{w}, {x}\,)\,{\rm P}(
  \,{w}, {x}\,) - {\rm M}(\,{w}, {x}\,)\,{\rm N}(\,{w}, {x}\,)}} , $ \\
\>\> $ \,0 , \, - \,{\displaystyle \frac {1}{2}}\,{\displaystyle \frac {
  \sqrt {2}}{{\rm S}(\,{w}, {x}\,)}} \! \! \left. {\vrule 
  height0.84em width0em depth0.84em} \right] $ \\
\> ${n}^{{a}}= $\>$ \left[ {\vrule 
  height0.84em width0em depth0.84em} \right. {\displaystyle 
  \frac {1}{2}}\,{\displaystyle \frac {{\rm P}(\,{w}, {x}\,)\,
  \sqrt {2}}{{\rm L}(\,{w}, {x}\,)\,{\rm P}(\,{w}, {x}\,) - {\rm M}
  (\,{w}, {x}\,)\,{\rm N}(\,{w}, {x}\,)}} ,
  -\,{\displaystyle \frac {1}{2}}\,{\displaystyle \frac {
  {\rm N}(\,{w}, {x}\,)\,\sqrt {2}}{{\rm L}(\,{w}, {x}\,)\,{\rm P}(
  \,{w}, {x}\,) - {\rm M}(\,{w}, {x}\,)\,{\rm N}(\,{w}, {x}\,)}} ,$ \\
\>\> $\,0 , \,{\displaystyle \frac {1}{2}}\,{\displaystyle \frac {\sqrt {2}}{
  {\rm S}(\,{w}, {x}\,)}} \! \! \left. {\vrule 
  height0.84em width0em depth0.84em} \right] $ \\
\> ${m}^{{a}}= $\>$ \left[ {\vrule 
  height0.84em width0em depth0.84em} \right. \! \!  - \,
  {\displaystyle \frac {1}{2}}\,{\displaystyle \frac {{\rm M}(\,{w}
  , {x}\,)\,\sqrt {2}}{{\rm L}(\,{w}, {x}\,)\,{\rm P}(\,{w}, {x}\,)
  - {\rm M}(\,{w}, {x}\,)\,{\rm N}(\,{w}, {x}\,)}} , 
   {\displaystyle \frac {1}{2}}\,{\displaystyle \frac {{\rm L}(
  \,{w}, {x}\,)\,\sqrt {2}}{{\rm L}(\,{w}, {x}\,)\,{\rm P}(\,{w}, {
  x}\,) - {\rm M}(\,{w}, {x}\,)\,{\rm N}(\,{w}, {x}\,)}}\, , $ \\
\>\> $ -\,{\displaystyle \frac {1}{2}}\,{\displaystyle \frac {{I}\,\sqrt {2
  }}{{\rm R}(\,{w}, {x}\,)}}\, , 0 \! \! \left. {\vrule 
  height0.84em width0em depth0.84em} \right] $ \\
\> $\overline{m}^a= $\>$\left[ {\vrule 
  height0.84em width0em depth0.84em} \right. \! \!  - \,
  {\displaystyle \frac {1}{2}}\,{\displaystyle \frac {{\rm M}(\,{w}
  , {x}\,)\,\sqrt {2}}{{\rm L}(\,{w}, {x}\,)\,{\rm P}(\,{w}, {x}\,)
  - {\rm M}(\,{w}, {x}\,)\,{\rm N}(\,{w}, {x}\,)}}  , 
  {\displaystyle \frac {1}{2}}\,{\displaystyle \frac {{\rm L}(
  \,{w}, {x}\,)\,\sqrt {2}}{{\rm L}(\,{w}, {x}\,)\,{\rm P}(\,{w}, {
  x}\,) - {\rm M}(\,{w}, {x}\,)\,{\rm N}(\,{w}, {x}\,)}}\, ,$ \\
\>\> $  {\displaystyle \frac {1}{2}}\,{\displaystyle \frac {{I}\,\sqrt {2
  }}{{\rm R}(\,{w}, {x}\,)}}\, , 0 \! \! \left. {\vrule 
  height0.84em width0em depth0.84em} \right] $ \\
\vspace{1mm} \> Corresponding line element: \\
\> $\lefteqn{{\it \:ds}^{2}=(\,{\rm L}(\,{w}, {x}\,) - {\rm N}(\,{w}
  , {x}\,)\,)\,(\,{\rm L}(\,{w}, {x}\,) + {\rm N}(\,{w}, {x}\,)\,)
  \,{\it \:d}\,{u}^{{\it 2\:}}} $ \\
\> $ \mbox{} + 2\,(\, - {\rm P}(\,{w}, {x}\,)\,{\rm N}(\,{w}, {x}
  \,) + {\rm M}(\,{w}, {x}\,)\,{\rm L}(\,{w}, {x}\,)\,)\,{\it \:d}
  \,{u}^{{\:}}\,{\it d\:}\,{v}^{{\:}} $ \\
\> $ \mbox{} + (\,{\rm M}(\,{w}, {x}\,) - {\rm P}(\,{w}, {x}\,)\,
  )\,(\,{\rm M}(\,{w}, {x}\,) + {\rm P}(\,{w}, {x}\,)\,)\,{\it \:d}
  \,{v}^{{\it 2\:}} - {\rm R}(\,{w}, {x}\,)^{2}\,{\it \:d}\,{w}^{
  {\it 2\:}} - {\rm S}(\,{w}, {x}\,)^{2}\,{\it \:d}\,{x}^{{\it 2\:}}$ \\
\rule{\linewidth}{.05mm}\\
%------------------------------------------------------------------------------
Input file\hspace{10mm} \> Covariant tetrad \\
\rule[.5\baselineskip]{\linewidth}{.05mm}\\
{\tt npdnsdmt.mpl} \> ${{l}_{{a}}}= $\>$\left[ \! \,{\displaystyle
  \frac {1}{2}}\,\sqrt {2}
  \,{\rm L}(\,{w}, {x}\,) , \,{\displaystyle \frac {1}{2}}\,\sqrt {2}
  \,{\rm M}(\,{w}, {x}\,) , \,0\, , {\displaystyle \frac {1}{2}}\,{\rm S}
  (\,{w}, {x}\,)\,\sqrt {2}\, \!  \right] $ \\
\> ${{n}_{{a}}}=$\>$ \left[ \! \,{\displaystyle \frac {1}{2}}\,\sqrt {2}
  \,{\rm L}(\,{w}, {x}\,)\, , {\displaystyle \frac {1}{2}}\,\sqrt {2}
  \,{\rm M}(\,{w}, {x}\,)\, , 0 , \, - \,{\displaystyle \frac {1}{2}}\,
  {\rm S}(\,{w}, {x}\,)\,\sqrt {2}\, \!  \right] $ \\
\> ${{m}_{{a}}}=$\>$ \left[ \! \, - \,{\displaystyle \frac {1}{2}}\,
  \sqrt {2}\,{\rm N}(\,{w}, {x}\,)\,  , -\,{\displaystyle \frac {1}{2
  }}\,\sqrt {2}\,{\rm P}(\,{w}, {x}\,)\, , {\displaystyle \frac {1}{2
  }}\,{I}\,{\rm R}(\,{w}, {x}\,)\,\sqrt {2}\, , 0\, \!  \right] $ \\
\> $\overline{m}_a =$\>$ \left[ \! \, - \,{\displaystyle \frac {1}{2}}
  \,\sqrt {2}\,{\rm N}(\,{w}, {x}\,)\, ,  - \,{\displaystyle \frac {1
  }{2}}\,\sqrt {2}\,{\rm P}(\,{w}, {x}\,)\, ,  - \,{\displaystyle 
  \frac {1}{2}}\,{I}\,{\rm R}(\,{w}, {x}\,)\,\sqrt {2}\, , 0\, \! 
  \right] $ \\
\vspace{1mm} \> Corresponding line element: \\
\> $\lefteqn{{\it \:ds}^{2}=(\,{\rm L}(\,{w}, {x}\,) - {\rm N}(\,{w}
  , {x}\,)\,)\,(\,{\rm L}(\,{w}, {x}\,) + {\rm N}(\,{w}, {x}\,)\,)
  \,{\it \:d}\,{u}^{{\it 2\:}}} $ \\
\> $ \mbox{} + 2\,(\, - {\rm P}(\,{w}, {x}\,)\,{\rm N}(\,{w}, {x}
  \,) + {\rm M}(\,{w}, {x}\,)\,{\rm L}(\,{w}, {x}\,)\,)\,{\it \:d}
  \,{u}^{{\:}}\,{\it d\:}\,{v}^{{\:}} $ \\
\> $ \mbox{} + (\,{\rm M}(\,{w}, {x}\,) - {\rm P}(\,{w}, {x}\,)\,
  )\,(\,{\rm M}(\,{w}, {x}\,) + {\rm P}(\,{w}, {x}\,)\,)\,{\it \:d}
  \,{v}^{{\it 2\:}} - {\rm R}(\,{w}, {x}\,)^{2}\,{\it \:d}\,{w}^{
  {\it 2\:}} - {\rm S}(\,{w}, {x}\,)^{2}\,{\it \:d}\,{x}^{{\it 2\:}}$ \\
%
%== Debever-McLenaghan-Tariq (modified) =======================================
\rule{\linewidth}{.5mm}\\
{\large Spacetime: Debever-McLenaghan-Tariq (modified) ({\bf DMT2})
\cite{debever/mclenaghan/tariq:1979}}\\
\rule{\linewidth}{.05mm}\\
Input file\hspace{10mm} \> Contravariant tetrad \\
\rule[.5\baselineskip]{\linewidth}{.05mm}\\
{\tt npupdmt.mpl} \> ${l}^{{a}}= $\>$\left[ {\vrule 
  height0.84em width0em depth0.84em} \right. \! \! {\displaystyle 
  \frac {1}{2}}\,{\displaystyle \frac {{\rm P}(\,{u}, {v}, {w}, {x}
  \,)\,\sqrt {2}}{{\rm L}(\,{u}, {v}, {w}, {x}\,)\,{\rm P}(\,{u}, {
  v}, {w}, {x}\,) - {\rm M}(\,{u}, {v}, {w}, {x}\,)\,{\rm N}(\,{u}
  , {v}, {w}, {x}\,)}} , $ \\
\>\> $ - \,{\displaystyle \frac {1}{2}}\,{\displaystyle \frac {
  {\rm N}(\,{u}, {v}, {w}, {x}\,)\,\sqrt {2}}{{\rm L}(\,{u}, {v}, {
  w}, {x}\,)\,{\rm P}(\,{u}, {v}, {w}, {x}\,) - {\rm M}(\,{u}, {v}
  , {w}, {x}\,)\,{\rm N}(\,{u}, {v}, {w}, {x}\,)}}\, , 0 , 
  - \,{\displaystyle \frac {1}{2}}\,{\displaystyle \frac {
  \sqrt {2}}{{\rm S}(\,{w}, {x}\,)}} \! \! \left. {\vrule 
  height0.84em width0em depth0.84em} \right] $ \\
\> ${n}^{{a}}=$\>$ \left[ {\vrule 
  height0.84em width0em depth0.84em} \right. \! \! {\displaystyle 
  \frac {1}{2}}\,{\displaystyle \frac {{\rm P}(\,{u}, {v}, {w}, {x}
  \,)\,\sqrt {2}}{{\rm L}(\,{u}, {v}, {w}, {x}\,)\,{\rm P}(\,{u}, {
  v}, {w}, {x}\,) - {\rm M}(\,{u}, {v}, {w}, {x}\,)\,{\rm N}(\,{u}
  , {v}, {w}, {x}\,)}} , $ \\
\>\> $ - \,{\displaystyle \frac {1}{2}}\,{\displaystyle \frac {
  {\rm N}(\,{u}, {v}, {w}, {x}\,)\,\sqrt {2}}{{\rm L}(\,{u}, {v}, {
  w}, {x}\,)\,{\rm P}(\,{u}, {v}, {w}, {x}\,) - {\rm M}(\,{u}, {v}
  , {w}, {x}\,)\,{\rm N}(\,{u}, {v}, {w}, {x}\,)}}\, , 0 ,
  {\displaystyle \frac {1}{2}}\,{\displaystyle \frac {\sqrt {2
  }}{{\rm S}(\,{w}, {x}\,)}} \! \! \left. {\vrule 
  height0.84em width0em depth0.84em} \right] $ \\
\> ${m}^{{a}}= $\>$ \left[ {\vrule 
  height0.84em width0em depth0.84em} \right. \! \!  - \,
  {\displaystyle \frac {1}{2}}\,{\displaystyle \frac {{\rm M}(\,{u}
  , {v}, {w}, {x}\,)\,\sqrt {2}}{{\rm L}(\,{u}, {v}, {w}, {x}\,)\,
  {\rm P}(\,{u}, {v}, {w}, {x}\,) - {\rm M}(\,{u}, {v}, {w}, {x}\,)
  \,{\rm N}(\,{u}, {v}, {w}, {x}\,)}} , $ \\
\>\> ${\displaystyle \frac {1}{2}}\,{\displaystyle \frac {{\rm L}(
  \,{u}, {v}, {w}, {x}\,)\,\sqrt {2}}{{\rm L}(\,{u}, {v}, {w}, {x}
  \,)\,{\rm P}(\,{u}, {v}, {w}, {x}\,) - {\rm M}(\,{u}, {v}, {w}, {
  x}\,)\,{\rm N}(\,{u}, {v}, {w}, {x}\,)}}\, - \,{\displaystyle 
  \frac {1}{2}}\,{\displaystyle \frac {{I}\,\sqrt {2}}{{\rm R}(\,{w
  }, {x}\,)}} , 0\left. {\vrule height0.84em width0em depth0.84em} \right] $\\
\> $\overline{m}^a= $\>$ \left[ {\vrule 
  height0.84em width0em depth0.84em} \right. \! \!  - \,
  {\displaystyle \frac {1}{2}}\,{\displaystyle \frac {{\rm M}(\,{u}
  , {v}, {w}, {x}\,)\,\sqrt {2}}{{\rm L}(\,{u}, {v}, {w}, {x}\,)\,
  {\rm P}(\,{u}, {v}, {w}, {x}\,) - {\rm M}(\,{u}, {v}, {w}, {x}\,)
  \,{\rm N}(\,{u}, {v}, {w}, {x}\,)}} ,  $ \\
\>\> $ {\displaystyle \frac {1}{2}}\,{\displaystyle \frac {{\rm L}(
  \,{u}, {v}, {w}, {x}\,)\,\sqrt {2}}{{\rm L}(\,{u}, {v}, {w}, {x}
  \,)\,{\rm P}(\,{u}, {v}, {w}, {x}\,) - {\rm M}(\,{u}, {v}, {w}, {
  x}\,)\,{\rm N}(\,{u}, {v}, {w}, {x}\,)}}\, , {\displaystyle \frac {1
  }{2}}\,{\displaystyle \frac {{I}\,\sqrt {2}}{{\rm R}(\,{w}, {x}\,
  )}}\, , 0 \left. {\vrule height0.84em width0em depth0.84em} \right] $ \\
\vspace{1mm} \> Corresponding line element: \\
\> $\lefteqn{{\it \:ds}^{2}= - (\,{\rm N}(\,{u}, {v}, {w}, {x}\,) - 
  {\rm L}(\,{u}, {v}, {w}, {x}\,)\,)\,(\,{\rm N}(\,{u}, {v}, {w}, {
  x}\,) + {\rm L}(\,{u}, {v}, {w}, {x}\,)\,)\,{\it \:d}\,{u}^{{\it 
  2\:}}}$ \\
\> $ \mbox{} + 2\,(\, - {\rm P}(\,{u}, {v}, {w}, {x}\,)\,{\rm N}(
  \,{u}, {v}, {w}, {x}\,) + {\rm M}(\,{u}, {v}, {w}, {x}\,)\,{\rm L
  }(\,{u}, {v}, {w}, {x}\,)\,)\,{\it \:d}\,{u}^{{\:}}\,{\it d\:}\,{
  v}^{{\:}}$ \\
\> $ \mbox{} - (\,{\rm P}(\,{u}, {v}, {w}, {x}\,) - {\rm M}(\,{u}
  , {v}, {w}, {x}\,)\,)\,(\,{\rm P}(\,{u}, {v}, {w}, {x}\,) + {\rm 
  M}(\,{u}, {v}, {w}, {x}\,)\,)\,{\it \:d}\,{v}^{{\it 2\:}}  $ \\
\>\> $\mbox{} - {\rm R}(\,{w}, {x}\,)^{2}\,{\it \:d}\,{w}^{{\it 2
  \:}} - {\rm S}(\,{w}, {x}\,)^{2}\,{\it \:d}\,{x}^{{\it 2\:}}$\\
\rule{\linewidth}{.05mm}\\
%------------------------------------------------------------------------------
Input file\hspace{10mm} \> Covariant tetrad \\
\rule[.5\baselineskip]{\linewidth}{.05mm}\\
{\tt npdndmt.mpl} \> ${{l}_{{a}}}=$\>$ \left[ \! \,{\displaystyle 
  \frac {1}{2}}\,\sqrt {2}
  \,{\rm L}(\,{u}, {v}, {w}, {x}\,)\, , {\displaystyle \frac {1}{2}}\,
  \sqrt {2}\,{\rm M}(\,{u}, {v}, {w}, {x}\,)\, , 0 , \,{\displaystyle 
  \frac {1}{2}}\,{\rm S}(\,{w}, {x}\,)\,\sqrt {2}\, \!  \right] $ \\
\> ${{n}_{{a}}}= $\>$ \left[ \! \,{\displaystyle \frac {1}{2}}\,\sqrt {2}
  \,{\rm L}(\,{u}, {v}, {w}, {x}\,) , \,{\displaystyle \frac {1}{2}}\,
  \sqrt {2}\,{\rm M}(\,{u}, {v}, {w}, {x}\,)\, , 0 , \, - \,
  {\displaystyle \frac {1}{2}}\,{\rm S}(\,{w}, {x}\,)\,\sqrt {2}\,
  \!  \right] $ \\
\> ${{m}_{{a}}}=$\>$ \left[ \! \, - \,{\displaystyle \frac {1}{2}}\,
  \sqrt {2}\,{\rm N}(\,{u}, {v}, {w}, {x}\,)\, ,  - \,{\displaystyle 
  \frac {1}{2}}\,\sqrt {2}\,{\rm P}(\,{u}, {v}, {w}, {x}\,)\, , 
  {\displaystyle \frac {1}{2}}\,{I}\,{\rm R}(\,{w}, {x}\,)\,\sqrt {
  2}\, , 0\, \!  \right] $ \\
\> $\overline{m}_a =$\>$ \left[ \! \, - \,{\displaystyle \frac {1}{2}}
  \,\sqrt {2}\,{\rm N}(\,{u}, {v}, {w}, {x}\,)\,  , - \,
  {\displaystyle \frac {1}{2}}\,\sqrt {2}\,{\rm P}(\,{u}, {v}, {w}
  , {x}\,)\,  , - \,{\displaystyle \frac {1}{2}}\,{I}\,{\rm R}(\,{w}, 
  {x}\,)\,\sqrt {2}\, , 0\, \!  \right] $ \\
\vspace{1mm} \> Corresponding line element: \\
\>$\lefteqn{{\it \:ds}^{2}= - (\,{\rm N}(\,{u}, {v}, {w}, {x}\,) - 
  {\rm L}(\,{u}, {v}, {w}, {x}\,)\,)\,(\,{\rm N}(\,{u}, {v}, {w}, {
  x}\,) + {\rm L}(\,{u}, {v}, {w}, {x}\,)\,)\,{\it \:d}\,{u}^{{\it 
  2\:}}}$ \\
\> $ \mbox{} + 2\,(\, - {\rm P}(\,{u}, {v}, {w}, {x}\,)\,{\rm N}(
  \,{u}, {v}, {w}, {x}\,) + {\rm M}(\,{u}, {v}, {w}, {x}\,)\,{\rm L
  }(\,{u}, {v}, {w}, {x}\,)\,)\,{\it \:d}\,{u}^{{\:}}\,{\it d\:}\,{
  v}^{{\:}} $ \\
\> $ \mbox{} + (\,{\rm M}(\,{u}, {v}, {w}, {x}\,) - {\rm P}(\,{u}
  , {v}, {w}, {x}\,)\,)\,(\,{\rm P}(\,{u}, {v}, {w}, {x}\,) + {\rm 
  M}(\,{u}, {v}, {w}, {x}\,)\,)\,{\it \:d}\,{v}^{{\it 2\:}} $ \\
\> $ \mbox{} - {\rm R}(\,{w}, {x}\,)^{2}\,{\it \:d}\,{w}^{{\it 2
  \:}} - {\rm S}(\,{w}, {x}\,)^{2}\,{\it \:d}\,{x}^{{\it 2\:}} $ \\
%
%== Kerr-Newman (Euclidean coordinates) =======================================
\rule{\linewidth}{.5mm}\\
{\large Spacetime: Kerr-Newman ($x,y,z,t$) ({\bf KN-Euc1})}\\
\rule{\linewidth}{.05mm}\\
Input file\hspace{10mm} \> Contravariant tetrad \\
\rule[.5\baselineskip]{\linewidth}{.05mm}\\
{\tt npupkn1.mpl} \> $ {l}^{{a}}= $ \> $ \left[ {\vrule 
  height0.89em width0em depth0.89em} \right. \! \! {\displaystyle 
  \frac {1}{2}}\,{\displaystyle \frac {(\,{x}^{2} + {a}^{2}\,)\,
  \sqrt {2}}{\sqrt {{x}^{2} - 2\,{M}\,{x} + {a}^{2} + {Q}^{2}}\,
  \sqrt {{x}^{2} + {a}^{2}\,{y}^{2}}}}\, , {\displaystyle \frac {1}{2
  }}\,{\displaystyle \frac {\sqrt {2}\,{a}}{\sqrt {{x}^{2} + {a}^{2
  }\,{y}^{2}}\,\sqrt {{x}^{2} - 2\,{M}\,{x} + {a}^{2} + {Q}^{2}}}} ,  $ \\
\>\> $ {\displaystyle \frac {1}{2}}\,{\displaystyle \frac {\sqrt {2
  }\,\sqrt {{x}^{2} - 2\,{M}\,{x} + {a}^{2} + {Q}^{2}}}{\sqrt {{x}
  ^{2} + {a}^{2}\,{y}^{2}}}}\, , 0 \! \! \left. {\vrule 
  height0.89em width0em depth0.89em} \right] \mbox{\hspace{188pt}} $ \\
\> ${n}^{{a}}= $\>$ \left[ {\vrule 
  height0.89em width0em depth0.89em} \right. \! \! {\displaystyle 
  \frac {1}{2}}\,{\displaystyle \frac {(\,{x}^{2} + {a}^{2}\,)\,
  \sqrt {2}}{\sqrt {{x}^{2} - 2\,{M}\,{x} + {a}^{2} + {Q}^{2}}\,
  \sqrt {{x}^{2} + {a}^{2}\,{y}^{2}}}}\, , {\displaystyle \frac {1}{2
  }}\,{\displaystyle \frac {\sqrt {2}\,{a}}{\sqrt {{x}^{2} + {a}^{2
  }\,{y}^{2}}\,\sqrt {{x}^{2} - 2\,{M}\,{x} + {a}^{2} + {Q}^{2}}}} , $ \\
\>\> $  - \,{\displaystyle \frac {1}{2}}\,{\displaystyle \frac {
  \sqrt {2}\,\sqrt {{x}^{2} - 2\,{M}\,{x} + {a}^{2} + {Q}^{2}}}{
  \sqrt {{x}^{2} + {a}^{2}\,{y}^{2}}}}\, , 0 \! \! \left. {\vrule 
  height0.89em width0em depth0.89em} \right] \mbox{\hspace{184pt}} $\\
\> ${m}^{{a}}=$\>$  \left[ \! \,{\displaystyle \frac {1}{2}}\,
  {\displaystyle \frac {\sqrt {2}\,{a}\,\sqrt {1 - {y}^{2}}}{
  \sqrt {{x}^{2} + {a}^{2}\,{y}^{2}}}}\, , {\displaystyle \frac {1}{2
  }}\,{\displaystyle \frac {\sqrt {2}}{\sqrt {{x}^{2} + {a}^{2}\,{y
  }^{2}}\,\sqrt {1 - {y}^{2}}}}\, , 0\, , - \,{\displaystyle \frac {1}{2
  }}\,{\displaystyle \frac {{I}\,\sqrt {2}\,\sqrt {1 - {y}^{2}}}{
  \sqrt {{x}^{2} + {a}^{2}\,{y}^{2}}}}\, \!  \right] $ \\
\> $\overline{m}^a = $\>$ \left[ \! \,{\displaystyle \frac {1}{2}}\,
  {\displaystyle \frac {\sqrt {2}\,{a}\,\sqrt {1 - {y}^{2}}}{
  \sqrt {{x}^{2} + {a}^{2}\,{y}^{2}}}}\, , {\displaystyle \frac {1}{2
  }}\,{\displaystyle \frac {\sqrt {2}}{\sqrt {{x}^{2} + {a}^{2}\,{y
  }^{2}}\,\sqrt {1 - {y}^{2}}}}\, , 0 , \,{\displaystyle \frac {1}{2}}\,
  {\displaystyle \frac {{I}\,\sqrt {2}\,\sqrt {1 - {y}^{2}}}{
  \sqrt {{x}^{2} + {a}^{2}\,{y}^{2}}}}\, \!  \right] $\\
\vspace{1mm} \> Corresponding line element: \\
\> $\lefteqn{{\it \:ds}^{2}={\displaystyle \frac {(\,{a}^{2}\,{y}^{2}
  + {x}^{2} - 2\,{M}\,{x} + {Q}^{2}\,)\,{\it \:d}\,{t}^{{\it 2\:}}
  }{{x}^{2} + {a}^{2}\,{y}^{2}}}
  + 2\,{\displaystyle \frac {{a}\,(\, - 2\,{M}\,{x} + 
  {Q}^{2}\,)\,(\, - 1 + {y}\,)\,(\,{y} + 1\,)\,{\it \:d}\,{t}^{{\:}
  }\,{\it d\:}\,{z}^{{\:}}}{{x}^{2} + {a}^{2}\,{y}^{2}}}} $  \\
\> $ + (\,{x}^{4} + {x}^{2}\,{a}^{2} + 2\,{a}^{2}\,{M}\,{x} - {a}^{
  2}\,{Q}^{2} + {a}^{2}\,{y}^{2}\,{x}^{2} - 2\,{a}^{2}\,{y}^{2}\,{M
  }\,{x} + {a}^{4}\,{y}^{2} + {a}^{2}\,{y}^{2}\,{Q}^{2}\,) $ \\
\> $ (\, - 1 + {y}\,)\,(\,{y} + 1\,)\,{\it \:d}\,{z}^{{\it 2\:}}
  \left/ {\vrule height0.37em width0em depth0.37em} \right. \! \! 
  (\,{x}^{2} + {a}^{2}\,{y}^{2}\,)\mbox{} - {\displaystyle \frac {(
  \,{x}^{2} + {a}^{2}\,{y}^{2}\,)\,{\it \:d}\,{x}^{{\it 2\:}}}{{x}
  ^{2} - 2\,{M}\,{x} + {a}^{2} + {Q}^{2}}}
  + {\displaystyle \frac {(\,{x}^{2} + {a}^{2}\,{y}^{2
  }\,)\,{\it \:d}\,{y}^{{\it 2\:}}}{(\, - 1 + {y}\,)\,(\,{y} + 1\,) }} $\\
\rule{\linewidth}{.05mm}\\
%------------------------------------------------------------------------------
Input file\hspace{10mm} \> Covariant tetrad \\
\rule[.5\baselineskip]{\linewidth}{.05mm}\\
{\tt npdnkn1.mpl} \> $ {{l}_{{a}}}= $\>$ \left[ {\vrule 
  height0.89em width0em depth0.89em} \right. \! \! {\displaystyle 
  \frac {1}{2}}\,{\displaystyle \frac {\sqrt {2}\,\sqrt {{x}^{2} - 
  2\,{M}\,{x} + {a}^{2} + {Q}^{2}}}{\sqrt {{x}^{2} + {a}^{2}\,{y}^{
  2}}}}\, ,  -\,{\displaystyle \frac {1}{2}}\,{\displaystyle \frac {
  \sqrt {2}\,{a}\,\sqrt {{x}^{2} - 2\,{M}\,{x} + {a}^{2} + {Q}^{2}}
  \,(\,1 - {y}^{2}\,)}{\sqrt {{x}^{2} + {a}^{2}\,{y}^{2}}}} , $ \\
\>\> $ - \,{\displaystyle \frac {1}{2}}\,{\displaystyle \frac {
  \sqrt {2}\,\sqrt {{x}^{2} + {a}^{2}\,{y}^{2}}}{\sqrt {{x}^{2} - 2
  \,{M}\,{x} + {a}^{2} + {Q}^{2}}}}\, , 0 \! \! \left. {\vrule 
  height0.89em width0em depth0.89em} \right] \mbox{\hspace{178pt}} $ \\
\> ${{n}_{{a}}}= $\>$ \left[ {\vrule 
  height0.89em width0em depth0.89em} \right. \! \! {\displaystyle 
  \frac {1}{2}}\,{\displaystyle \frac {\sqrt {2}\,\sqrt {{x}^{2} - 
  2\,{M}\,{x} + {a}^{2} + {Q}^{2}}}{\sqrt {{x}^{2} + {a}^{2}\,{y}^{
  2}}}}\, , -\,{\displaystyle \frac {1}{2}}\,{\displaystyle \frac {
  \sqrt {2}\,{a}\,\sqrt {{x}^{2} - 2\,{M}\,{x} + {a}^{2} + {Q}^{2}}
  \,(\,1 - {y}^{2}\,)}{\sqrt {{x}^{2} + {a}^{2}\,{y}^{2}}}} , $ \\
\>\> $ {\displaystyle \frac {1}{2}}\,{\displaystyle \frac {\sqrt {2
  }\,\sqrt {{x}^{2} + {a}^{2}\,{y}^{2}}}{\sqrt {{x}^{2} - 2\,{M}\,{
  x} + {a}^{2} + {Q}^{2}}}}\, , 0 \! \! \left. {\vrule 
  height0.89em width0em depth0.89em} \right] \mbox{\hspace{187pt}} $\\
\> ${m}_{{a}} = $\>$ \left[ \! \,{\displaystyle \frac {1}{2}}\,
  {\displaystyle \frac {\sqrt {2}\,{a}\,\sqrt {1 - {y}^{2}}}{
  \sqrt {{x}^{2} + {a}^{2}\,{y}^{2}}}}\, , -\,{\displaystyle \frac {
  1}{2}}\,{\displaystyle \frac {\sqrt {2}\,\sqrt {1 - {y}^{2}}\,(\,
  {x}^{2} + {a}^{2}\,)}{\sqrt {{x}^{2} + {a}^{2}\,{y}^{2}}}}\, , 0\, , 
   - \,{\displaystyle \frac {1}{2}}\,{\displaystyle \frac {{I}\,
  \sqrt {2}\,\sqrt {{x}^{2} + {a}^{2}\,{y}^{2}}}{\sqrt {1 - {y}^{2}
  }}}\, \!  \right] $ \\
\> $\overline{m}_a = $\>$\left[ \! \,{\displaystyle \frac {1}{2}}\,
  {\displaystyle \frac {\sqrt {2}\,{a}\,\sqrt {1 - {y}^{2}}}{
  \sqrt {{x}^{2} + {a}^{2}\,{y}^{2}}}}\, , -\,{\displaystyle \frac {
  1}{2}}\,{\displaystyle \frac {\sqrt {2}\,\sqrt {1 - {y}^{2}}\,(\,
  {x}^{2} + {a}^{2}\,)}{\sqrt {{x}^{2} + {a}^{2}\,{y}^{2}}}}\, , 0\, ,
  {\displaystyle \frac {1}{2}}\,{\displaystyle \frac {{I}\,\sqrt {2
  }\,\sqrt {{x}^{2} + {a}^{2}\,{y}^{2}}}{\sqrt {1 - {y}^{2}}}}\,
  \!  \right] $ \\
\vspace{1mm} \> Corresponding line element: \\
\> $\lefteqn{{\it \:ds}^{2}={\displaystyle \frac {(\,{a}^{2}\,{y}^{2}
  + {x}^{2} - 2\,{M}\,{x} + {Q}^{2}\,)\,{\it \:d}\,{t}^{{\it 2\:}}
  }{{x}^{2} + {a}^{2}\,{y}^{2}}}
  + 2\,{\displaystyle \frac {{a}\,(\, - 2\,{M}\,{x} + 
  {Q}^{2}\,)\,(\, - 1 + {y}\,)\,(\,{y} + 1\,)\,{\it \:d}\,{t}^{{\:}
  }\,{\it d\:}\,{z}^{{\:}}}{{x}^{2} + {a}^{2}\,{y}^{2}}}} $ \\
\> $ + (\,{x}^{4} + {x}^{2}\,{a}^{2} + 2\,{a}^{2}\,{M}\,{x} - {a}^{
  2}\,{Q}^{2} + {a}^{2}\,{y}^{2}\,{x}^{2} - 2\,{a}^{2}\,{y}^{2}\,{M
  }\,{x} + {a}^{4}\,{y}^{2} + {a}^{2}\,{y}^{2}\,{Q}^{2}\,) $ \\
\> $ (\, - 1 + {y}\,)\,(\,{y} + 1\,)\,{\it \:d}\,{z}^{{\it 2\:}}
  \left/ {\vrule height0.37em width0em depth0.37em} \right. \! \! 
  (\,{x}^{2} + {a}^{2}\,{y}^{2}\,)\mbox{} - {\displaystyle \frac {(
  \,{x}^{2} + {a}^{2}\,{y}^{2}\,)\,{\it \:d}\,{x}^{{\it 2\:}}}{{x}
  ^{2} - 2\,{M}\,{x} + {a}^{2} + {Q}^{2}}}
  + {\displaystyle \frac {(\,{x}^{2} + {a}^{2}\,{y}^{2
  }\,)\,{\it \:d}\,{y}^{{\it 2\:}}}{(\, - 1 + {y}\,)\,(\,{y} + 1\,)
  }} $ \\
%
%== Kerr-Newman (Euclidean coordinates 2) =====================================
\rule{\linewidth}{.5mm}\\
{\large Spacetime: Kerr-Newman ($x,y,z,t$) ({\bf KN-Euc2})}\\
\rule{\linewidth}{.05mm}\\
Input file\hspace{10mm} \> Contravariant tetrad \\
\rule[.5\baselineskip]{\linewidth}{.05mm}\\
{\tt npupkn2.mpl} \> ${l}^{{a}}= $\>$\left[ {\vrule 
  height0.89em width0em depth0.89em} \right. \! \! {\displaystyle 
  \frac {1}{2}}\,{\displaystyle \frac {(\,{x}^{2} + {a}^{2}\,)\,
  \sqrt {2}}{\sqrt {{x}^{2} - 2\,{M}\,{x} + {a}^{2} + {Q}^{2}}\,
  \sqrt {{x}^{2} + {a}^{2}\,{y}^{2}}}}\, , {\displaystyle \frac {1}{2
  }}\,{\displaystyle \frac {\sqrt {2}\,{a}}{\sqrt {{x}^{2} + {a}^{2
  }\,{y}^{2}}\,\sqrt {{x}^{2} - 2\,{M}\,{x} + {a}^{2} + {Q}^{2}}}} ,  $ \\
\>\> $ {\displaystyle \frac {1}{2}}\,{\displaystyle \frac {\sqrt {2
  }\,\sqrt {{x}^{2} - 2\,{M}\,{x} + {a}^{2} + {Q}^{2}}}{\sqrt {{x}
  ^{2} + {a}^{2}\,{y}^{2}}}}\, , 0 \! \! \left. {\vrule 
  height0.89em width0em depth0.89em} \right] \mbox{\hspace{188pt}} $ \\
\> ${n}^{{a}}= $\>$\left[ {\vrule 
  height0.89em width0em depth0.89em} \right. \! \! {\displaystyle 
  \frac {1}{2}}\,{\displaystyle \frac {(\,{x}^{2} + {a}^{2}\,)\,
  \sqrt {2}}{\sqrt {{x}^{2} - 2\,{M}\,{x} + {a}^{2} + {Q}^{2}}\,
  \sqrt {{x}^{2} + {a}^{2}\,{y}^{2}}}}\, , {\displaystyle \frac {1}{2
  }}\,{\displaystyle \frac {\sqrt {2}\,{a}}{\sqrt {{x}^{2} + {a}^{2
  }\,{y}^{2}}\,\sqrt {{x}^{2} - 2\,{M}\,{x} + {a}^{2} + {Q}^{2}}}} ,  $ \\
\>\> $ - \,{\displaystyle \frac {1}{2}}\,{\displaystyle \frac {
  \sqrt {2}\,\sqrt {{x}^{2} - 2\,{M}\,{x} + {a}^{2} + {Q}^{2}}}{
  \sqrt {{x}^{2} + {a}^{2}\,{y}^{2}}}}\, , 0 \! \! \left. {\vrule 
  height0.89em width0em depth0.89em} \right] \mbox{\hspace{184pt}} $\\
\> ${m}^{{a}}=$\>$ \left[ \! \,{\displaystyle \frac {1}{2}}\,
  {\displaystyle \frac {\sqrt {2}\,{a}\,\sqrt {1 - {y}^{2}}}{
  \sqrt {{x}^{2} + {a}^{2}\,{y}^{2}}}}\, , {\displaystyle \frac {1}{2
  }}\,{\displaystyle \frac {\sqrt {2}}{\sqrt {{x}^{2} + {a}^{2}\,{y
  }^{2}}\,\sqrt {1 - {y}^{2}}}}\, , 0 , \, - \,{\displaystyle \frac {1}{2
  }}\,{\displaystyle \frac {\sqrt {2}\,\sqrt { - 1 + {y}^{2}}}{
  \sqrt {{x}^{2} + {a}^{2}\,{y}^{2}}}}\, \!  \right] $ \\
\> $\overline{m}^a = $\>$\left[ \! \,{\displaystyle \frac {1}{2}}\,
  {\displaystyle \frac {\sqrt {2}\,{a}\,\sqrt {1 - {y}^{2}}}{
  \sqrt {{x}^{2} + {a}^{2}\,{y}^{2}}}}\, , {\displaystyle \frac {1}{2
  }}\,{\displaystyle \frac {\sqrt {2}}{\sqrt {{x}^{2} + {a}^{2}\,{y
  }^{2}}\,\sqrt {1 - {y}^{2}}}}\, , 0 , \,{\displaystyle \frac {1}{2}}\,
  {\displaystyle \frac {\sqrt {2}\,\sqrt { - 1 + {y}^{2}}}{\sqrt {{
  x}^{2} + {a}^{2}\,{y}^{2}}}}\, \!  \right] $ \\
\vspace{1mm} \> Corresponding line element: \\
\> $\lefteqn{{\it \:ds}^{2}={\displaystyle \frac {(\,{a}^{2}\,{y}^{2}
  + {x}^{2} - 2\,{M}\,{x} + {Q}^{2}\,)\,{\it \:d}\,{t}^{{\it 2\:}}
  }{{x}^{2} + {a}^{2}\,{y}^{2}}}
  + 2\,{\displaystyle \frac {{a}\,(\, - 2\,{M}\,{x} + 
  {Q}^{2}\,)\,(\, - 1 + {y}\,)\,(\,{y} + 1\,)\,{\it \:d}\,{t}^{{\:}
  }\,{\it d\:}\,{z}^{{\:}}}{{x}^{2} + {a}^{2}\,{y}^{2}}}} $ \\
\> $ + (\,{x}^{4} + {x}^{2}\,{a}^{2} + 2\,{a}^{2}\,{M}\,{x} - {a}^{
  2}\,{Q}^{2} + {a}^{2}\,{y}^{2}\,{x}^{2} - 2\,{a}^{2}\,{y}^{2}\,{M
  }\,{x} + {a}^{4}\,{y}^{2} + {a}^{2}\,{y}^{2}\,{Q}^{2}\,) $ \\
\> $ (\, - 1 + {y}\,)\,(\,{y} + 1\,)\,{\it \:d}\,{z}^{{\it 2\:}}
  \left/ {\vrule height0.37em width0em depth0.37em} \right. \! \! 
  (\,{x}^{2} + {a}^{2}\,{y}^{2}\,)\mbox{} - {\displaystyle \frac {(
  \,{x}^{2} + {a}^{2}\,{y}^{2}\,)\,{\it \:d}\,{x}^{{\it 2\:}}}{{x}
  ^{2} - 2\,{M}\,{x} + {a}^{2} + {Q}^{2}}}
  + {\displaystyle \frac {(\,{x}^{2} + {a}^{2}\,{y}^{2
  }\,)\,{\it \:d}\,{y}^{{\it 2\:}}}{(\, - 1 + {y}\,)\,(\,{y} + 1\,)}} $\\
\rule{\linewidth}{.05mm}\\
%------------------------------------------------------------------------------
Input file\hspace{10mm} \> Covariant tetrad \\
\rule[.5\baselineskip]{\linewidth}{.05mm}\\
{\tt npdnkn2.mpl} \> $ {{l}_{{a}}}= $\>$ \left[ {\vrule 
  height0.89em width0em depth0.89em} \right. \! \! {\displaystyle 
  \frac {1}{2}}\,{\displaystyle \frac {\sqrt {2}\,\sqrt {{x}^{2} - 
  2\,{M}\,{x} + {a}^{2} + {Q}^{2}}}{\sqrt {{x}^{2} + {a}^{2}\,{y}^{
  2}}}}\, , -\,{\displaystyle \frac {1}{2}}\,{\displaystyle \frac {
  \sqrt {2}\,{a}\,\sqrt {{x}^{2} - 2\,{M}\,{x} + {a}^{2} + {Q}^{2}}
  \,(\,1 - {y}^{2}\,)}{\sqrt {{x}^{2} + {a}^{2}\,{y}^{2}}}} , $ \\
\>\> $  - \,{\displaystyle \frac {1}{2}}\,{\displaystyle \frac {
  \sqrt {2}\,\sqrt {{x}^{2} + {a}^{2}\,{y}^{2}}}{\sqrt {{x}^{2} - 2
  \,{M}\,{x} + {a}^{2} + {Q}^{2}}}}\, , 0 \! \! \left. {\vrule 
  height0.89em width0em depth0.89em} \right] \mbox{\hspace{178pt}} $ \\
\>$ {{n}_{{a}}} = $\>$ \left[ {\vrule 
  height0.89em width0em depth0.89em} \right. \! \! {\displaystyle 
  \frac {1}{2}}\,{\displaystyle \frac {\sqrt {2}\,\sqrt {{x}^{2} - 
  2\,{M}\,{x} + {a}^{2} + {Q}^{2}}}{\sqrt {{x}^{2} + {a}^{2}\,{y}^{
  2}}}}\, , -\,{\displaystyle \frac {1}{2}}\,{\displaystyle \frac {
  \sqrt {2}\,{a}\,\sqrt {{x}^{2} - 2\,{M}\,{x} + {a}^{2} + {Q}^{2}}
  \,(\,1 - {y}^{2}\,)}{\sqrt {{x}^{2} + {a}^{2}\,{y}^{2}}}} , $ \\
\>\> ${\displaystyle \frac {1}{2}}\,{\displaystyle \frac {\sqrt {2
  }\,\sqrt {{x}^{2} + {a}^{2}\,{y}^{2}}}{\sqrt {{x}^{2} - 2\,{M}\,{
  x} + {a}^{2} + {Q}^{2}}}}\, , 0 \! \! \left. {\vrule 
  height0.89em width0em depth0.89em} \right] \mbox{\hspace{187pt}} $ \\
\> $ {{m}_{{a}}}= $\>$ \left[ \! \,{\displaystyle \frac {1}{2}}\,
  {\displaystyle \frac {\sqrt {2}\,{a}\,\sqrt {1 - {y}^{2}}}{
  \sqrt {{x}^{2} + {a}^{2}\,{y}^{2}}}}\, , -\,{\displaystyle \frac {
  1}{2}}\,{\displaystyle \frac {\sqrt {2}\,\sqrt {1 - {y}^{2}}\,(\,
  {x}^{2} + {a}^{2}\,)}{\sqrt {{x}^{2} + {a}^{2}\,{y}^{2}}}}\, , 0\, ,
  {\displaystyle \frac {1}{2}}\,{\displaystyle \frac {\sqrt {2}\,
  \sqrt {{x}^{2} + {a}^{2}\,{y}^{2}}}{\sqrt { - 1 + {y}^{2}}}}\,
  \!  \right] $ \\
\> $\overline{m}_a = $\>$ \left[ \! \,{\displaystyle \frac {1}{2}}\,
  {\displaystyle \frac {\sqrt {2}\,{a}\,\sqrt {1 - {y}^{2}}}{
  \sqrt {{x}^{2} + {a}^{2}\,{y}^{2}}}}\, , -\,{\displaystyle \frac {
  1}{2}}\,{\displaystyle \frac {\sqrt {2}\,\sqrt {1 - {y}^{2}}\,(\,
  {x}^{2} + {a}^{2}\,)}{\sqrt {{x}^{2} + {a}^{2}\,{y}^{2}}}}\, , 0\, ,
  - \,{\displaystyle \frac {1}{2}}\,{\displaystyle \frac {\sqrt {2
  }\,\sqrt {{x}^{2} + {a}^{2}\,{y}^{2}}}{\sqrt { - 1 + {y}^{2}}}}\,
  \!  \right] $ \\
\vspace{1mm} \> Corresponding line element: \\
\> $\lefteqn{{\it \:ds}^{2}={\displaystyle \frac {(\,{a}^{2}\,{y}^{2}
  + {x}^{2} - 2\,{M}\,{x} + {Q}^{2}\,)\,{\it \:d}\,{t}^{{\it 2\:}}
  }{{x}^{2} + {a}^{2}\,{y}^{2}}}
  + 2\,{\displaystyle \frac {{a}\,(\, - 2\,{M}\,{x} + 
  {Q}^{2}\,)\,(\, - 1 + {y}\,)\,(\,{y} + 1\,)\,{\it \:d}\,{t}^{{\:}
  }\,{\it d\:}\,{z}^{{\:}}}{{x}^{2} + {a}^{2}\,{y}^{2}}}} $ \\
\> $ + (\,{x}^{4} + {x}^{2}\,{a}^{2} + 2\,{a}^{2}\,{M}\,{x} - {a}^{
  2}\,{Q}^{2} + {a}^{2}\,{y}^{2}\,{x}^{2} - 2\,{a}^{2}\,{y}^{2}\,{M
  }\,{x} + {a}^{4}\,{y}^{2} + {a}^{2}\,{y}^{2}\,{Q}^{2}\,) $ \\
\> $ (\, - 1 + {y}\,)\,(\,{y} + 1\,)\,{\it \:d}\,{z}^{{\it 2\:}}
  \left/ {\vrule height0.37em width0em depth0.37em} \right. \! \! 
  (\,{x}^{2} + {a}^{2}\,{y}^{2}\,)\mbox{} - {\displaystyle \frac {(
  \,{x}^{2} + {a}^{2}\,{y}^{2}\,)\,{\it \:d}\,{x}^{{\it 2\:}}}{{x}
  ^{2} - 2\,{M}\,{x} + {a}^{2} + {Q}^{2}}}
  + {\displaystyle \frac {(\,{x}^{2} + {a}^{2}\,{y}^{2
  }\,)\,{\it \:d}\,{y}^{{\it 2\:}}}{(\, - 1 + {y}\,)\,(\,{y} + 1\,)}} $\\
%
%== Kerr-Newman (Boyer-Lindquist coordinates) =================================
\rule{\linewidth}{.5mm}\\
{\large Spacetime: Kerr-Newman (Boyer-Lindquist coordinates) ({\bf KN-BL1})}
\cite{boyer/lindquist:1967} \\
\rule{\linewidth}{.05mm}\\
Input file\hspace{10mm} \> Contravariant tetrad \\
\rule[.5\baselineskip]{\linewidth}{.05mm}\\
{\tt npupkn3.mpl} \> $ {l}^{{a}}=$\>$ \left[ \! \,
  {\displaystyle \frac {{r}^{2} + {a}^{2}}{{
  r}^{2} - 2\,{M}\,{r} + {a}^{2} + {Q}^{2}}}\, , 1 , \, , 0\,{\displaystyle 
  \frac {{a}}{{r}^{2} - 2\,{M}\,{r} + {a}^{2} + {Q}^{2}}}\, \! \right] $ \\
\> ${n}^{{a}}=$\>$ \left[ \! \,{\displaystyle \frac {1}{2}}\,
  {\displaystyle \frac {{r}^{2} + {a}^{2}}{{r}^{2} + {a}^{2}\,{\rm 
  cos}(\,{ \theta}\,)^{2}}}\, - \,{\displaystyle \frac {1}{2}}\,
  {\displaystyle \frac {{r}^{2} - 2\,{M}\,{r} + {a}^{2} + {Q}^{2}}{
  {r}^{2} + {a}^{2}\,{\rm cos}(\,{ \theta}\,)^{2}}}\,0\,
  {\displaystyle \frac {1}{2}}\,{\displaystyle \frac {{a}}{{r}^{2}
  + {a}^{2}\,{\rm cos}(\,{ \theta}\,)^{2}}}\, \!  \right] $ \\
\> ${m}^{{a}}= $\>$ \left[ {\vrule 
  height0.92em width0em depth0.92em} \right. \! \! {\displaystyle 
  \frac {1}{2}}\,{\displaystyle \frac {(\,{I}\,{\rm sin}(\,{ \theta
  }\,)\,{r} + {\rm sin}(\,{ \theta}\,)\,{a}\,{\rm cos}(\,{ \theta}
  \,)\,)\,{a}\,\sqrt {2}}{{r}^{2} + {a}^{2}\,{\rm cos}(\,{ \theta}
  \,)^{2}}}\, , 0 , \,{\displaystyle \frac {1}{2}}\,{\displaystyle 
  \frac {(\,{r} - {I}\,{a}\,{\rm cos}(\,{ \theta}\,)\,)\,\sqrt {2}
  }{{r}^{2} + {a}^{2}\,{\rm cos}(\,{ \theta}\,)^{2}}} , $ \\
\>\> $ {\displaystyle \frac {1}{2}}\,{\displaystyle \frac {{I}\,(\,
  {r} - {I}\,{a}\,{\rm cos}(\,{ \theta}\,)\,)\,\sqrt {2}}{{\rm sin}
  (\,{ \theta}\,)\, \left( \! \,{r}^{2} + {a}^{2}\,{\rm cos}(\,{ 
  \theta}\,)^{2}\, \!  \right) }} \! \! \left. {\vrule 
  height0.92em width0em depth0.92em} \right] \mbox{\hspace{180pt}} $ \\
\> $\overline{m}^a = \left[ {\vrule 
  height0.92em width0em depth0.92em} \right. \! \! {\displaystyle 
  \frac {1}{2}}\,{\displaystyle \frac {(\, - {I}\,{\rm sin}(\,{ 
  \theta}\,)\,{r} + {\rm sin}(\,{ \theta}\,)\,{a}\,{\rm cos}(\,{ 
  \theta}\,)\,)\,{a}\,\sqrt {2}}{{r}^{2} + {a}^{2}\,{\rm cos}(\,{ 
  \theta}\,)^{2}}}\, , 0 , \,{\displaystyle \frac {1}{2}}\,
  {\displaystyle \frac {(\,{r} + {I}\,{a}\,{\rm cos}(\,{ \theta}\,)
  \,)\,\sqrt {2}}{{r}^{2} + {a}^{2}\,{\rm cos}(\,{ \theta}\,)^{2}}} , $ \\
\>\> $- \,{\displaystyle \frac {1}{2}}\,{\displaystyle \frac {{I}
  \,(\,{r} + {I}\,{a}\,{\rm cos}(\,{ \theta}\,)\,)\,\sqrt {2}}{
  {\rm sin}(\,{ \theta}\,)\, \left( \! \,{r}^{2} + {a}^{2}\,{\rm 
  cos}(\,{ \theta}\,)^{2}\, \!  \right) }} \! \! \left. {\vrule 
  height0.92em width0em depth0.92em} \right] \mbox{\hspace{196pt}}$ \\
\vspace{1mm} \> Corresponding line element: \\
\> $\lefteqn{{\it \:ds}^{2}={\displaystyle \frac { \left( \! \,{a}^{2
  }\,{\rm cos}(\,{ \theta}\,)^{2} + {r}^{2} + {Q}^{2} - 2\,{M}\,{r}
  \, \!  \right) \,{\it \:d}\,{t}^{{\it 2\:}}}{{r}^{2} + {a}^{2}\,
  {\rm cos}(\,{ \theta}\,)^{2}}}
  + 2\,{\displaystyle \frac {(\, - 1 + {\rm cos}(\,{ 
  \theta}\,)\,)\,(\,{\rm cos}(\,{ \theta}\,) + 1\,)\,(\,{Q}^{2} - 2
  \,{M}\,{r}\,)\,{a}\,{\it \:d}\,{t}^{{\:}}\,{\it d\:}\,{ \phi}^{{
  \:}}}{{r}^{2} + {a}^{2}\,{\rm cos}(\,{ \theta}\,)^{2}}}} $ \\
\> $\mbox{} - {\displaystyle \frac { \left( \! \,{r}^{2} + {a}^{
  2}\,{\rm cos}(\,{ \theta}\,)^{2}\, \!  \right) \,{\it \:d}\,{r}^{
  {\it 2\:}}}{{r}^{2} - 2\,{M}\,{r} + {a}^{2} + {Q}^{2}}} + 
  \left( \! \, - {r}^{2} - {a}^{2}\,{\rm cos}(\,{ \theta}\,)^{2}\,
  \!  \right) \,{\it \:d}\,{ \theta}^{{\it 2\:}} + (\, - 1 + {\rm 
  cos}(\,{ \theta}\,)\,) $\\
\> $(\,{\rm cos}(\,{ \theta}\,) + 1\,) \left( {\vrule 
  height0.44em width0em depth0.44em} \right. \! \!  - 2\,{a}^{2}\,
  {\rm cos}(\,{ \theta}\,)^{2}\,{M}\,{r} + {a}^{4}\,{\rm cos}(\,{ 
  \theta}\,)^{2} + {r}^{2}\,{a}^{2}\,{\rm cos}(\,{ \theta}\,)^{2}$ \\
\> $ \mbox{} + {a}^{2}\,{Q}^{2}\,{\rm cos}(\,{ \theta}\,)^{2} - {
  a}^{2}\,{Q}^{2} + 2\,{a}^{2}\,{M}\,{r} + {r}^{4} + {r}^{2}\,{a}^{
  2} \! \! \left. {\vrule height0.44em width0em depth0.44em}
  \right) {\it \:d}\,{ \phi}^{{\it 2\:}} \left/ {\vrule 
  height0.44em width0em depth0.44em} \right. \! \!  \left( \! \,{r}
  ^{2} + {a}^{2}\,{\rm cos}(\,{ \theta}\,)^{2}\, \!  \right) $ \\
\rule{\linewidth}{.05mm}\\
%------------------------------------------------------------------------------
Input file\hspace{10mm} \> Covariant tetrad \\
\rule[.5\baselineskip]{\linewidth}{.05mm}\\
{\tt npdnkn3.mpl} \> ${{l}_{{a}}}=$\>$ \left[ \! \,1\, ,  - \,
  {\displaystyle \frac {{r}^{2} + 
  {a}^{2}\,{\rm cos}(\,{ \theta}\,)^{2}}{{r}^{2} - 2\,{M}\,{r} + {a
  }^{2} + {Q}^{2}}}\, , 0 , \, - {a}\,{\rm sin}(\,{ \theta}\,)^{2}\, 
  \! \right] $\\
\> $ {{n}_{{a}}}=$\>$ \left[ \! \,{\displaystyle \frac {1}{2}}\,
  {\displaystyle \frac {{r}^{2} - 2\,{M}\,{r} + {a}^{2} + {Q}^{2}}{
  {r}^{2} + {a}^{2}\,{\rm cos}(\,{ \theta}\,)^{2}}}\, ,
  {\displaystyle \frac {1}{2}}\, , 0 , \, - \,{\displaystyle \frac {1}{2
  }}\,{\displaystyle \frac {{a}\,(\,{r}^{2} - 2\,{M}\,{r} + {a}^{2}
  + {Q}^{2}\,)\,{\rm sin}(\,{ \theta}\,)^{2}}{{r}^{2} + {a}^{2}\,
  {\rm cos}(\,{ \theta}\,)^{2}}}\, \!  \right] $\\
\> $ {{m}_{{a}}} = $ \> $ \left[ {\vrule 
  height0.92em width0em depth0.92em} \right. \! \! {\displaystyle 
  \frac {1}{2}}\,{\displaystyle \frac {(\,{I}\,{\rm sin}(\,{ \theta
  }\,)\,{r} + {\rm sin}(\,{ \theta}\,)\,{a}\,{\rm cos}(\,{ \theta}
  \,)\,)\,{a}\,\sqrt {2}}{{r}^{2} + {a}^{2}\,{\rm cos}(\,{ \theta}
  \,)^{2}}}\, , 0 , \, - \,{\displaystyle \frac {1}{2}}\,(\,{r} - {I}\,{a
  }\,{\rm cos}(\,{ \theta}\,)\,)\,\sqrt {2} ,  $ \\
\>\>$  - \,{\displaystyle \frac {1}{2}}\,{\displaystyle \frac {(\,
  {I}\,{\rm sin}(\,{ \theta}\,)\,{r} + {\rm sin}(\,{ \theta}\,)\,{a
  }\,{\rm cos}(\,{ \theta}\,)\,)\,(\,{r}^{2} + {a}^{2}\,)\,\sqrt {2
  }}{{r}^{2} + {a}^{2}\,{\rm cos}(\,{ \theta}\,)^{2}}} \! 
  \! \left. {\vrule height0.92em width0em depth0.92em} \right] 
  \mbox{\hspace{84pt}}$\\
\> $\overline{m}_a = $\>$ \left[ {\vrule 
  height0.92em width0em depth0.92em} \right. \! \! {\displaystyle
  \frac {1}{2}}\,{\displaystyle \frac {(\, - {I}\,{\rm sin}(\,{ 
  \theta}\,)\,{r} + {\rm sin}(\,{ \theta}\,)\,{a}\,{\rm cos}(\,{ 
  \theta}\,)\,)\,{a}\,\sqrt {2}}{{r}^{2} + {a}^{2}\,{\rm cos}(\,{ 
  \theta}\,)^{2}}}\, , 0 , \, - \,{\displaystyle \frac {1}{2}}\,(\,{r} + 
  {I}\,{a}\,{\rm cos}(\,{ \theta}\,)\,)\,\sqrt {2} $ \\
\>\> $  - \,{\displaystyle \frac {1}{2}}\,{\displaystyle \frac {(\,
  - {I}\,{\rm sin}(\,{ \theta}\,)\,{r} + {\rm sin}(\,{ \theta}\,)
  \,{a}\,{\rm cos}(\,{ \theta}\,)\,)\,(\,{r}^{2} + {a}^{2}\,)\,
  \sqrt {2}}{{r}^{2} + {a}^{2}\,{\rm cos}(\,{ \theta}\,)^{2}}} \! 
  \! \left. {\vrule height0.92em width0em depth0.92em} \right] 
  \mbox{\hspace{104pt}} $ \\
\vspace{1mm} \> Corresponding line element: \\
\>  $ \lefteqn{{\it \:ds}^{2}={\displaystyle \frac { \left( \! \,{a}^{2
  }\,{\rm cos}(\,{ \theta}\,)^{2} + {r}^{2} + {Q}^{2} - 2\,{M}\,{r}
  \, \!  \right) \,{\it \:d}\,{t}^{{\it 2\:}}}{{r}^{2} + {a}^{2}\,
  {\rm cos}(\,{ \theta}\,)^{2}}}
  + 2\,{\displaystyle \frac {(\, - 1 + {\rm cos}(\,{ 
  \theta}\,)\,)\,(\,{\rm cos}(\,{ \theta}\,) + 1\,)\,(\,{Q}^{2} - 2
  \,{M}\,{r}\,)\,{a}\,{\it \:d}\,{t}^{{\:}}\,{\it d\:}\,{ \phi}^{{
  \:}}}{{r}^{2} + {a}^{2}\,{\rm cos}(\,{ \theta}\,)^{2}}}} $ \\
\> $ \mbox{} - {\displaystyle \frac { \left( \! \,{r}^{2} + {a}^{
  2}\,{\rm cos}(\,{ \theta}\,)^{2}\, \!  \right) \,{\it \:d}\,{r}^{
  {\it 2\:}}}{{r}^{2} - 2\,{M}\,{r} + {a}^{2} + {Q}^{2}}} + 
  \left( \! \, - {r}^{2} - {a}^{2}\,{\rm cos}(\,{ \theta}\,)^{2}\,
  \!  \right) \,{\it \:d}\,{ \theta}^{{\it 2\:}} + (\, - 1 + {\rm 
  cos}(\,{ \theta}\,)\,) $ \\
\> $ (\,{\rm cos}(\,{ \theta}\,) + 1\,) \left( {\vrule 
  height0.44em width0em depth0.44em} \right. \! \!  - 2\,{a}^{2}\,
  {\rm cos}(\,{ \theta}\,)^{2}\,{M}\,{r} + {a}^{4}\,{\rm cos}(\,{ 
  \theta}\,)^{2} + {r}^{2}\,{a}^{2}\,{\rm cos}(\,{ \theta}\,)^{2} $ \\
\> $ \mbox{} + {a}^{2}\,{Q}^{2}\,{\rm cos}(\,{ \theta}\,)^{2} - {
  a}^{2}\,{Q}^{2} + 2\,{a}^{2}\,{M}\,{r} + {r}^{4} + {r}^{2}\,{a}^{
  2} \! \! \left. {\vrule height0.44em width0em depth0.44em}
  \right) {\it \:d}\,{ \phi}^{{\it 2\:}} \left/ {\vrule 
  height0.44em width0em depth0.44em} \right. \! \!  \left( \! \,{r}
  ^{2} + {a}^{2}\,{\rm cos}(\,{ \theta}\,)^{2}\, \!  \right) $ \\
%
%== Kerr-Newman (Eddington-Finklestein coordinates) ===========================
\rule{\linewidth}{.5mm}\\
{\large Spacetime: Kerr-Newman (Eddington-Finklestein coordinates) 
({\bf KN-EF1}) \cite{mtw} }\\
\rule{\linewidth}{.05mm}\\
Input file\hspace{10mm} \> Contravariant tetrad \\
\rule[.5\baselineskip]{\linewidth}{.05mm}\\
{\tt npupkn4.mpl} \> $ {l}^{{a}}= $ \> $ [\,0 , \,1 , \,0 , \,0\,] $ \\
\> $ {n}^{{a}}= $ \> $  \left[ \! \,{\displaystyle \frac {{r}^{2} + {a}^{2}}{{
  r}^{2} + {a}^{2}\,{\rm cos}(\,{ \theta}\,)^{2}}}\, , -\,
  {\displaystyle \frac {1}{2}}\,{\displaystyle \frac {{r}^{2} - 2\,
  {m}\,{r} + {a}^{2} + {Q}^{2}}{{r}^{2} + {a}^{2}\,{\rm cos}(\,{ 
  \theta}\,)^{2}}}\, , 0 , \,{\displaystyle \frac {{a}}{{r}^{2} + {a}^{2}
  \,{\rm cos}(\,{ \theta}\,)^{2}}}\, \!  \right] $ \\
\> $ {m}^{{a}}= $ \> $ \left[ \! \,{\displaystyle \frac {1}{2}}\,
  {\displaystyle \frac {{I}\,\sqrt {2}\,{a}\,{\rm sin}(\,{ \theta}
  \,)}{{r} + {I}\,{a}\,{\rm cos}(\,{ \theta}\,)}}\, , 0 , \,
  {\displaystyle \frac {1}{2}}\,{\displaystyle \frac {\sqrt {2}}{{r
  } + {I}\,{a}\,{\rm cos}(\,{ \theta}\,)}}\, , {\displaystyle \frac {1
  }{2}}\,{\displaystyle \frac {{I}\,\sqrt {2}}{(\,{r} + {I}\,{a}\,
  {\rm cos}(\,{ \theta}\,)\,)\,{\rm sin}(\,{ \theta}\,)}}\, \! 
  \right] $ \\
\> $ \overline{m}^a = $\>$ \left[ \! \, -\,{\displaystyle \frac {1}{2}}\,
  {\displaystyle \frac {{I}\,\sqrt {2}\,{a}\,{\rm sin}(\,{ \theta}
  \,)}{{r} - {I}\,{a}\,{\rm cos}(\,{ \theta}\,)}}\, , 0\, ,
  {\displaystyle \frac {1}{2}}\,{\displaystyle \frac {\sqrt {2}}{{r
  } - {I}\,{a}\,{\rm cos}(\,{ \theta}\,)}}\, , -\,{\displaystyle 
  \frac {1}{2}}\,{\displaystyle \frac {{I}\,\sqrt {2}}{(\,{r} - {I}
  \,{a}\,{\rm cos}(\,{ \theta}\,)\,)\,{\rm sin}(\,{ \theta}\,)}}\,
  \!  \right] $ \\
\vspace{1mm} \> Corresponding line element: \\
\> $\lefteqn{{\it \:ds}^{2}={\displaystyle \frac { \left( \! \,{a}^{2
  }\,{\rm cos}(\,{ \theta}\,)^{2} + {r}^{2} - 2\,{m}\,{r} + {Q}^{2}
  \, \!  \right) \,{\it \:d}\,{u}^{{\it 2\:}}}{{r}^{2} + {a}^{2}\,
  {\rm cos}(\,{ \theta}\,)^{2}}} + 2\,{\it \:d}\,{u}^{{\:}}\,{\it d
  \:}\,{r}^{{\:}}} $ \\
\> $ \mbox{} + 2\,{\displaystyle \frac {{a}\,(\, - 1 + {\rm cos}(
  \,{ \theta}\,)\,)\,(\,{\rm cos}(\,{ \theta}\,) + 1\,)\,(\, - 2\,{
  m}\,{r} + {Q}^{2}\,)\,{\it \:d}\,{u}^{{\:}}\,{\it d\:}\,{ \phi}^{
  {\:}}}{{r}^{2} + {a}^{2}\,{\rm cos}(\,{ \theta}\,)^{2}}} $ \\
\> $ \mbox{} + 2\,(\, - 1 + {\rm cos}(\,{ \theta}\,)\,)\,(\,{\rm 
  cos}(\,{ \theta}\,) + 1\,)\,{a}\,{\it \:d}\,{r}^{{\:}}\,{\it d\:}
  \,{ \phi}^{{\:}} +  \left( \! \, - {r}^{2} - {a}^{2}\,{\rm cos}(
  \,{ \theta}\,)^{2}\, \!  \right) \,{\it \:d}\,{ \theta}^{{\it 2\:
  }} + $  \\
\> $ (\, - 1 + {\rm cos}(\,{ \theta}\,)\,)\,(\,{\rm cos}(\,{ 
  \theta}\,) + 1\,) \left( {\vrule 
  height0.44em width0em depth0.44em} \right. \! \! {r}^{2}\,{a}^{2}
  \,{\rm cos}(\,{ \theta}\,)^{2} - 2\,{a}^{2}\,{m}\,{r}\,{\rm cos}(
  \,{ \theta}\,)^{2} + {a}^{4}\,{\rm cos}(\,{ \theta}\,)^{2} $ \\
\> $ \mbox{} + {a}^{2}\,{Q}^{2}\,{\rm cos}(\,{ \theta}\,)^{2} + 2
  \,{a}^{2}\,{m}\,{r} + {r}^{4} - {a}^{2}\,{Q}^{2} + {r}^{2}\,{a}^{
  2} \! \! \left. {\vrule height0.44em width0em depth0.44em}
  \right) {\it \:d}\,{ \phi}^{{\it 2\:}} \left/ {\vrule 
  height0.44em width0em depth0.44em} \right. \! \!  \left( \! \,{r}
  ^{2} + {a}^{2}\,{\rm cos}(\,{ \theta}\,)^{2}\, \!  \right) $\\
\rule{\linewidth}{.05mm}\\
%------------------------------------------------------------------------------
Input file\hspace{10mm} \> Covariant tetrad \\
\rule[.5\baselineskip]{\linewidth}{.05mm}\\
{\tt npdnkn4.mpl} \> $ {{l}_{{a}}}=$ \> $ [\,1 , \,0 , \,0\, ,
  (\, - 1 + {\rm cos}(\,{ \theta}\,)\,)\,(
  \,{\rm cos}(\,{ \theta}\,) + 1\,)\,{a}\,] $ \\
\> $ {{n}_{{a}}}= $ \> $  \left[ {\vrule 
  height0.87em width0em depth0.87em} \right. \! \! {\displaystyle 
  \frac {1}{2}}\,{\displaystyle \frac {{r}^{2} - 2\,{m}\,{r} + {a}
  ^{2} + {Q}^{2}}{{r}^{2} + {a}^{2}\,{\rm cos}(\,{ \theta}\,)^{2}}}
  \, , 1\, , 0 , $ \\
\>\> $ {\displaystyle \frac {1}{2}}\,{\displaystyle \frac {{a}\,(\,
  - 1 + {\rm cos}(\,{ \theta}\,)\,)\,(\,{\rm cos}(\,{ \theta}\,)
  + 1\,)\,(\,{r}^{2} - 2\,{m}\,{r} + {a}^{2} + {Q}^{2}\,)}{{r}^{2}
  + {a}^{2}\,{\rm cos}(\,{ \theta}\,)^{2}}} \! \! \left. {\vrule 
  height0.87em width0em depth0.87em} \right] $ \\
\> $ {{m}_{{a}}}= $ \> $ \left[ \! \,{\displaystyle \frac {1}{2}}\,
  {\displaystyle \frac {{I}\,\sqrt {2}\,{a}\,{\rm sin}(\,{ \theta}
  \,)}{{r} + {I}\,{a}\,{\rm cos}(\,{ \theta}\,)}}\, , 0\, ,
  {\displaystyle \frac {1}{2}}\,(\, - {r} + {I}\,{a}\,{\rm cos}(\,{
  \theta}\,)\,)\,\sqrt {2}\,  , -\,{\displaystyle \frac {1}{2}}\,
  {\displaystyle \frac {{I}\,\sqrt {2}\,{\rm sin}(\,{ \theta}\,)\,(
  \,{r}^{2} + {a}^{2}\,)}{{r} + {I}\,{a}\,{\rm cos}(\,{ \theta}\,)
  }}\, \!  \right] $ \\
\> $ \overline{m}_a = $ \> $ \left[ {\vrule 
  height0.84em width0em depth0.84em} \right.
  {\displaystyle \frac {1}{2}}\,{\displaystyle \frac {{I}\,
  \sqrt {2}\,{a}\,{\rm sin}(\,{ \theta}\,)}{ - {r} + {I}\,{a}\,
  {\rm cos}(\,{ \theta}\,)}}\, , 0 , \, -\,{\displaystyle \frac {1}{2}}
  \,(\,{r} + {I}\,{a}\,{\rm cos}(\,{ \theta}\,)\,)\,\sqrt {2}\, 
  , -\,{\displaystyle \frac {1}{2}}\,{\displaystyle \frac {{I}\,
  \sqrt {2}\,{\rm sin}(\,{ \theta}\,)\,(\,{r}^{2} + {a}^{2}\,)}{ - 
  {r} + {I}\,{a}\,{\rm cos}(\,{ \theta}\,)}}
  \left. {\vrule height0.84em width0em depth0.84em}\right] $ \\
\vspace{1mm} \> Corresponding line element: \\
\> $\lefteqn{{\it \:ds}^{2}={\displaystyle \frac { \left( \! \,{a}^{2
  }\,{\rm cos}(\,{ \theta}\,)^{2} + {r}^{2} - 2\,{m}\,{r} + {Q}^{2}
  \, \!  \right) \,{\it \:d}\,{u}^{{\it 2\:}}}{{r}^{2} + {a}^{2}\,
  {\rm cos}(\,{ \theta}\,)^{2}}} + 2\,{\it \:d}\,{u}^{{\:}}\,{\it d
  \:}\,{r}^{{\:}}} $ \\
\> $ \mbox{} + 2\,{\displaystyle \frac {{a}\,(\, - 1 + {\rm cos}(
  \,{ \theta}\,)\,)\,(\,{\rm cos}(\,{ \theta}\,) + 1\,)\,(\, - 2\,{
  m}\,{r} + {Q}^{2}\,)\,{\it \:d}\,{u}^{{\:}}\,{\it d\:}\,{ \phi}^{
  {\:}}}{{r}^{2} + {a}^{2}\,{\rm cos}(\,{ \theta}\,)^{2}}} $ \\
\> $\mbox{} + 2\,(\, - 1 + {\rm cos}(\,{ \theta}\,)\,)\,(\,{\rm 
  cos}(\,{ \theta}\,) + 1\,)\,{a}\,{\it \:d}\,{r}^{{\:}}\,{\it d\:}
  \,{ \phi}^{{\:}} +  \left( \! \, - {r}^{2} - {a}^{2}\,{\rm cos}(
  \,{ \theta}\,)^{2}\, \!  \right) \,{\it \:d}\,{ \theta}^{{\it 2\:
  }} + $  \\
\> $ (\, - 1 + {\rm cos}(\,{ \theta}\,)\,)\,(\,{\rm cos}(\,{ 
  \theta}\,) + 1\,) \left( {\vrule 
  height0.44em width0em depth0.44em} \right. \! \! {r}^{2}\,{a}^{2}
  \,{\rm cos}(\,{ \theta}\,)^{2} - 2\,{a}^{2}\,{m}\,{r}\,{\rm cos}(
  \,{ \theta}\,)^{2} + {a}^{4}\,{\rm cos}(\,{ \theta}\,)^{2} $ \\
\> $ \mbox{} + {a}^{2}\,{Q}^{2}\,{\rm cos}(\,{ \theta}\,)^{2} + 2
  \,{a}^{2}\,{m}\,{r} + {r}^{4} - {a}^{2}\,{Q}^{2} + {r}^{2}\,{a}^{
  2} \! \! \left. {\vrule height0.44em width0em depth0.44em}
  \right) {\it \:d}\,{ \phi}^{{\it 2\:}} \left/ {\vrule 
  height0.44em width0em depth0.44em} \right. \! \!  \left( \! \,{r}
 ^{2} + {a}^{2}\,{\rm cos}(\,{ \theta}\,)^{2}\, \!  \right) $ \\
%
%== Kerr-Newman (Boyer-Lindquist coordinates) =================================
\rule{\linewidth}{.5mm}\\
{\large Spacetime: Kerr-Newman (Boyer-Lindquist coordinates, 
$u=a\cos\theta$) ({\bf KN-BL2})}
\cite{boyer/lindquist:1967} \\
\rule{\linewidth}{.05mm}\\
Input file\hspace{10mm} \> Contravariant tetrad \\
\rule[.5\baselineskip]{\linewidth}{.05mm}\\
{\tt npupkn5.mpl} \> ${l}^{{a}}=$\>$ \left[ \! \,
  {\displaystyle \frac {{r}^{2} + {a}^{2}}{{
  r}^{2} - 2\,{M}\,{r} + {a}^{2} + {Q}^{2}}}\, , 1\, , 0 , \,{\displaystyle 
  \frac {{a}}{{r}^{2} - 2\,{M}\,{r} + {a}^{2} + {Q}^{2}}}\, \! \right] $\\
\> ${n}^{{a}}= $ \> $ \left[ \! \,{\displaystyle \frac {1}{2}}\,
  {\displaystyle \frac {{r}^{2} + {a}^{2}}{{r}^{2} + {u}^{2}}}\, , 
  -\,{\displaystyle \frac {1}{2}}\,{\displaystyle \frac {{r}^{2} - 2
  \,{M}\,{r} + {a}^{2} + {Q}^{2}}{{r}^{2} + {u}^{2}}}\, , 0 , \,
  {\displaystyle \frac {1}{2}}\,{\displaystyle \frac {{a}}{{r}^{2}
  + {u}^{2}}}\, \!  \right] $ \\
\> $ {m}^{{a}}= $ \> $ \left[ {\vrule 
  height0.89em width0em depth0.89em} \right. \! \! {\displaystyle 
  \frac {1}{2}}\,{\displaystyle \frac { \left( \! \,{I}\,\sqrt {{a}
  ^{2} - {u}^{2}}\,{r} + \sqrt {{a}^{2} - {u}^{2}}\,{u}\, \! 
  \right) \,\sqrt {2}}{{r}^{2} + {u}^{2}}}\, , 0 , \,
   -\,{\displaystyle \frac {1}{2}}\,{\displaystyle \frac {\sqrt {{a}^{2
  } - {u}^{2}}\,(\,{r} - {I}\,{u}\,)\,\sqrt {2}}{{r}^{2} + {u}^{2}}} , $ \\
\>\> $ {\displaystyle \frac {1}{2}}\,{\displaystyle \frac {{I}\,(\,
  {r} - {I}\,{u}\,)\,{a}\,\sqrt {2}}{\sqrt {{a}^{2} - {u}^{2}}\,(\,
  {r}^{2} + {u}^{2}\,)}} \! \! \left. {\vrule 
  height0.89em width0em depth0.89em} \right] \mbox{\hspace{195pt}} $ \\
\> $ \overline{m}^a = $ \> $ \left[ {\vrule 
  height0.89em width0em depth0.89em} \right. \! \! {\displaystyle 
  \frac {1}{2}}\,{\displaystyle \frac { \left( \! \, - {I}\,\sqrt {
  {a}^{2} - {u}^{2}}\,{r} + \sqrt {{a}^{2} - {u}^{2}}\,{u}\, \! 
  \right) \,\sqrt {2}}{{r}^{2} + {u}^{2}}}\, , 0 , \,
  -\,{\displaystyle \frac {1}{2}}\,{\displaystyle \frac {\sqrt {{a}^{2
  } - {u}^{2}}\,(\,{r} + {I}\,{u}\,)\,\sqrt {2}}{{r}^{2} + {u}^{2}}}, $ \\
\>\> $  - \,{\displaystyle \frac {1}{2}}\,{\displaystyle \frac {{I}
  \,{a}\,(\,{r} + {I}\,{u}\,)\,\sqrt {2}}{\sqrt {{a}^{2} - {u}^{2}}
  \,(\,{r}^{2} + {u}^{2}\,)}} \! \! \left. {\vrule 
  height0.89em width0em depth0.89em} \right] \mbox{\hspace{212pt}} $ \\
\vspace{1mm} \> Corresponding line element: \\
\> $ \lefteqn{{\it \:ds}^{2}={\displaystyle \frac {(\,{u}^{2} - 2\,{M}
  \,{r} + {r}^{2} + {Q}^{2}\,)\,{\it \:d}\,{t}^{{\it 2\:}}}{{r}^{2}
  + {u}^{2}}} - 2\,{\displaystyle \frac {(\,{Q}^{2} - 2\,{M}\,{r}
  \,)\,(\,{a} - {u}\,)\,(\,{a} + {u}\,)\,{\it \:d}\,{t}^{{\:}}\,
  {\it d\:}\,{ \phi}^{{\:}}}{{a}\,(\,{r}^{2} + {u}^{2}\,)}}} $ \\
\> $ \mbox{} - {\displaystyle \frac {(\,{r}^{2} + {u}^{2}\,)\,
  {\it \:d}\,{r}^{{\it 2\:}}}{{r}^{2} - 2\,{M}\,{r} + {a}^{2} + {Q}
  ^{2}}} + {\displaystyle \frac {(\,{I}\,{r} + {u}\,)\,(\,{I}\,{r}
  - {u}\,)\,{\it \:d}\,{u}^{{\it 2\:}}}{(\,{a} - {u}\,)\,(\,{a} + 
  {u}\,)}} + $  \\
\> $ (\,2\,{u}^{2}\,{M}\,{r} - {a}^{2}\,{u}^{2} - {r}^{2}\,{u}^{2
  } - {Q}^{2}\,{u}^{2} - {r}^{4} - {r}^{2}\,{a}^{2} + {a}^{2}\,{Q}
  ^{2} - 2\,{a}^{2}\,{M}\,{r}\,)\,(\,{a} - {u}\,) $ \\
\> $  (\,{a} + {u}\,)\,{\it \:d}\,{ \phi}^{{\it 2\:}} \left/ 
  {\vrule height0.37em width0em depth0.37em} \right. \! \! (\,(\,{r
  }^{2} + {u}^{2}\,)\,{a}^{2}\,)\mbox{\hspace{189pt}} $ \\
\rule{\linewidth}{.05mm}\\
%------------------------------------------------------------------------------
Input file\hspace{10mm} \> Covariant tetrad \\
\rule[.5\baselineskip]{\linewidth}{.05mm}\\
{\tt npdnkn5.mpl} \> ${{l}_{{a}}}= $\>$\left[ \! \,1\, , 
  -\,{\displaystyle \frac {{r}^{2} + 
  {u}^{2}}{{r}^{2} - 2\,{M}\,{r} + {a}^{2} + {Q}^{2}}}\, , 0 , \,
  -\,{\displaystyle \frac {{a}^{2} - {u}^{2}}{{a}}}\, \!  \right] $ \\
\> $ {{n}_{{a}}}= $ \> $ \left[ \! \,{\displaystyle \frac {1}{2}}\,
  {\displaystyle \frac {{r}^{2} - 2\,{M}\,{r} + {a}^{2} + {Q}^{2}}{
  {r}^{2} + {u}^{2}}}\, , {\displaystyle \frac {1}{2}}\, , 0 , \, 
  -\,{\displaystyle \frac {1}{2}}\,{\displaystyle \frac {(\,{r}^{2} - 
  2\,{M}\,{r} + {a}^{2} + {Q}^{2}\,)\,(\,{a}^{2} - {u}^{2}\,)}{{a}
  \,(\,{r}^{2} + {u}^{2}\,)}}\, \!  \right] $ \\
\> $ {m}_{{a}}= $ \> $ \left[ {\vrule 
  height0.89em width0em depth0.89em} \right. \! \! {\displaystyle 
  \frac {1}{2}}\,{\displaystyle \frac {\sqrt {{a}^{2} - {u}^{2}}\,(
  \,{I}\,{r} + {u}\,)\,\sqrt {2}}{{r}^{2} + {u}^{2}}}\, , 0\, , 
  {\displaystyle \frac {1}{2}}\,{\displaystyle \frac {(\,{r}^{2} + 
  {u}^{2}\,)\,\sqrt {2}}{\sqrt {{a}^{2} - {u}^{2}}\,(\,{r} + {I}\,{
  u}\,)}} , $ \\
\>\> $  -\,{\displaystyle \frac {1}{2}}\,{\displaystyle \frac {(\,
  {I}\,{r}^{3} + {u}\,{r}^{2} + {I}\,{a}^{2}\,{r} + {u}\,{a}^{2}\,)
  \,\sqrt {{a}^{2} - {u}^{2}}\,\sqrt {2}}{{a}\,(\,{r}^{2} + {u}^{2}
  \,)}} \! \! \left. {\vrule height0.89em width0em depth0.89em}
  \right] \mbox{\hspace{22pt}} $ \\
\> $ \overline{m}_a = $ \> $ \left[ {\vrule 
  height0.89em width0em depth0.89em} \right. \! \!  - \,
  {\displaystyle \frac {1}{2}}\,{\displaystyle \frac {\sqrt {{a}^{2
  } - {u}^{2}}\,(\,{I}\,{r} - {u}\,)\,\sqrt {2}}{{r}^{2} + {u}^{2}
  }}\, , 0\, ,  -\,{\displaystyle \frac {1}{2}}\,{\displaystyle \frac {(
  \,{r}^{2} + {u}^{2}\,)\,\sqrt {2}}{\sqrt {{a}^{2} - {u}^{2}}\,(\,
  - {r} + {I}\,{u}\,)}} , $ \\
\>\> $ {\displaystyle \frac {1}{2}}\,{\displaystyle \frac {(\,{I}\,
  {r}^{3} - {u}\,{r}^{2} + {I}\,{a}^{2}\,{r} - {u}\,{a}^{2}\,)\,
  \sqrt {{a}^{2} - {u}^{2}}\,\sqrt {2}}{{a}\,(\,{r}^{2} + {u}^{2}\,
  )}} \! \! \left. {\vrule height0.89em width0em depth0.89em}
  \right] \mbox{\hspace{67pt}} $ \\
\vspace{1mm} \> Corresponding line element: \\
\> $\lefteqn{{\it \:ds}^{2}={\displaystyle \frac {(\,{u}^{2} - 2\,{M}
  \,{r} + {r}^{2} + {Q}^{2}\,)\,{\it \:d}\,{t}^{{\it 2\:}}}{{r}^{2}
  + {u}^{2}}} - 2\,{\displaystyle \frac {(\,{Q}^{2} - 2\,{M}\,{r}
  \,)\,(\,{a} - {u}\,)\,(\,{a} + {u}\,)\,{\it \:d}\,{t}^{{\:}}\,
  {\it d\:}\,{ \phi}^{{\:}}}{{a}\,(\,{r}^{2} + {u}^{2}\,)}}} $\\
\> $ \mbox{} - {\displaystyle \frac {(\,{r}^{2} + {u}^{2}\,)\,
  {\it \:d}\,{r}^{{\it 2\:}}}{{r}^{2} - 2\,{M}\,{r} + {a}^{2} + {Q}
  ^{2}}} + {\displaystyle \frac {(\,{I}\,{r} + {u}\,)\,(\,{I}\,{r}
  - {u}\,)\,{\it \:d}\,{u}^{{\it 2\:}}}{(\,{a} - {u}\,)\,(\,{a} + 
  {u}\,)}} + $ \\
\> $ (\,2\,{u}^{2}\,{M}\,{r} - {a}^{2}\,{u}^{2} - {r}^{2}\,{u}^{2
  } - {Q}^{2}\,{u}^{2} - {r}^{4} - {r}^{2}\,{a}^{2} + {a}^{2}\,{Q}
  ^{2} - 2\,{a}^{2}\,{M}\,{r}\,)\,(\,{a} - {u}\,) $ \\
\> $ (\,{a} + {u}\,)\,{\it \:d}\,{ \phi}^{{\it 2\:}} \left/ 
  {\vrule height0.37em width0em depth0.37em} \right. \! \! (\,(\,{r
  }^{2} + {u}^{2}\,)\,{a}^{2}\,)\mbox{\hspace{189pt}} $ \\
%
%== Kerr-Newman (Eddington-Finklestein coordinates) ===========================
\rule{\linewidth}{.5mm}\\
{\large Spacetime: Kerr-Newman (Eddington-Finklestein coordinates) 
({\bf KN-EF2}) \cite{mtw} } \\
\rule{\linewidth}{.05mm}\\
Input file\hspace{10mm} \> Contravariant tetrad \\
\rule[.5\baselineskip]{\linewidth}{.05mm}\\
{\tt npupkn6.mpl} \> $ {l}^{{a}}= $ \> $ [\,0 , \,1 , \,0 , \,0\,] $ \\
\> $ {n}^{{a}}= $ \> $ \left[ \! \,{\displaystyle \frac {{r}^{2} + {a}^{2}}{{
  r}^{2} + {a}^{2}\,{\rm cos}(\,{ \theta}\,)^{2}}}\, , 
  -\,{\displaystyle \frac {1}{2}}\,{\displaystyle \frac {{r}^{2} - 2\,
  {m}\,{r} + {a}^{2} + {Q}^{2}}{{r}^{2} + {a}^{2}\,{\rm cos}(\,{ 
  \theta}\,)^{2}}}\, , 0 , \,{\displaystyle \frac {{a}}{{r}^{2} + {a}^{2}
  \,{\rm cos}(\,{ \theta}\,)^{2}}}\, \!  \right] $ \\
\> $ {m}^{{a}}= $ \> $ \left[ \! \,{\displaystyle \frac {1}{2}}\,
  {\displaystyle \frac {{I}\,\sqrt {2}\,{a}\,{\rm sin}(\,{ \theta}
  \,)}{{r} + {I}\,{a}\,{\rm cos}(\,{ \theta}\,)}}\, , 0\, ,
  {\displaystyle \frac {1}{2}}\,{\displaystyle \frac {{I}\,\sqrt {2
  }}{{I}\,{r} - {a}\,{\rm cos}(\,{ \theta}\,)}}\, , 
  -\,{\displaystyle \frac {1}{2}}\,{\displaystyle \frac {\sqrt {2}}{(
  \,{I}\,{r} - {a}\,{\rm cos}(\,{ \theta}\,)\,)\,{\rm sin}(\,{ 
  \theta}\,)}}\, \!  \right] $ \\
\> $ \overline{m}^a = $ \> $ \left[ \! \, - \,{\displaystyle \frac {1}{2}}\,
  {\displaystyle \frac {{I}\,\sqrt {2}\,{a}\,{\rm sin}(\,{ \theta}
  \,)}{{r} - {I}\,{a}\,{\rm cos}(\,{ \theta}\,)}}\, , 0\, ,
  {\displaystyle \frac {1}{2}}\,{\displaystyle \frac {{I}\,\sqrt {2
  }}{{I}\,{r} + {a}\,{\rm cos}(\,{ \theta}\,)}}\, , {\displaystyle 
  \frac {1}{2}}\,{\displaystyle \frac {\sqrt {2}}{(\,{I}\,{r} + {a}
  \,{\rm cos}(\,{ \theta}\,)\,)\,{\rm sin}(\,{ \theta}\,)}}\, \! \right] $\\
\vspace{1mm} \> Corresponding line element: \\
\> $\lefteqn{{\it \:ds}^{2}={\displaystyle \frac { \left( \! \,{a}^{2
  }\,{\rm cos}(\,{ \theta}\,)^{2} + {r}^{2} - 2\,{m}\,{r} + {Q}^{2}
  \, \!  \right) \,{\it \:d}\,{u}^{{\it 2\:}}}{{r}^{2} + {a}^{2}\,
  {\rm cos}(\,{ \theta}\,)^{2}}} + 2\,{\it \:d}\,{u}^{{\:}}\,{\it d
  \:}\,{r}^{{\:}}} $\\
\> $  \mbox{} + 2\,{\displaystyle \frac {{a}\,(\, - 1 + {\rm cos}(
  \,{ \theta}\,)\,)\,(\,{\rm cos}(\,{ \theta}\,) + 1\,)\,(\, - 2\,{
  m}\,{r} + {Q}^{2}\,)\,{\it \:d}\,{u}^{{\:}}\,{\it d\:}\,{ \phi}^{
  {\:}}}{{r}^{2} + {a}^{2}\,{\rm cos}(\,{ \theta}\,)^{2}}} $ \\
\> $ \mbox{} + 2\,(\, - 1 + {\rm cos}(\,{ \theta}\,)\,)\,(\,{\rm 
  cos}(\,{ \theta}\,) + 1\,)\,{a}\,{\it \:d}\,{r}^{{\:}}\,{\it d\:}
  \,{ \phi}^{{\:}} +  \left( \! \, - {r}^{2} - {a}^{2}\,{\rm cos}(
  \,{ \theta}\,)^{2}\, \!  \right) \,{\it \:d}\,{ \theta}^{{\it 2\:
  }} + $ \\
\> $ (\, - 1 + {\rm cos}(\,{ \theta}\,)\,)\,(\,{\rm cos}(\,{ 
  \theta}\,) + 1\,) \left( {\vrule 
  height0.44em width0em depth0.44em} \right. \! \! {r}^{2}\,{a}^{2}
  \,{\rm cos}(\,{ \theta}\,)^{2} - 2\,{a}^{2}\,{m}\,{r}\,{\rm cos}(
  \,{ \theta}\,)^{2} + {a}^{4}\,{\rm cos}(\,{ \theta}\,)^{2} $ \\
\> $ \mbox{} + {a}^{2}\,{Q}^{2}\,{\rm cos}(\,{ \theta}\,)^{2} + 2
  \,{a}^{2}\,{m}\,{r} + {r}^{4} - {a}^{2}\,{Q}^{2} + {r}^{2}\,{a}^{
  2} \! \! \left. {\vrule height0.44em width0em depth0.44em}
  \right) {\it \:d}\,{ \phi}^{{\it 2\:}} \left/ {\vrule 
  height0.44em width0em depth0.44em} \right. \! \!  \left( \! \,{r}
  ^{2} + {a}^{2}\,{\rm cos}(\,{ \theta}\,)^{2}\, \!  \right) $ \\
\rule{\linewidth}{.05mm}\\
%------------------------------------------------------------------------------
Input file\hspace{10mm} \> Covariant tetrad \\
\rule[.5\baselineskip]{\linewidth}{.05mm}\\
{\tt npdnkn6.mpl} \> $ {{l}_{{a}}}= $ \> $ [\,1 , \,0 , \,0 , \,(\, 
  - 1 + {\rm cos}(\,{ \theta}\,)\,)\,( \,{\rm cos}(\,{ \theta}\,) 
  + 1\,)\,{a}\,] $ \\
\> $ {{n}_{{a}}}= $ \> $ \left[ {\vrule 
  height0.87em width0em depth0.87em} \right. \! \! {\displaystyle 
  \frac {1}{2}}\,{\displaystyle \frac {{r}^{2} - 2\,{m}\,{r} + {a}
  ^{2} + {Q}^{2}}{{r}^{2} + {a}^{2}\,{\rm cos}(\,{ \theta}\,)^{2}}}
  \, , 1\, , 0 , $ \\
\> $ {\displaystyle \frac {1}{2}}\,{\displaystyle \frac {{a}\,(\,
  - 1 + {\rm cos}(\,{ \theta}\,)\,)\,(\,{\rm cos}(\,{ \theta}\,)
  + 1\,)\,(\,{r}^{2} - 2\,{m}\,{r} + {a}^{2} + {Q}^{2}\,)}{{r}^{2}
  + {a}^{2}\,{\rm cos}(\,{ \theta}\,)^{2}}} \! \! \left. {\vrule 
  height0.87em width0em depth0.87em} \right] $ \\
\> $ {{m}_{{a}}}= $ \> $ \left[ \! \,{\displaystyle \frac {1}{2}}\,
  {\displaystyle \frac {{I}\,\sqrt {2}\,{a}\,{\rm sin}(\,{ \theta}
  \,)}{{r} + {I}\,{a}\,{\rm cos}(\,{ \theta}\,)}}\, , 0\, ,
  {\displaystyle \frac {1}{2}}\,(\, - {r} + {I}\,{a}\,{\rm cos}(\,{
  \theta}\,)\,)\,\sqrt {2} , \,{\displaystyle \frac {1}{2}}\,
  {\displaystyle \frac {\sqrt {2}\,{\rm sin}(\,{ \theta}\,)\,(\,{r}
  ^{2} + {a}^{2}\,)}{{I}\,{r} - {a}\,{\rm cos}(\,{ \theta}\,)}}\,
  \!  \right] $ \\
\> $ \overline{m}_a = $\>$ \left[ \! \,{\displaystyle \frac {1}{2}}\,
  {\displaystyle 
  \frac {{I}\,\sqrt {2}\,{a}\,{\rm sin}(\,{ \theta}\,)}{ - {r} + {I
  }\,{a}\,{\rm cos}(\,{ \theta}\,)}}\, , 0\, ,  -\,{\displaystyle 
  \frac {1}{2}}\,(\,{r} + {I}\,{a}\,{\rm cos}(\,{ \theta}\,)\,)\,
  \sqrt {2}\, , {\displaystyle \frac {1}{2}}\,{\displaystyle \frac {
  \sqrt {2}\,{\rm sin}(\,{ \theta}\,)\,(\,{r}^{2} + {a}^{2}\,)}{ - 
  {I}\,{r} - {a}\,{\rm cos}(\,{ \theta}\,)}}\, \!  \right] $ \\
\vspace{1mm} \> Corresponding line element: \\
\> $\lefteqn{{\it \:ds}^{2}={\displaystyle \frac { \left( \! \,{a}^{2
  }\,{\rm cos}(\,{ \theta}\,)^{2} + {r}^{2} - 2\,{m}\,{r} + {Q}^{2}
  \, \!  \right) \,{\it \:d}\,{u}^{{\it 2\:}}}{{r}^{2} + {a}^{2}\,
  {\rm cos}(\,{ \theta}\,)^{2}}} + 2\,{\it \:d}\,{u}^{{\:}}\,{\it d
  \:}\,{r}^{{\:}}} $ \\
\> $  + 2\,{\displaystyle \frac {{a}\,(\, - 1 + {\rm cos}(
  \,{ \theta}\,)\,)\,(\,{\rm cos}(\,{ \theta}\,) + 1\,)\,(\, - 2\,{
  m}\,{r} + {Q}^{2}\,)\,{\it \:d}\,{u}^{{\:}}\,{\it d\:}\,{ \phi}^{
  {\:}}}{{r}^{2} + {a}^{2}\,{\rm cos}(\,{ \theta}\,)^{2}}} $ \\
\> $ \mbox{} + 2\,(\, - 1 + {\rm cos}(\,{ \theta}\,)\,)\,(\,{\rm 
  cos}(\,{ \theta}\,) + 1\,)\,{a}\,{\it \:d}\,{r}^{{\:}}\,{\it d\:}
  \,{ \phi}^{{\:}} +  \left( \! \, - {r}^{2} - {a}^{2}\,{\rm cos}(
  \,{ \theta}\,)^{2}\, \!  \right) \,{\it \:d}\,{ \theta}^{{\it 2\:
  }} + $  \\
\> $ (\, - 1 + {\rm cos}(\,{ \theta}\,)\,)\,(\,{\rm cos}(\,{ 
  \theta}\,) + 1\,) \left( {\vrule 
  height0.44em width0em depth0.44em} \right. \! \! {r}^{2}\,{a}^{2}
  \,{\rm cos}(\,{ \theta}\,)^{2} - 2\,{a}^{2}\,{m}\,{r}\,{\rm cos}(
  \,{ \theta}\,)^{2} + {a}^{4}\,{\rm cos}(\,{ \theta}\,)^{2} $ \\
\> $ \mbox{} + {a}^{2}\,{Q}^{2}\,{\rm cos}(\,{ \theta}\,)^{2} + 2
  \,{a}^{2}\,{m}\,{r} + {r}^{4} - {a}^{2}\,{Q}^{2} + {r}^{2}\,{a}^{
  2} \! \! \left. {\vrule height0.44em width0em depth0.44em}
  \right) {\it \:d}\,{ \phi}^{{\it 2\:}} \left/ {\vrule 
  height0.44em width0em depth0.44em} \right. \! \!  \left( \! \,{r}
  ^{2} + {a}^{2}\,{\rm cos}(\,{ \theta}\,)^{2}\, \!  \right) $ \\
\rule{\linewidth}{.5mm}\\
%==============================================================================
\end{tabbing}

%==============================================================================
\subsection{Mixmaster spacetime (Table 3)}
%==============================================================================
For this set of tests, not only are the basis vectors varied, but also
their inner product. The input used for these tests is listed below.

\begin{tabbing}
%== Mixmaster =================================================================
\rule{\linewidth}{.5mm}\\
{\large Spacetime: Mixmaster ({\bf Mix}) \cite{mtw2} } \\
\rule{\linewidth}{.05mm}\\
Input file\hspace{10mm} \= Frame \\
\rule[.5\baselineskip]{\linewidth}{.05mm}\\
{\tt mix.mpl} \>  Inner product of basis vectors: \\
\> $ \eta^{(a)(b)} = \text{diag} \left(
  {\rm e}^{(\,2\,{\rm a}(\,{t}\,)\,)},
  {\rm e}^{(\,2\,{\rm b}(\,{t}\,)\,)},
  {\rm e}^{(\,2\,{\rm c}(\,{t}\,)\,)},
  -{\rm e}^{(\,2\,{\rm a}(\,{t}\,) + 2\,{\rm b}(\,{t}
  \,) + 2\,{\rm c}(\,{t}\,)\,)}
  \right) $ \\
\> Basis vectors: \\
\> $ {{ \omega }_{{1a}}}= $ \= $ [\,{\rm cos}(\,{ \psi}\,)\, , 
  {\rm sin}(\,{ \psi}\,)\,{\rm sin}(\,{ \theta}\,)\, , 0\, , 0\,] $ \\
\> $ {{ \omega }_{{2a}}}=[\,{\rm sin}(\,{ \psi}\,)\, ,  -{\rm cos}(\,{ 
  \psi}\,)\,{\rm sin}(\,{ \theta}\,)\, , 0\, , 0 \,] $ \\
\> $ {{ \omega }_{{3a}}}=$\>$[\,0 , \,{\rm cos}(\,{ \theta}\,)\, ,
  1\, , 0\,]$ \\
\> $ {{ \omega }_{{4a}}}=[\,0 , \,0 , \,0 , \,1\,]$ \\
\vspace{1mm} \> Corresponding line element: \\
\> $ \lefteqn{{\it \:ds}^{2}= \left( \! \,{\rm cos}(\,{ \psi}\,)^{2}\,
  (\,{\rm e}^{{\rm a}(\,{t}\,)}\,)^{2} + {\rm sin}(\,{ \psi}\,)^{2}
  \,(\,{\rm e}^{{\rm b}(\,{t}\,)}\,)^{2}\, \!  \right) \,{\it \:d}
  \,{ \theta}^{{\it 2\:}}} $ \\
\> $ \mbox{} - 2\,{\rm sin}(\,{ \psi}\,)\,{\rm cos}(\,{ \psi}\,)
  \,{\rm sin}(\,{ \theta}\,)\,(\,{\rm e}^{{\rm b}(\,{t}\,)} - {\rm 
  e}^{{\rm a}(\,{t}\,)}\,)\,(\,{\rm e}^{{\rm b}(\,{t}\,)} + {\rm e}
  ^{{\rm a}(\,{t}\,)}\,)\,{\it \:d}\,{ \theta}^{{\:}}\,{\it d\:}\,{
  \phi}^{{\:}} + $ \\
\> $\left( \! \,{\rm sin}(\,{ \psi}\,)^{2}\,{\rm sin}(\,{ 
  \theta}\,)^{2}\,(\,{\rm e}^{{\rm a}(\,{t}\,)}\,)^{2} + {\rm cos}(
  \,{ \psi}\,)^{2}\,{\rm sin}(\,{ \theta}\,)^{2}\,(\,{\rm e}^{{\rm 
  b}(\,{t}\,)}\,)^{2} + {\rm cos}(\,{ \theta}\,)^{2}\,(\,{\rm e}^{
  {\rm c}(\,{t}\,)}\,)^{2}\, \!  \right) \,{\it \:d} $ \\
\> $ { \phi}^{{\it 2\:}}\mbox{} + 2\,(\,{\rm e}^{{\rm c}(\,{t}\,)
  }\,)^{2}\,{\rm cos}(\,{ \theta}\,)\,{\it \:d}\,{ \phi}^{{\:}}\,
  {\it d\:}\,{ \psi}^{{\:}} + (\,{\rm e}^{{\rm c}(\,{t}\,)}\,)^{2}
  \,{\it \:d}\,{ \psi}^{{\it 2\:}}
  - (\,{\rm e}^{{\rm a}(\,{t}\,)}\,)^{2}\,(\,{\rm e}^{
  {\rm b}(\,{t}\,)}\,)^{2}\,(\,{\rm e}^{{\rm c}(\,{t}\,)}\,)^{2}\,
  {\it \:d}\,{t}^{{\it 2\:}} $ \\
%
%== Mixmaster 1 ===============================================================
\rule{\linewidth}{.5mm}\\
{\large Spacetime: Mixmaster ({\bf Mix1}) \cite{mtw2} } \\
\rule{\linewidth}{.05mm}\\
Input file\hspace{10mm} \> Frame \\
\rule[.5\baselineskip]{\linewidth}{.05mm}\\
{\tt mix1.mpl} \>  Inner product of basis vectors: \\
\> $ \eta^{(a)(b)} = \text{diag} \left( {\rm e}^{(\,2\,{\rm a}(\,{T}\,)\,)},
  {\rm e}^{(\,2\,{\rm b}(\,{T}\,)\,)},
  {\rm e}^{(\,2\,{\rm c}(\,{T}\,)\,)},
  - {\rm e}^{(\,2\,{\rm a}(\,{T}\,) + 2\,{\rm b}(\,{T}
  \,) + 2\,{\rm c}(\,{T}\,)\,)} \right) $ \\
\> Basis vectors: \\
\> ${{ \omega }_{{1a}}}= $\>$ \left[ \! \, -\,{\displaystyle \frac {
  \sqrt {1 - { \Psi}^{2}}}{\sqrt {1 - { \Theta}^{2}}}}\, , { \Psi}\,
  \sqrt {1 - { \Theta}^{2}}\, , 0\, , 0\, \!  \right] $ \\
\> ${{ \omega }_{{2a}}}= $\>$ \left[ \! \, - \,{\displaystyle \frac {{ 
  \Psi}}{\sqrt {1 - { \Theta}^{2}}}}\,  , - \sqrt {1 - { \Theta}^{2}}
  \,\sqrt {1 - { \Psi}^{2}}\, , 0\, , 0\, \!  \right] $ \\
\> ${{ \omega }_{{3a}}}= $\>$ \left[ \! \,0 , \,{ \Theta}\, , {\displaystyle 
  \frac {1}{\sqrt {1 - { \Psi}^{2}}}}\, , 0\, \!  \right] $ \\
\> ${{ \omega }_{{4a}}}= $\>$ [\,0 , \,0 , \,0 , \,1\,] $\\
\vspace{1mm} \> Corresponding line element: \\
\> $\lefteqn{{\it \:ds}^{2}= - \,{\displaystyle \frac { \left( \! \,(
  \,{\rm e}^{{\rm a}(\,{T}\,)}\,)^{2} - (\,{\rm e}^{{\rm a}(\,{T}\,
  )}\,)^{2}\,{ \Psi}^{2} + { \Psi}^{2}\,(\,{\rm e}^{{\rm b}(\,{T}\,
  )}\,)^{2}\, \!  \right) \,{\it \:d}\,{ \Theta}^{{\it 2\:}}}{(\,{ 
  \Theta} - 1\,)\,(\,{ \Theta} + 1\,)}}} $ \\
\> $ \mbox{} + 2\,{ \Psi}\,\sqrt { - (\,{ \Psi} - 1\,)\,(\,{ \Psi
  } + 1\,)}\,(\,{\rm e}^{{\rm b}(\,{T}\,)} - {\rm e}^{{\rm a}(\,{T}
  \,)}\,)\,(\,{\rm e}^{{\rm b}(\,{T}\,)} + {\rm e}^{{\rm a}(\,{T}\,
  )}\,)\,{\it \:d}\,{ \Theta}^{{\:}}\,{\it d\:}\,{ \Phi}^{{\:}} + 
  \left( {\vrule height0.44em width0em depth0.44em} \right. \! \! $ \\
\> $ (\,{\rm e}^{{\rm a}(\,{T}\,)}\,)^{2}\,{ \Psi}^{2} - (\,{\rm 
  e}^{{\rm a}(\,{T}\,)}\,)^{2}\,{ \Psi}^{2}\,{ \Theta}^{2} + (\,
  {\rm e}^{{\rm b}(\,{T}\,)}\,)^{2} - { \Psi}^{2}\,(\,{\rm e}^{
  {\rm b}(\,{T}\,)}\,)^{2} - (\,{\rm e}^{{\rm b}(\,{T}\,)}\,)^{2}\,
  { \Theta}^{2} $ \\
\> $ \mbox{} + (\,{\rm e}^{{\rm b}(\,{T}\,)}\,)^{2}\,{ \Theta}^{2
  }\,{ \Psi}^{2} + { \Theta}^{2}\,(\,{\rm e}^{{\rm c}(\,{T}\,)}\,)
  ^{2} \! \! \left. {\vrule height0.44em width0em depth0.44em}
  \right) {\it \:d}\,{ \Phi}^{{\it 2\:}}\mbox{} + 2\,
  {\displaystyle \frac {{ \Theta}\,(\,{\rm e}^{{\rm c}(\,{T}\,)}\,)
  ^{2}\,{\it \:d}\,{ \Phi}^{{\:}}\,{\it d\:}\,{ \Psi}^{{\:}}}{
  \sqrt { - (\,{ \Psi} - 1\,)\,(\,{ \Psi} + 1\,)}}} $ \\
\> $ \mbox{} - {\displaystyle \frac {(\,{\rm e}^{{\rm c}(\,{T}\,)
  }\,)^{2}\,{\it \:d}\,{ \Psi}^{{\it 2\:}}}{(\,{ \Psi} - 1\,)\,(\,{
  \Psi} + 1\,)}} - (\,{\rm e}^{{\rm a}(\,{T}\,)}\,)^{2}\,(\,{\rm e
  }^{{\rm b}(\,{T}\,)}\,)^{2}\,(\,{\rm e}^{{\rm c}(\,{T}\,)}\,)^{2}
  \,{\it \:d}\,{T}^{{\it 2\:}} $ \\
%
%== Mixmaster 2 ===============================================================
\rule{\linewidth}{.5mm}\\
{\large Spacetime: Mixmaster ({\bf Mix2}) \cite{mtw2} } \\
\rule{\linewidth}{.05mm}\\
Input file\hspace{10mm} \= Frame \\
\rule[.5\baselineskip]{\linewidth}{.05mm}\\
{\tt mix2.mpl} \>  Inner product of basis vectors: \\
\> $\eta^{(a)(b)} = \text{diag} ( 1, 1, 1, -1 ) $ \\
\> Basis vectors: \\
\> $ {{ \omega }_{{1a}}}=[\,{\rm e}^{{\rm a}(\,{t}\,)}\,{\rm cos}(\,{ 
  \psi}\,) , \,{\rm e}^{{\rm a}(\,{t}\,)}\,{\rm sin}(\,{ \psi}\,)\,
  {\rm sin}(\,{ \theta}\,)\, , 0\, , 0\,] $ \\
\> ${{ \omega }_{{2a}}}= $\>$ [\,{\rm e}^{{\rm b}(\,{t}\,)}\,{\rm sin}(\,{ 
  \psi}\,)\, , -{\rm e}^{{\rm b}(\,{t}\,)}\,{\rm cos}(\,{ \psi}\,)\,
  {\rm sin}(\,{ \theta}\,)\, , 0\, , 0\,] $ \\
\> ${{ \omega }_{{3a}}}=$\>$[\,0 , \,{\rm e}^{{\rm c}(\,{t}\,)}\,{\rm cos}(
  \,{ \theta}\,)\, , {\rm e}^{{\rm c}(\,{t}\,)}\, , 0\,] $ \\
\> ${{ \omega }_{{4a}}}=$\>$[\,0 , \,0 , \,0 , \,{\rm e}^{(\,{\rm a}(\,{t}\,)
  + {\rm b}(\,{t}\,) + {\rm c}(\,{t}\,)\,)}\,] $ \\
\vspace{1mm} \> Corresponding line element: \\
\> $\lefteqn{{\it \:ds}^{2}= \left( \! \,{\rm cos}(\,{ \psi}\,)^{2}\,
  (\,{\rm e}^{{\rm a}(\,{t}\,)}\,)^{2} + {\rm sin}(\,{ \psi}\,)^{2}
  \,(\,{\rm e}^{{\rm b}(\,{t}\,)}\,)^{2}\, \!  \right) \,{\it \:d}
  \,{ \theta}^{{\it 2\:}}} $ \\
\> $ \mbox{} - 2\,{\rm sin}(\,{ \psi}\,)\,{\rm cos}(\,{ \psi}\,)
  \,{\rm sin}(\,{ \theta}\,)\,(\,{\rm e}^{{\rm b}(\,{t}\,)} - {\rm 
  e}^{{\rm a}(\,{t}\,)}\,)\,(\,{\rm e}^{{\rm b}(\,{t}\,)} + {\rm e}
  ^{{\rm a}(\,{t}\,)}\,)\,{\it \:d}\,{ \theta}^{{\:}}\,{\it d\:}\,{
  \phi}^{{\:}} + $ \\
\> $ \left( \! \,{\rm sin}(\,{ \psi}\,)^{2}\,{\rm sin}(\,{ 
  \theta}\,)^{2}\,(\,{\rm e}^{{\rm a}(\,{t}\,)}\,)^{2} + {\rm cos}(
  \,{ \psi}\,)^{2}\,{\rm sin}(\,{ \theta}\,)^{2}\,(\,{\rm e}^{{\rm 
  b}(\,{t}\,)}\,)^{2} + {\rm cos}(\,{ \theta}\,)^{2}\,(\,{\rm e}^{
  {\rm c}(\,{t}\,)}\,)^{2}\, \!  \right) \,{\it \:d} $ \\
\> $ { \phi}^{{\it 2\:}}\mbox{} + 2\,(\,{\rm e}^{{\rm c}(\,{t}\,)
  }\,)^{2}\,{\rm cos}(\,{ \theta}\,)\,{\it \:d}\,{ \phi}^{{\:}}\,
  {\it d\:}\,{ \psi}^{{\:}} + (\,{\rm e}^{{\rm c}(\,{t}\,)}\,)^{2}
  \,{\it \:d}\,{ \psi}^{{\it 2\:}}
  - (\,{\rm e}^{{\rm a}(\,{t}\,)}\,)^{2}\,(\,{\rm e}^{
  {\rm b}(\,{t}\,)}\,)^{2}\,(\,{\rm e}^{{\rm c}(\,{t}\,)}\,)^{2}\,
  {\it \:d}\,{t}^{{\it 2\:}} $ \\
%
%== Mixmaster 3 ===============================================================
\rule{\linewidth}{.5mm}\\
{\large Spacetime: Mixmaster ({\bf Mix3}) \cite{mtw2} } \\
\rule{\linewidth}{.05mm}\\
Input file\hspace{10mm} \= Frame \\
\rule[.5\baselineskip]{\linewidth}{.05mm}\\
{\tt mix2.mpl} \>  Inner product of basis vectors: \\
\> $\eta^{(a)(b)} = \text{diag} ( 1, 1, 1, -1 ) $ \\
\> Basis vectors: \\
\> ${{ \omega }_{{1a}}}= $\>$\left[ \! \, -\,{\displaystyle \frac {{\rm 
  e}^{{\rm a}(\,{T}\,)}\,\sqrt {1 - { \Psi}^{2}}}{\sqrt {1 - { 
  \Theta}^{2}}}}\,, {\rm e}^{{\rm a}(\,{T}\,)}\,{ \Psi}\,\sqrt {1 - {
  \Theta}^{2}}\, , 0\, , 0\, \!  \right] $ \\
\> ${{ \omega }_{{2a}}}=$\>$ \left[ \! \, - \,{\displaystyle \frac {{\rm 
  e}^{{\rm b}(\,{T}\,)}\,{ \Psi}}{\sqrt {1 - { \Theta}^{2}}}}\, ,
  -{\rm e}^{{\rm b}(\,{T}\,)}\,\sqrt {1 - { \Theta}^{2}}\,\sqrt {1
  - { \Psi}^{2}}\, , 0\, , 0\, \!  \right] $ \\
\> ${{ \omega }_{{3a}}}= \left[ \! \,0\, , { \Theta}\,{\rm e}^{{\rm c}(
  \,{T}\,)} , \,{\displaystyle \frac {{\rm e}^{{\rm c}(\,{T}\,)}}{
  \sqrt {1 - { \Psi}^{2}}}}\, , 0\, \!  \right] $ \\
\> $ {{ \omega }_{{4a}}}=[\,0 , \, 0\, , 0 , \,{\rm e}^{(\,{\rm a}(\,{T}\,) + 
  {\rm b}(\,{T}\,) + {\rm c}(\,{T}\,)\,)}\,] $ \\
\vspace{1mm} \> Corresponding line element: \\
\> $\lefteqn{{\it \:ds}^{2}={\displaystyle \frac { \left( \! \, - (\,
  {\rm e}^{{\rm a}(\,{T}\,)}\,)^{2} + (\,{\rm e}^{{\rm a}(\,{T}\,)}
  \,)^{2}\,{ \Psi}^{2} - { \Psi}^{2}\,(\,{\rm e}^{{\rm b}(\,{T}\,)}
  \,)^{2}\, \!  \right) \,{\it \:d}\,{ \Theta}^{{\it 2\:}}}{(\,{ 
  \Theta} - 1\,)\,(\,{ \Theta} + 1\,)}}} $ \\
\> $ \mbox{} - 2\,{ \Psi}\,\sqrt { - (\,{ \Psi} - 1\,)\,(\,{ \Psi
  } + 1\,)}\,(\,{\rm e}^{{\rm a}(\,{T}\,)} - {\rm e}^{{\rm b}(\,{T}
  \,)}\,)\,(\,{\rm e}^{{\rm b}(\,{T}\,)} + {\rm e}^{{\rm a}(\,{T}\,
  )}\,)\,{\it \:d}\,{ \Theta}^{{\:}}\,{\it d\:}\,{ \Phi}^{{\:}} + 
  \left( {\vrule height0.44em width0em depth0.44em} \right. \! \! $ \\
\> $ (\,{\rm e}^{{\rm a}(\,{T}\,)}\,)^{2}\,{ \Psi}^{2} - (\,{\rm 
  e}^{{\rm a}(\,{T}\,)}\,)^{2}\,{ \Psi}^{2}\,{ \Theta}^{2} + (\,
  {\rm e}^{{\rm b}(\,{T}\,)}\,)^{2} - { \Psi}^{2}\,(\,{\rm e}^{
  {\rm b}(\,{T}\,)}\,)^{2} - (\,{\rm e}^{{\rm b}(\,{T}\,)}\,)^{2}\,
  { \Theta}^{2} $ \\
\> $ \mbox{} + (\,{\rm e}^{{\rm b}(\,{T}\,)}\,)^{2}\,{ \Theta}^{2
  }\,{ \Psi}^{2} + { \Theta}^{2}\,(\,{\rm e}^{{\rm c}(\,{T}\,)}\,)
  ^{2} \! \! \left. {\vrule height0.44em width0em depth0.44em}
  \right) {\it \:d}\,{ \Phi}^{{\it 2\:}}\mbox{} + 2\,
  {\displaystyle \frac {{ \Theta}\,(\,{\rm e}^{{\rm c}(\,{T}\,)}\,)
  ^{2}\,{\it \:d}\,{ \Phi}^{{\:}}\,{\it d\:}\,{ \Psi}^{{\:}}}{
  \sqrt { - (\,{ \Psi} - 1\,)\,(\,{ \Psi} + 1\,)}}} $ \\
\> $ \mbox{} - {\displaystyle \frac {(\,{\rm e}^{{\rm c}(\,{T}\,)
  }\,)^{2}\,{\it \:d}\,{ \Psi}^{{\it 2\:}}}{(\,{ \Psi} - 1\,)\,(\,{
  \Psi} + 1\,)}} - (\,{\rm e}^{{\rm a}(\,{T}\,)}\,)^{2}\,(\,{\rm e
  }^{{\rm b}(\,{T}\,)}\,)^{2}\,(\,{\rm e}^{{\rm c}(\,{T}\,)}\,)^{2}
  \,{\it \:d}\,{T}^{{\it 2\:}} $ \\
\end{tabbing}
\rule{\linewidth}{.5mm}\\
%==============================================================================
% Appendix C: Sample input file npupkn5.mpl and sample of output from kne1.ms
%==============================================================================
\pagebreak
%------------------------------------------------------------------------------
\section{Sample input file {\tt npupkn5.mpl} and sample of output from 
{\tt kne1.ms}}\label{app:C}
%------------------------------------------------------------------------------
The following is the input file:
\begin{center}
\begin{minipage}{.8\linewidth}
\begin{verbatim}
Ndim_ :=    4   :
x1_   :=   t   :
x2_   :=   r   :
x3_   :=   u   :
x4_   :=   phi   :
eta12_   :=   1   :
eta34_   :=   -1   :
b11_   :=   (r^2+a^2)/(r^2-2*M*r+a^2+Q^2)   :
b12_   :=   1   :
b14_   :=   a/(r^2-2*M*r+a^2+Q^2)   :
b21_   :=   1/2*(r^2+a^2)/(r^2+u^2)   :
b22_   :=   -1/2*(r^2-2*M*r+a^2+Q^2)/(r^2+u^2)   :
b24_   :=   1/2*a/(r^2+u^2)   :
b31_   :=   1/2*(I*(a^2-u^2)^(1/2)*r+(a^2-u^2)^(1/2)*u)*2^(1/2)/(r^2+u^2)   :
b33_   :=   1/2*(-(a^2-u^2)^(1/2))*(r-I*u)*2^(1/2)/(r^2+u^2)   :
b34_   :=   1/2*I*(r-I*u)*a/(a^2-u^2)^(1/2)*2^(1/2)/(r^2+u^2):
b41_   :=   1/2*(-I*(a^2-u^2)^(1/2)*r+(a^2-u^2)^(1/2)*u)*2^(1/2)/(r^2+u^2)   :
b43_   :=   1/2*(-(a^2-u^2)^(1/2))*(r+I*u)*2^(1/2)/(r^2+u^2)   :
b44_   :=   -1/2*I*a*(r+I*u)/(a^2-u^2)^(1/2)*2^(1/2)/(r^2+u^2):
Info_:=`Contravariant NPtetrad for Kerr-Newman metric 
        (u=a*cos(theta) to Boyer-Lindquist coordinates)`:
\end{verbatim}
\end{minipage}
\end{center}
The following is the annotated input/output of {\tt kne1.ms}.

\begin{mapleinput}
restart:
\end{mapleinput}
\begin{mapleinput}
readlib(grii):
\end{mapleinput}
\begin{mapleinput}
grtensor();
\end{mapleinput}
\begin{maplelatex}
\[
{\it GRTensorII\:Version\:1.26a}
\]
\end{maplelatex}
\begin{maplelatex}
\[
{\it December\:1,\:1995}
\]
\end{maplelatex}
\begin{maplelatex}
\[
{\it Developed\:by\:Peter\:Musgrave,\:Denis\:Pollney\:and\:Kayll
\:Lake}
\]
\end{maplelatex}
\begin{maplelatex}
\[
{\it Copyright\:1994-1995\:by\:the\:authors.}
\]
\end{maplelatex}
\begin{maplelatex}
\[
{\it Latest\:version\:available\:from:\:http://astro.queensu.ca/
\char'176GRHome.html}
\]
\end{maplelatex}
\begin{maplelatex}
\[
{\it To\:initiate\:help\:type\:?grtensor}
\]
\end{maplelatex}
\begin{maplelatex}
\[
{\it Defaults\:read\:from\:C:\backslash MAPLEV3\backslash
lib/grtensor.ini}
\]
\end{maplelatex}
We now load the contravariant tetrad,

calculate the scalars, and factor them.

\begin{mapleinput}
qload(npupkn5);
\end{mapleinput}
\begin{maplelatex}
\[
{\it Default\:spacetime}={\it npupkn5}
\]
\end{maplelatex}
\begin{maplelatex}
\[
{\it For\:the\:npupkn5\:spacetime:}
\]
\end{maplelatex}
\begin{maplelatex}
\[
{\it Coordinates}
\]
\end{maplelatex}
\begin{maplelatex}
\[
{\it x\:}^{{a}}=[\,{t}\,{r}\,{u}\,{ \phi}\,]
\]
\end{maplelatex}
\begin{maplelatex}
\[
{\it Basis\:inner\:product}
\]
\end{maplelatex}
\begin{maplelatex}
\[
{ \eta}^{{\it (a)}}\,{}^{{\it (b)}}= \left[ 
{\begin{array}{rrrr}
0 & 1 & 0 & 0 \\
1 & 0 & 0 & 0 \\
0 & 0 & 0 & -1 \\
0 & 0 & -1 & 0
\end{array}}
 \right] 
\]
\end{maplelatex}
\begin{maplelatex}
\[
{\it Null\:tetrad\:(contravariant\:components)}
\]
\end{maplelatex}
\begin{maplelatex}
\[
{l}^{{a}}= \left[ \! \,{\displaystyle \frac {{r}^{2} + {a}^{2}}{{
r}^{2} - 2\,{M}\,{r} + {a}^{2} + {Q}^{2}}}\,1\,0\,{\displaystyle 
\frac {{a}}{{r}^{2} - 2\,{M}\,{r} + {a}^{2} + {Q}^{2}}}\, \! 
 \right] 
\]
\end{maplelatex}
\begin{maplelatex}
\[
{n}^{{a}}= \left[ \! \,{\displaystyle \frac {1}{2}}\,
{\displaystyle \frac {{r}^{2} + {a}^{2}}{{r}^{2} + {u}^{2}}}\, - 
\,{\displaystyle \frac {1}{2}}\,{\displaystyle \frac {{r}^{2} - 2
\,{M}\,{r} + {a}^{2} + {Q}^{2}}{{r}^{2} + {u}^{2}}}\,0\,
{\displaystyle \frac {1}{2}}\,{\displaystyle \frac {{a}}{{r}^{2}
 + {u}^{2}}}\, \!  \right] 
\]
\end{maplelatex}
\begin{maplelatex}
\begin{eqnarray*}
\lefteqn{{m}^{{a}}= \left[ {\vrule 
height0.89em width0em depth0.89em} \right. \! \! {\displaystyle 
\frac {1}{2}}\,{\displaystyle \frac { \left( \! \,{I}\,\sqrt {{a}
^{2} - {u}^{2}}\,{r} + \sqrt {{a}^{2} - {u}^{2}}\,{u}\, \! 
 \right) \,\sqrt {2}}{{r}^{2} + {u}^{2}}}\,0\, - \,
{\displaystyle \frac {1}{2}}\,{\displaystyle \frac {\sqrt {{a}^{2
} - {u}^{2}}\,(\,{r} - {I}\,{u}\,)\,\sqrt {2}}{{r}^{2} + {u}^{2}
}}} \\
 & & {\displaystyle \frac {1}{2}}\,{\displaystyle \frac {{I}\,(\,
{r} - {I}\,{u}\,)\,{a}\,\sqrt {2}}{\sqrt {{a}^{2} - {u}^{2}}\,(\,
{r}^{2} + {u}^{2}\,)}} \! \! \left. {\vrule 
height0.89em width0em depth0.89em} \right] \mbox{\hspace{195pt}}
\end{eqnarray*}
\end{maplelatex}
\begin{maplelatex}
\begin{eqnarray*}
\lefteqn{{\it mbar}^{{a}}= \left[ {\vrule 
height0.89em width0em depth0.89em} \right. \! \! {\displaystyle 
\frac {1}{2}}\,{\displaystyle \frac { \left( \! \, - {I}\,\sqrt {
{a}^{2} - {u}^{2}}\,{r} + \sqrt {{a}^{2} - {u}^{2}}\,{u}\, \! 
 \right) \,\sqrt {2}}{{r}^{2} + {u}^{2}}}\,0\, - \,
{\displaystyle \frac {1}{2}}\,{\displaystyle \frac {\sqrt {{a}^{2
} - {u}^{2}}\,(\,{r} + {I}\,{u}\,)\,\sqrt {2}}{{r}^{2} + {u}^{2}
}}} \\
 & &  - \,{\displaystyle \frac {1}{2}}\,{\displaystyle \frac {{I}
\,{a}\,(\,{r} + {I}\,{u}\,)\,\sqrt {2}}{\sqrt {{a}^{2} - {u}^{2}}
\,(\,{r}^{2} + {u}^{2}\,)}} \! \! \left. {\vrule 
height0.89em width0em depth0.89em} \right] \mbox{\hspace{212pt}}
\end{eqnarray*}
\end{maplelatex}
\begin{maplelatex}
\begin{eqnarray*}
\lefteqn{{\it Contravariant\:NPtetrad\:for\:Kerr-Newman\:metric\:
(u=a*cos(theta)\:to\:Boye\backslash}} \\
 & & {\it r-Lindquist\:coordinates\:(J.\:Math.\:Phys.\:8\:265))}
\mbox{\hspace{94pt}}
\end{eqnarray*}
\end{maplelatex}
\begin{mapleinput}
grcalc(RicciSc,WeylSc);
\end{mapleinput}
\begin{mapleinput}
gralter(_,radical,factor);
\end{mapleinput}
\begin{mapleinput}
grdisplay(_);
\end{mapleinput}
\begin{maplelatex}
\[
{\it For\:the\:npupkn5\:spacetime:}
\]
\end{maplelatex}
\begin{maplelatex}
\[
{ \Phi 00}=0
\]
\end{maplelatex}
\begin{maplelatex}
\[
{ \Phi 01}=0
\]
\end{maplelatex}
\begin{maplelatex}
\[
{ \Phi 02}=0
\]
\end{maplelatex}
\begin{maplelatex}
\[
{ \Phi 11}={\displaystyle \frac {1}{2}}\,{\displaystyle \frac {{Q
}^{2}}{(\,{r} + {I}\,{u}\,)^{2}\,(\,{r} - {I}\,{u}\,)^{2}}}
\]
\end{maplelatex}
\begin{maplelatex}
\[
{ \Phi 12}=0
\]
\end{maplelatex}
\begin{maplelatex}
\[
{ \Phi 22}=0
\]
\end{maplelatex}
\begin{maplelatex}
\[
{ \Lambda}=0
\]
\end{maplelatex}
\begin{maplelatex}
\[
{ \Psi 0}=0
\]
\end{maplelatex}
\begin{maplelatex}
\[
{ \Psi 1}=0
\]
\end{maplelatex}
\begin{maplelatex}
\[
{ \Psi 2}={\displaystyle \frac {{Q}^{2} - {M}\,{r} - {I}\,{u}\,{M
}}{(\,{r} + {I}\,{u}\,)\,(\,{r} - {I}\,{u}\,)^{3}}}
\]
\end{maplelatex}
\begin{maplelatex}
\[
{ \Psi 3}=0
\]
\end{maplelatex}
\begin{maplelatex}
\[
{ \Psi 4}=0
\]
\end{maplelatex}
We now generate the metric and simplify it.

\begin{mapleinput}
grcalc(g(dn,dn));
\end{mapleinput}
\begin{mapleinput}
gralter(_,simplify,factor);
\end{mapleinput}
\begin{mapleinput}
grcalc(ds);
\end{mapleinput}
\begin{mapleinput}
grdisplay(_);
\end{mapleinput}
\begin{maplelatex}
\[
{\it For\:the\:npupkn5\:spacetime:}
\]
\end{maplelatex}
\begin{maplelatex}
\[
{\it Line\:element}
\]
\end{maplelatex}
\begin{maplelatex}
\begin{eqnarray*}
\lefteqn{{\it \:ds}^{2}={\displaystyle \frac {(\,{r}^{2} - 2\,{M}
\,{r} + {u}^{2} + {Q}^{2}\,)\,{\it \:d}\,{t}^{{\it 2\:}}}{{r}^{2}
 + {u}^{2}}} - 2\,{\displaystyle \frac {(\, - 2\,{M}\,{r} + {Q}^{
2}\,)\,(\,{a} - {u}\,)\,(\,{a} + {u}\,)\,{\it \:d}\,{t}^{{\:}}\,
{\it d\:}\,{ \phi}^{{\:}}}{{a}\,(\,{r}^{2} + {u}^{2}\,)}}} \\
 & & \mbox{} - {\displaystyle \frac {(\,{r}^{2} + {u}^{2}\,)\,
{\it \:d}\,{r}^{{\it 2\:}}}{{r}^{2} - 2\,{M}\,{r} + {a}^{2} + {Q}
^{2}}} - {\displaystyle \frac {(\,{r} - {I}\,{u}\,)\,(\,{r} + {I}
\,{u}\,)\,{\it \:d}\,{u}^{{\it 2\:}}}{(\,{a} - {u}\,)\,(\,{a} + {
u}\,)}} +  \\
 & & (\, - {r}^{4} - {u}^{2}\,{r}^{2} - {r}^{2}\,{a}^{2} + 2\,{u}
^{2}\,{M}\,{r} - 2\,{r}\,{M}\,{a}^{2} + {Q}^{2}\,{a}^{2} - {u}^{2
}\,{a}^{2} - {u}^{2}\,{Q}^{2}\,)\,(\,{a} - {u}\,) \\
 & & (\,{a} + {u}\,)\,{\it \:d}\,{ \phi}^{{\it 2\:}} \left/ 
{\vrule height0.37em width0em depth0.37em} \right. \! \! (\,(\,{r
}^{2} + {u}^{2}\,)\,{a}^{2}\,)\mbox{\hspace{196pt}}
\end{eqnarray*}
\end{maplelatex}
The coordinate components of Ricci and Weyl are calculated and simplified.

(Display deleted for this appendix.)

\begin{mapleinput}
grcalc(g(up,up));
\end{mapleinput}
\begin{mapleinput}
gralter(_,radical,expand,factor);
\end{mapleinput}
\begin{mapleinput}
grcalc(R(dn,dn),Ricciscalar);
\end{mapleinput}
\begin{mapleinput}
gralter(_,expand,factor);
\end{mapleinput}
\begin{mapleinput}
grcalc(C(dn,dn,dn,dn));
\end{mapleinput}
\begin{mapleinput}
gralter(_,expand,factor);
\end{mapleinput}

The following lines define the Weyl Scalars.

\begin{mapleinput}
grdefine(`P0`,{},`-C{a b c d}*NPl{^a}*NPm{^b}*NPl{^c}*NPm{^d}`);
\end{mapleinput}
\begin{maplettyout}
Created definition for P0

\end{maplettyout}
\begin{mapleinput}
grdefine(`P1`,{},`-C{a b c d}*NPl{^a}*NPn{^b}*NPl{^c}*NPm{^d}`);
\end{mapleinput}
\begin{maplettyout}
Created definition for P1

\end{maplettyout}
\begin{mapleinput}
grdefine(`P2`,{},`-C{a b c d}*(NPl{^a}*NPn{^b}*NPl{^c}*NPn{^d}-NPl{^a}*NPn{^b}*NPm{^c}*NPmbar{^d})/2`);
\end{mapleinput}
\begin{maplettyout}
Created definition for P2

\end{maplettyout}
\begin{mapleinput}
grdefine(`P3`,{},`-C{a b c d}*NPn{^a}*NPl{^b}*NPn{^c}*NPmbar{^d}`);
\end{mapleinput}
\begin{maplettyout}
Created definition for P3

\end{maplettyout}
\begin{mapleinput}
grdefine(`P4`,{},`-C{a b c d}*NPn{^a}*NPmbar{^b}*NPn{^c}*NPmbar{^d}`);
\end{mapleinput}
\begin{maplettyout}
Created definition for P4

\end{maplettyout}
The following lines define the Ricci Scalars.

\begin{mapleinput}
grdefine(`P00`,{},`R{a b}*NPl{^a}*NPl{^b}/2`);
\end{mapleinput}
\begin{maplettyout}
Created definition for P00

\end{maplettyout}
\begin{mapleinput}
grdefine(`P01`,{},`R{a b}*NPl{^a}*NPm{^b}/2`);
\end{mapleinput}
\begin{maplettyout}
Created definition for P01

\end{maplettyout}
\begin{mapleinput}
grdefine(`P02`,{},`R{a b}*NPm{^a}*NPm{^b}/2`);
\end{mapleinput}
\begin{maplettyout}
Created definition for P02

\end{maplettyout}
\begin{mapleinput}
grdefine(`P11`,{},`R{a b}*(NPl{^a}*NPn{^b}+NPm{^a}*NPmbar{^b})/4`);
\end{mapleinput}
\begin{maplettyout}
Created definition for P11

\end{maplettyout}
\begin{mapleinput}
grdefine(`P12`,{},`R{a b}*NPn{^a}*NPm{^b}/2`);
\end{mapleinput}
\begin{maplettyout}
Created definition for P12

\end{maplettyout}
\begin{mapleinput}
grdefine(`P22`,{},`R{a b}*NPn{^a}*NPn{^b}/2`);
\end{mapleinput}
\begin{maplettyout}
Created definition for P22

\end{maplettyout}
We calculate and simplify the Weyl scalars.

\begin{mapleinput}
grcalc(P0,P1,P2,P3,P4);
\end{mapleinput}
\begin{mapleinput}
gralter(_,factor);
\end{mapleinput}
\begin{mapleinput}
grdisplay(_);
\end{mapleinput}
\begin{maplelatex}
\[
{\it For\:the\:npupkn5\:spacetime:}
\]
\end{maplelatex}
\begin{maplelatex}
\[
{\it P0}=0
\]
\end{maplelatex}
\begin{maplelatex}
\[
{\it P1}=0
\]
\end{maplelatex}
\begin{maplelatex}
\[
{\it P2}={\displaystyle \frac {{Q}^{2} - {M}\,{r} - {I}\,{u}\,{M}
}{(\,{r} + {I}\,{u}\,)\,(\,{r} - {I}\,{u}\,)^{3}}}
\]
\end{maplelatex}
\begin{maplelatex}
\[
{\it P3}=0
\]
\end{maplelatex}
\begin{maplelatex}
\[
{\it P4}=0
\]
\end{maplelatex}
We calculate and simplify the Ricci scalars.

\begin{mapleinput}
grcalc(P00,P01,P02,P11,P12,P22);
\end{mapleinput}
\begin{mapleinput}
gralter(_,factor);
\end{mapleinput}
\begin{mapleinput}
grdisplay(_);
\end{mapleinput}
\begin{maplelatex}
\[
{\it For\:the\:npupkn5\:spacetime:}
\]
\end{maplelatex}
\begin{maplelatex}
\[
{\it P00}=0
\]
\end{maplelatex}
\begin{maplelatex}
\[
{\it P01}=0
\]
\end{maplelatex}
\begin{maplelatex}
\[
{\it P02}=0
\]
\end{maplelatex}
\begin{maplelatex}
\[
{\it P11}={\displaystyle \frac {1}{2}}\,{\displaystyle \frac {{Q}
^{2}}{(\,{r}^{2} + {u}^{2}\,)^{2}}}
\]
\end{maplelatex}
\begin{maplelatex}
\[
{\it P12}=0
\]
\end{maplelatex}
\begin{maplelatex}
\[
{\it P22}=0
\]
\end{maplelatex}
Unfortunately, the output obtained from each of these methods are
not in exactly the same form. Since {\tt P2} is
smaller as measured in Maple {\em words}, this must be considered
the fully simplified form. In fact, by performing the MapleV
operation {\tt factor} on the denominators of $\Psi_2$
(an operation requiring less than $.1$ CPU seconds), it
can be reduced to the required fully simplified form. In all tests listed
in Appendix \ref{app:A} the final results of calculation are either
presented in exactly equivalent forms, or equivalent
within some simplification operation requiring a negligible amount of
time.
%==============================================================================
\end{sloppypar}
\end{document}